\documentclass[aps,prd,showpacs,notitlepage,nofootinbib,preprintnumbers,amsmath,amssymb]{revtex4-2}

\usepackage[dvipsnames]{xcolor}
\usepackage{graphics,graphicx}
\usepackage{dcolumn}
\usepackage{bm}
\usepackage{epsfig}
\usepackage{hyperref} 
\usepackage{mathbbol}
\usepackage{epstopdf}
\usepackage{simplewick}
\usepackage[utf8]{inputenc} 
\usepackage{slashed}
\usepackage{stackrel}
\usepackage{mathrsfs}
\usepackage{xcolor}
\usepackage{gensymb}

\usepackage{tikz}
\usetikzlibrary{decorations.pathmorphing, decorations.markings, shapes.geometric, shapes.misc, calc, math}
\tikzstyle{gluon}=[decorate, decoration={coil,aspect=0.8, amplitude=1.5pt,  segment length=3pt}]

\usepackage{stackrel}
\usepackage[most]{tcolorbox}

\def\eq#1{{Eq.~(\ref{#1})}}
\def\fig#1{{Fig.~\ref{#1}}}
\newcommand{\ben}{\begin{eqnarray*}}
\newcommand{\een}{\end{eqnarray*}}
\newcommand{\un}[1]{\underline{#1}}

\newcommand{\pd}{\partial}

\newcommand{\tr}{\mbox{tr}}
\newcommand{\thalf}{\tfrac{1}{2}}
\newcommand{\llangle}{\Big\langle \!\! \Big\langle}
\newcommand{\rrangle}{\Big\rangle \!\! \Big\rangle}
\newcommand{\cc}{\mbox{\mbox{c.c.}}}

\newcommand{\as}{\alpha_s}

\newcommand{\dhd}{{\textstyle d}
\lower.03ex\hbox{\kern-0.38em$^{\scriptstyle-}$}\kern-0.05em{}}
\newcommand{\dbar}{{\textstyle \delta}
\lower.03ex\hbox{\kern-0.38em$^{\scriptstyle-}$}\kern-0.05em{}}
\newcommand{\half}{{1\over 2}}

\newcommand{\bra}[1]{\left\langle #1 \right|}
\newcommand{\ket}[1]{\left| #1 \right\rangle}

\newcommand{\ul}[1]{\underline{#1}}

\newcommand{\tord}{\textrm{T} \:}
\newcommand{\atord}{\bar{\textrm{T}} \:}

\makeatletter
\DeclareRobustCommand{\cev}[1]{%
  {\mathpalette\do@cev{#1}}%
}
\newcommand{\do@cev}[2]{%
  \vbox{\offinterlineskip
    \sbox\z@{$\m@th#1 x$}%
    \ialign{##\cr
      \hidewidth\reflectbox{$\m@th#1\vec{}\mkern4mu$}\hidewidth\cr
      \noalign{\kern-\ht\z@}
      $\m@th#1#2$\cr
    }%
  }%
}
\makeatother

\begin{document}

\title{Weizs\"acker-Williams Gluon Helicity Distribution and Inclusive Dijet Production \\ in Longitudinally Polarized Electron-Proton Collisions}

\author{Yuri V. Kovchegov} 
         \email[Email: ]{kovchegov.1@osu.edu}
         \affiliation{Department of Physics, The Ohio State
           University, Columbus, OH 43210, USA}

\author{Ming Li} 
         \email[Email: ]{li.13449@osu.edu}
         \affiliation{Department of Physics, The Ohio State
           University, Columbus, OH 43210, USA}

\begin{abstract}
It is well-known that the back-to-back (correlation) limit of inclusive quark--antiquark dijet production in unpolarized high energy electron--proton collisions can probe the Weizs\"{a}cker-Williams (WW) gluon transverse momentum-dependent distribution (TMD) at small $x$ \cite{Dominguez:2010xd, Dominguez:2011wm}. In this paper, we consider a helicity-dependent version of the same process: we study the double-spin asymmetry for inclusive quark--antiquark dijet production in longitudinally polarized electron--proton 
scattering at high energies. We show that in the back-to-back limit this process  
probes the WW gluon \textit{helicity} TMD. Furthermore, we derive the small-$x$ evolution equation for the operator related to the WW  
gluon helicity distribution. 
We find that in the double-logarithmic approximation and in the large-$N_c$ limit, the small-$x$ asymptotics of the WW gluon helicity distribution is governed by exactly the same evolution equation as that for the \textit{dipole} gluon helicity distribution. 
The longitudinal double-spin asymmetry for inclusive dijet production in the longitudinally polarized electron--proton collisions can thus test the small-$x$ helicity evolution equations and facilitate constraining the initial conditions for phenomenology based on these equations. 
\end{abstract}


\maketitle

\tableofcontents

\section{Introduction}

Inclusive dijet (or di-hadron) production in high-energy unpolarized electron-proton/nucleus collisions is considered to be one of the key processes to study gluon saturation at the future Electron-Ion Collider (EIC)  \cite{Accardi:2012qut,Boer:2011fh,Proceedings:2020eah,AbdulKhalek:2021gbh}. Specifically, the back-to-back azimuthal correlations of dijets (or di-hadrons) are expected to be suppressed due to multiple gluon exchanges in the gluon saturation regime \cite{Kharzeev:2004bw, Dominguez:2010xd, Dominguez:2011wm}. In the back-to-back limit, it is also well known that inclusive dijet production in high-energy deep inelastic scattering (DIS) can probe the Weizs\"{a}cker-Williams (WW) gluon transverse momentum-dependent distribution (TMD) \cite{Dominguez:2010xd, Dominguez:2011wm}. Despite the WW gluon TMD having a clear physical interpretation of the gluon number density in (some sub-gauges of) the light-cone gauge of the target proton or nucleus \cite{Jalilian-Marian:1996mkd}, in the small Bjorken-$x$ regime the most common distribution that enters various production cross sections is the \textit{dipole} gluon TMD \cite{Braun:2000bh, Kovchegov:2001sc, Kharzeev:2003wz, Dominguez:2010xd, Dominguez:2011wm, Kovchegov:2012mbw}. These two gluon TMDs, the dipole one and the WW one, differ in the shape of the Wilson line staple in their definitions \cite{Dominguez:2010xd, Dominguez:2011wm}: the WW gluon TMD definition contains the future-pointing adjoint Wilson line staple while the dipole gluon TMD is defined using one future-pointing and one past-pointing fundamental gauge links. Furthermore, the dipole gluon TMD is related to the dipole scattering amplitude on the target, whose small-$x$ behavior at large $N_c$ \cite{tHooft:1973alw} (with $N_c$ the number of quark colors) is governed by the Balitsky-Kovchegov (BK) evolution equation \cite{Balitsky:1995ub,Balitsky:1998ya,Kovchegov:1999yj,Kovchegov:1999ua} and by the Jalilian-Marian--Iancu--McLerran--Weigert--Leonidov--Kovner
(JIMWLK)
\cite{Jalilian-Marian:1997dw,Jalilian-Marian:1997gr,Weigert:2000gi,Iancu:2001ad,Iancu:2000hn,Ferreiro:2001qy} functional evolution equation at all $N_c$ (see \cite{Gribov:1984tu, Iancu:2003xm, Weigert:2005us, JalilianMarian:2005jf, Gelis:2010nm, Albacete:2014fwa, Kovchegov:2012mbw, Morreale:2021pnn} for reviews of saturation physics). In contrast, the WW gluon TMD is related to a different operator made out of the light-cone Wilson lines, whose small-$x$ evolution equation is not closed even at large $N_c$ and contains different Wilson line structures on its right-hand side \cite{Dominguez:2011gc, Dominguez:2011br}. The Wilson-line operator related the WW gluon TMD can be obtained by an appropriate limit of the quadrupole scattering amplitude, whose evolution is known \cite{Jalilian-Marian:2004vhw, Dominguez:2011gc}: this evolution is closed in the large-$N_c$ limit. Recent developments extending inclusive dijet production in unpolarized DIS to the next-to-leading order in the QCD coupling constant can be found in \cite{Caucal:2021ent, Caucal:2022ulg, Caucal:2023nci, Caucal:2023fsf, Taels:2022tza}. 

Over the past decade, a significant amount of work in the small-$x$ community has been dedicated to understanding the sub-eikonal corrections to the distribution functions and cross sections \cite{Altinoluk:2014oxa,Balitsky:2015qba,Balitsky:2016dgz, Kovchegov:2017lsr, Kovchegov:2018znm, Chirilli:2018kkw, Jalilian-Marian:2018iui, Jalilian-Marian:2019kaf, Altinoluk:2020oyd, Boussarie:2020vzf, Boussarie:2020fpb, Kovchegov:2021iyc, Altinoluk:2021lvu, Kovchegov:2022kyy, Altinoluk:2022jkk, Altinoluk:2023qfr,Altinoluk:2023dww, Li:2023tlw, Altinoluk:2024dba}. The sub-eikonal corrections are suppressed by a power of the center-of-mass energy $s$ in the cross sections or by a multiplicative power of the longitudinal momentum fraction $x$ in the distribution functions. While the leading (eikonal) high-energy contributions to many observables does not couple to the longitudinal spin of the proton target, the sub-eikonal corrections are, indeed, sensitive to the longitudinal polarization of the proton. This sensitivity allowed for development of small-$x$ evolution equations for helicity parton distribution functions (helicity PDFs or hPDFs) and for helicity TMDs in \cite{Kovchegov:2015pbl, Hatta:2016aoc, Kovchegov:2016zex, Kovchegov:2016weo, Kovchegov:2017jxc, Kovchegov:2017lsr, Kovchegov:2018znm, Kovchegov:2019rrz, Boussarie:2019icw, Cougoulic:2019aja, Kovchegov:2020hgb, Cougoulic:2020tbc, Chirilli:2021lif, Kovchegov:2021lvz, Cougoulic:2022gbk, Borden:2023ugd, Adamiak:2023okq, Borden:2024bxa} (see also the earlier work  \cite{Bartels:1995iu, Bartels:1996wc} in the infrared evolution equations (IREE) framework \cite{Gorshkov:1966ht,Kirschner:1983di,Kirschner:1994rq,Kirschner:1994vc,Griffiths:1999dj, Blumlein:1996hb, Blumlein:1995jp}). The helicity PDFs and TMDs at small $x$ are related to the correlation functions of the so-called polarized Wilson lines, with the latter being the regular infinite light-cone Wilson lines with a single or double insertion of a sub-eikonal gluon or quark operators, respectively. Polarized dipole amplitudes are constructed out of a trace of a polarized Wilson line with a (conjugate of the) regular eikonal Wilson line \cite{Kovchegov:2016zex, Kovchegov:2018znm, Chirilli:2021lif, Cougoulic:2022gbk}: the hPDFs at small $x$ can be shown to depend on these structures \cite{Hatta:2016aoc, Kovchegov:2016zex, Cougoulic:2022gbk, Borden:2024bxa}.

Similarly to the unpolarized case, for the gluon helicity distribution in a longitudinally polarized proton/nucleus, one can also define two distributions: the WW gluon helicity TMD and the dipole gluon helicity TMD \cite{Hatta:2016aoc,Kovchegov:2017lsr}. The small-$x$ limit of the dipole gluon helicity TMD, which is related to the 
polarized dipole amplitude $G^i_{10}$ made out of the polarized Wilson line with the insertion of the $D^i - \cev{D}^i$ sub-eikonal operator and a regular light-cone Wilson line (with $D^i = \pd^i - i g A^i$ and $\cev{D}^i = \cev{\pd}^i + i g A^i$ the right- and left-acting fundamental covariant derivatives, $g$ the QCD coupling, $A^\mu$ the gluon field, and $i=1,2$ the transverse Lorentz index) has been widely studied in the past decade. Specifically, the small-$x$ evolution equations for $G^i_{10}$ have been derived in \cite{Kovchegov:2015pbl,Kovchegov:2018znm,Cougoulic:2022gbk, Borden:2024bxa}. In the large-$N_c$ limit and under the double-logarithmic approximation (DLA), which re-sums powers of $\alpha_s \ln^2(1/x)$ with $\as = g^2/4 \pi$, the small-$x$ evolution equations for $G^i_{10}$ mix with another polarized dipole amplitude $G_{10}$, obtained by using an insertion of the sub-eikonal gluon field-strength operator $F^{12}$. The large-$N_c$ DLA evolution of  $G^i_{10}$ and $G_{10}$ is a closed system of integral equations. 
The analytic solution to these equations has been derived in \cite{Borden:2023ugd} and found to agree with the small-$x$ large-$N_c$ limit of the helicity DGLAP equation for all three known loops \cite{Altarelli:1977zs,Dokshitzer:1977sg,Zijlstra:1993sh,Mertig:1995ny,Moch:1999eb,vanNeerven:2000uj,Vermaseren:2005qc,Moch:2014sna,Blumlein:2021ryt,Blumlein:2021lmf,Davies:2022ofz,Blumlein:2022gpp}. 
In the large-$N_c\& N_f$ limit (with $N_f$ the number of quark flavors) and also under DLA, the closed system of small-$x$ evolution equations was obtained in \cite{Borden:2024bxa} for a slightly expanded set of polarized dipole amplitudes (see also \cite{Kovchegov:2015pbl, Kovchegov:2018znm, Chirilli:2021lif,  Cougoulic:2022gbk}). Phenomenologically, the dipole gluon helicity TMD and the gluon hPDF contribute to various observables involving longitudinally polarized proton/nucleus at high collisional energies: the $g_1$ structure function
\cite{Cougoulic:2022gbk}, the $g_1^h$ structure function in the semi-inclusive DIS (SIDIS) \cite{Adamiak:2023yhz}, the double-spin asymmetry for hadron (gluon) production in the longitudinally polarized proton-proton collisions \cite{Kovchegov:2024aus}, the single-spin asymmetry in elastic dijet production in the electron--proton collisions with a longitudinally polarized proton \cite{Hatta:2016aoc, Kovchegov:2024wjs}, and the double-spin asymmetry in elastic dijet production in the longitudinally polarized electron--proton collisions \cite{Hatta:2016aoc,Bhattacharya:2022vvo, Bhattacharya:2023hbq, Bhattacharya:2024sck, Kovchegov:2024wjs}.

On the other hand, the WW gluon helicity TMD in the small-$x$ regime has been rarely studied in the past (see, e.g., \cite{Hatta:2016aoc, Kovchegov:2017lsr}). The goal of this paper is to identify a potential scattering process that probes the WW gluon helicity TMD, to derive the small-$x$ helicity evolution equations for the WW gluon helicity TMD, and obtain its small-$x$ asymptotics. Below, working in the small-$x$ limit, we show that the WW gluon helicity TMD is related to the correlator containing a polarized Wilson line with the insertion of the $D^i - \cev{D}^i$ sub-eikonal operator, along with several eikonal light-cone Wilson lines. We then consider longitudinally polarized electron scattering on a longitudinally polarized proton/nucleus at high collisional energies. We focus on the double-spin asymmetry for inclusive quark-antiquark dijet production, particularly its back-to-back (correlation) limit in which the mean transverse momentum of the quark-antiquark pair $p_T$  is much larger than their momentum imbalance $\Delta_{\perp}$. 
The double spin asymmetry is defined by 
\begin{equation}
A_{LL} = \frac{d\sigma(++) - d\sigma(+-)}{d\sigma (++) + d\sigma (+-)},
\end{equation}
in which $(++)$ indicates that both the electron and the proton have positive helicities while $(+-)$ indicates that the helicity state of the proton is negative. Below we calculate the numerator of $A_{LL}$, keeping only the leading contribution appearing at the sub-eikonal ($\sim 1/s$) order. We find that, at the leading order in $\Delta_{\perp}/p_T$, the numerator of $A_{LL}$, averaged over the angles of the electron's transverse momentum in the ``dipole" frame, uniquely probes the small $x$ WW gluon helicity TMD. (There is no contribution from the linearly polarized gluon distribution \cite{Metz:2011wb, Dominguez:2011br} in longitudinally polarized proton at the sub-eikonal order.) The denominator of $A_{LL}$ can be well-approximated by the differential cross section for the same process in the eikonal unpolarized collisions which depends on both the WW gluon TMD and the linearly polarized WW gluon TMD at the leading order \cite{Metz:2011wb, Dominguez:2011br}.  At the sub-leading order in $\Delta_{\perp}/p_T$, the numerator of $A_{LL}$ receives contributions from the correlator containing the polarized Wilson line with an $F^{12}$ insertion and another correlator with the insertions of the quark axial current and the eikonal gluon field strength $F^{+i}$.

Experimental measurements of longitudinal double-spin asymmetry for inclusive quark-antiquark dijet production in deep inelastic muon proton/nucleus collisions were performed by the COMPASS collaboration \cite{COMPASS:2012mpe, COMPASS:2012pfa, COMPASS:2015pim}. A feasibility study using double-spin asymmetry for dijet production in DIS at the future EIC to access the gluon helicity PDF was performed in \cite{Page:2019gbf, AbdulKhalek:2021gbh}. Although a feasibility study of extracting WW gluon helicity TMD from EIC inclusive dijet $A_{LL}$ data is not a goal of the present paper, the above works make the authors hopeful that such extraction will be possible in the future. Further, our calculation of dijet production in longitudinally polarized electron--proton collisions does not stand alone. On the theoretical side, NLO dijet production in longitudinally polarized DIS was computed in \cite{Borsa:2021afb} in the collinear factorization framework. It was proposed that longitudinal double-spin asymmetry for \textit{exclusive} quark-antiquark dijet production in DIS can probe gluon and quark orbital angular momentum distributions \cite{Hatta:2016aoc,Bhattacharya:2022vvo, Bhattacharya:2023hbq, Bhattacharya:2024sck, Kovchegov:2024wjs}. It is also worth noting that the subeikonal-order contributions to inclusive dijet production in unpolarized DIS were recently computed in \cite{Altinoluk:2022jkk, Altinoluk:2023qfr, Agostini:2024xqs,Altinoluk:2024zom, Altinoluk:2024tyx}.

Below, we also derive the small-$x$ evolution equation for the WW gluon helicity distribution using the light-cone operator treatment (LCOT) technique \cite{Kovchegov:2018znm, Cougoulic:2022gbk} in alignment with the spirit of Wilson's renormalization group approach \cite{Peskin:1995ev} and the background field method \cite{Abbott:1981ke}.  As a warm-up, we use this operator method to reproduce the known small-$x$ evolution equation for unpolarized WW gluon distribution, which was previously obtained from the JIMWLK evolution equation for quadrupole by taking spatial derivatives and collapsing four transverse positions pairwise into two \cite{Dominguez:2011gc}. The advantage of the operator treatment approach is that one can identify the associated Feynman diagram for each term in the evolution equation, which makes the physical meaning of each term more transparent. 

We then proceed by deriving the small-$x$ evolution equation for the operator governing the WW gluon helicity TMD. The evolution equation receives both double-logarithmic (powers of $\alpha_s \ln^2(1/x)$) and single-logarithmic (powers of $\alpha_s \ln(1/x)$) contributions. Concentrating on terms that are double-logarithmic and assuming weak fields (no saturation corrections), the evolution equation for WW gluon helicity distribution simplifies dramatically and is reduced to the same equation as for the dipole amplitude $G^i_{10}$, related to the dipole gluon helicity TMD, as mentioned above. 
 
One therefore concludes that the WW gluon helicity distribution and the dipole gluon helicity distribution follow the same small-$x$ evolution equation in the DLA and the linearized (weak background field) limit. As a result, the small-$x$ asymptotics for the dipole gluon helicity distribution obtained in \cite{Borden:2023ugd} equally applies to the WW gluon helicity distribution. On the other hand, our work demonstrates that the longitudinal double-spin asymmetry $A_{LL}$ for dijet production in polarized DIS can help test and further constrain the small-$x$ evolution of the WW gluon helicity TMD, and, due to their equivalence, the evolution for the dipole gluon helicity TMD.

While no complete small-$x$ helicity evolution exists at the single-logarithmic level (see \cite{Kovchegov:2021iyc} for an incomplete evolution), we also analyze the single-logarithmic terms obtained in our calculation of the evolution equations for the WW helicity gluon TMD. In the large-$N_c$ limit, we find that these terms in the equation cannot be expressed in terms of the same correlator driving the WW gluon helicity TMD: the single-logarithmic equation for WW distribution is not closed, similar to the unpolarized case \cite{Dominguez:2011gc, Dominguez:2011br}. In addition to the $D^i - \cev{D}^i$ gluon dipole correlator and the $F^{12}$ gluon dipole correlator, there appear polarized gluon tri-poles that depend on three different transverse coordinates. This appears to pose a challenge for direct application of the evolution equations at the single logarithm. In the future, if the corrected helicity JIMWLK equation is derived, perhaps by building on the existing result \cite{Cougoulic:2019aja}, and extended to the single logarithmic level, the small-$x$ evolution equation for WW gluon helicity distribution can be obtained from the evolution equation of the $D^i - \cev{D}^i$-type gluon quadrupole, again similar to the unpolarized case \cite{Dominguez:2011gc}. 
 
The paper is organized as follows. In Sec.~\ref{sec:two_distributions}, we derive the operator expressions for the small $x$ limit of the two gluon helicity distributions: the WW gluon helicity TMD and the dipole gluon helicity TMD. Sec.~\ref{eq:computing_ALL_dijet} contains a general discussion of particle productions in longitudinally polarized electron--proton collisions. A detailed derivation of the numerator of the double-spin asymmetry $A_{LL}$ for inclusive quark-antiquark dijet production in the longitudinally polarized electron--proton collisions (averaged over the angle of the electron's transverse momentum) is presented in  Sec.~\ref{sec:dijet_production} with the final result given in \eq{XS_total}. The back-to-back (correlation) limit of the dijet production cross section is subsequently analyzed in Sec.~\ref{sec:expansion}, where we show explicitly in \eq{DSA106} that the cross-section depends on the WW gluon helicity TMD. The small $x$ helicity evolution equation for the operator governing the WW gluon helicity distribution is derived in Sec.~\ref{sec:deriving_helicity_evolution} and analyzed in the double-logarithmic approximation in the linearized regime. Conclusions and outlook are given in Sec.~\ref{sec:conclusions}.

\section{A Tale of Two Gluon Helicity Distributions}
\label{sec:two_distributions}

The transverse momentum-dependent gluon helicity distribution can be extracted from the general gauge-invariant gluon--gluon field strength correlator in the longitudinally polarized proton/nucleus state \cite{Mulders:2000sh, Meissner:2007rx}
\begin{align}
\Gamma_L^{ij}(x, \un{k}) \equiv &\frac{4}{xP^+}\frac{1}{2}\sum_{S_L} S_L\int \frac{d\xi^- d^2\xi}{(2\pi)^3}e^{ixP^+\xi^-} e^{-i\un{k}\cdot\un{\xi}}\left\langle P, S_L| \mathrm{Tr}\left[F^{+i}(0) \mathcal{U}[0, \xi]F^{+j}(\xi) \mathcal{U}'[\xi, 0]\right] |P, S_L\right\rangle\Big|_{\xi^+=0} 
\notag \\
=& \, i\epsilon^{ij}  g^G_{1L}(x, k_T^2) + \frac{(\epsilon^{li} k^j + \epsilon^{lj} k^i) k^l}{k_T^2}  h^{\perp G}_{1L}(x, k_T^2) . \label{subeq:decomposition}
\end{align} 
Here the transverse momentum vector is denoted by $\un k = (k^1, k^2)$ with its magnitude $k_T = |\un k|$, light-cone variables are defined by $v^\pm = (v^0 \pm v^3)/\sqrt{2}$ with $P^+$ the large momentum of the proton, $\epsilon^{ij}$ is the two-dimensional Levi-Civita symbol ($\epsilon^{12} = 1$), Latin indices label transverse components $i,j,l = 1,2$, trasverse position vectors are $\un \xi = (\xi^1, \xi^2)$, $S_L = \pm 1$ is the proton's helicity, and $F^{\mu\nu}$ is the gluon field strength tensor in the fundamental representation.
In \eq{subeq:decomposition}, the gluon field-field correlator is decomposed into the gluon helicity TMD $g^G_{1L}(x, k_T^2)$ and the linearly polarized gluon TMD in the longitudinally polarized proton $h_{1L}^{\perp G}(x, k_T^2)$ \cite{Anselmino:2005sh}. Note that our convention for the definitions of both TMDs is slightly different from those in \cite{Mulders:2000sh, Meissner:2007rx}, with our definition of the linearly polarized TMD following that in \cite{Metz:2011wb}. It is worth noting that both $g^G_{1L}(x, k_T^2)$ and $h_{1L}^{\perp G}(x, k_T^2)$  are real functions \cite{Mulders:2000sh, Meissner:2007rx}. Here $\Gamma_L^{ij}(x, \un{k})$ depends on gauge links $\mathcal{U}[0, \xi]$ and $\mathcal{U}'[\xi, 0]$, which are, in general, different. Their explicit expressions are scattering process-dependent. The two conventional options are the future-pointing ($[+]$) and past-pointing ($[-]$) gauge links 
\begin{equation}\label{eq:U+_U-}
\mathcal{U}^{[+]}[\zeta, \xi] = V_{\un{\zeta}}[\zeta^-, \infty]V_{\un{\xi}}[\infty, \xi^-], \quad \mathcal{U}^{[-]}[\zeta, \xi] = V_{\un{\zeta}}[\zeta^-, -\infty] V_{\un{\xi}}[-\infty, \xi^-] 
\end{equation}
with the light-cone Wilson line in the fundamental representation defined by
\begin{equation}\label{eq:V_def}
V_{\un{\xi}}[\xi_f^-, \xi_i^-] = \mathcal{P} \mathrm{exp}\left\{ig \int_{\xi_i^-}^{\xi_f^-} d\xi^- A^+(0^+, \xi^-, \un{\xi})\right\},
\end{equation}
where $\mathcal{P}$ denotes path ordering and $A^\mu$ is the gluon field. Note that we will work in $A^- =0$ light-cone gauge of the projectile, in which the transverse gluon field $A^i$ vanishes at $\xi^-, \zeta^-=\pm \infty$ so that the transverse segments of the gauge links drop out of Eqs.~\eqref{eq:U+_U-}. Below, we will suppress the $0^+$ argument of the operators for brevity, unless it is needed explicitly. 

In \eq{subeq:decomposition}, when the two gauge links are $\mathcal{U} = \mathcal{U}^{[+]}$ and $\mathcal{U}' = \mathcal{U}^{[-]}$, the corresponding gluon helicity distribution $g^G_{1L}(x, k_T^2)$ is known as the dipole gluon helicity TMD. On the other hand, when the two gauge links are the same $\mathcal{U}=\mathcal{U}' = \mathcal{U}^{[+]}$ or $\mathcal{U}=\mathcal{U}' = \mathcal{U}^{[-]}$, the gluon helicity distribution is called the WW gluon helicity TMD with the SIDIS or Drell-Yan (DY) gauge link, respectively. It should be noted that the dipole gluon helicity TMD and the WW gluon helicity TMD reduce to the same gluon helicity PDF after integrating over the transverse momentum in \eq{subeq:decomposition}. As mentioned in the Introduction, the dipole gluon helicity TMD, particularly its small-$x$ asymptotics has been extensively studied in the past decade or so \cite{Kovchegov:2015pbl, Kovchegov:2018znm, Cougoulic:2022gbk, Borden:2024bxa}. In contrast to that, the small-$x$ regime of the WW gluon helicity TMD is almost unexplored. In this Section, we express the WW gluon helicity TMD at small $x$ in terms of polarized Wilson line correlators.

\subsection{Weizs\"{a}cker-Williams gluon helicity TMD}\label{sec:WW_TMD}

Employing $\mathcal{U}=\mathcal{U}' = \mathcal{U}^{[+]}$ in Eq.~\eqref{subeq:decomposition}, we consider
\begin{align}\label{eq:Gammaij}
\Gamma^{ij}_{WW}(x, \un{k})  = & \ \frac{4}{xP^+ V^-} \frac{1}{(2\pi)^3} \frac{1}{2} \sum_{S_L}S_L \int d\xi^- d^2 \xi d\zeta^- d^2\zeta \, e^{ixP^+(\xi^--\zeta^-)} e^{-i\un{k}\cdot(\un{\xi}-\un{\zeta})} \\
&\qquad \times \left\langle P, S_L\left| \mathrm{tr}\left[F^{+i}(\zeta) \mathcal{U}^{[+]}[\zeta, \xi] F^{+j}(\xi) \mathcal{U}^{[+]}[\xi, \zeta]\right] \right|P, S_L\right\rangle_{\xi^+=\zeta^+=0} 
\notag \\
=& \ \frac{4}{xP^+ V^-} \frac{1}{(2\pi)^3} \frac{1}{2} \sum_{S_L}S_L \int d^2 \xi  d^2\zeta \,  e^{-i\un{k}\cdot(\un{\xi}-\un{\zeta})}\, \langle P, S_L| \mathrm{tr}\Big[[E^i(x, \un{\zeta})]^{\dagger} E^j(x, \un{\xi})\Big] |P, S_L\rangle  \notag
\\
=& \ i \epsilon^{ij}  g^{G \, WW}_{1L}(x, k_T^2) + \frac{(\epsilon^{li} k^j + \epsilon^{lj} k^i) k^l}{k_T^2}  h^{\perp G \, WW}_{1L}(x, k_T^2) .\notag 
\end{align}
Here $V^- = \int d^2\zeta d\zeta^-$.  
In Eq.~\eqref{eq:Gammaij}, we have defined (cf. \cite{Cougoulic:2022gbk})
\begin{align}\label{eq:Ej_full}
E^j(x, \un{\xi}) = &\int\limits_{-\infty}^\infty d\xi^- e^{ixP^+ \xi^-} V_{\un{\xi}}[\infty, \xi^-] F^{+j}(\xi^-, \un{\xi}) V_{\un{\xi}}[\xi^-, \infty]\\
=&-\int\limits_{-\infty}^\infty d\xi^- e^{ixP^+ \xi^-}V_{\un{\xi}}[\infty, \xi^-] \Big(ixP^+ A^j + \partial^j A^+ \Big) V_{\un{\xi}}[\xi^-, \infty]. \notag
\end{align}
To arrive at the second equality in Eq.~\eqref{eq:Ej_full}, we have integrated by parts over the longitudinal coordinate using the boundary condition $A^i(\xi^-=\pm \infty, \un{\xi}) =0$. 

From the definition Eq.~\eqref{eq:Gammaij}, one can readily see that the hermiticity condition is satisfied \cite{Mulders:2000sh}, such that
\begin{equation}
\left(\Gamma^{ij}_{WW}(x, \un{k})\right)^{\ast} = \Gamma^{ji}_{WW}(x, \un{k}).
\end{equation}
One therefore concludes that the real part of $\Gamma^{ij}_{WW}(x, \un{k})$ is symmetric with respect to the spatial indices exchange $i\leftrightarrow j$ while the imaginary part of $\Gamma^{ij}_{WW}(x, \un{k})$ is antisymmetric. From Eq.~\eqref{eq:Gammaij}, one can see that $ g^{G \, WW}_{1L}(x, k_T^2)$ represents the imaginary part of the gluon field-field correlator $\Gamma^{ij}_{WW}(x, \un{k})$ while $h^{\perp G \, WW}_{1L}(x, k_T^2)$ characterizes the real part of $\Gamma^{ij}_{WW}(x, \un{k})$. We also see that both TMDs are indeed real. 

Our goal is to derive an expression for $\Gamma^{ij}_{WW}(x, \un{k})$ at small $x$. 
To do so, expanding the phase in Eq.~\eqref{eq:Ej_full} in powers of $x$, we obtain (cf. \cite{Cougoulic:2022gbk})
\begin{align}\label{Ej_exp}
    E^j(x, \un{\xi}) = \int\limits_{-\infty}^\infty d \xi^- \, \ V_{\un \xi} [\infty, \xi^-] \,
    \left[ \pd^j A^+ + i x P^+ \left(\xi^- \, \pd^j A^+ + A^j \right) + {\cal O} (x^2) \right] \, V_{\un \xi} [\xi^-, \infty] .
\end{align}
This simplifies to 
\begin{align}\label{Exp}
    E^j(x, \un{\xi}) = \frac{1}{i g} \, \left( \pd^j V_{\un \xi} \right) \, V_{\un \xi}^\dagger - \frac{x s}{g} \, V_{\un \xi}^{j \, \textrm{G} [2]} \, V_{\un \xi}^\dagger + {\cal O} (x^2)
\end{align}
with $V_{\un{\xi}}\equiv V_{\un{\xi}}[\infty, -\infty]$ and 
\begin{equation}\label{VjG2}
\begin{split}
V_{\un{\xi}}^{j \, \textrm{G} [2]} \equiv &\frac{P^+}{2s} \int\limits_{-\infty}^{\infty} d\xi^-\, V_{\un{\xi}}[\infty, \xi^-] \left[{D}^j - \cev{D}^j\right] V_{\un{\xi}}[\xi^-, -\infty]\\
=&\frac{-igP^+}{s} \int\limits_{-\infty}^{\infty} d\xi^- \, V_{\un{\xi}}[\infty, \xi^-] (\xi^-\partial^j A^+ + A^j)V_{\un{\xi}}[\xi^-, -\infty].\\
\end{split}
\end{equation}
As defined above, the left covariant derivative and right covariant derivative in the fundamental representation are ${D}^j = \partial^j - igA^j$ and $\cev{D}^j = \overleftarrow{\partial}^j + igA^j$, respectively. The variable $s$ denotes the center of mass energy squared for the projectile--target scattering. It is easy to verify that 
\begin{equation}
\left(V_{\un{\xi}}^{j \, \textrm{G} [2]} V_{\un{\xi}}^{\dagger}\right)^{\dagger} = V_{\un{\xi}} V_{\un{\xi}}^{j \, \textrm{G}[2]\dagger} = - V_{\un{\xi}}^{j \, \textrm{G}[2]} V_{\un{\xi}}^{\dagger}.
\end{equation}
Substituting Eq.~\eqref{Ej_exp} into Eq.~\eqref{eq:Gammaij}, and defining the single- and double-angle brackets denoting the saturation physics averaging by
\begin{equation}\label{CGC_ave}
\begin{split}
\Big\langle \ldots \Big\rangle \equiv\frac{1}{2}\sum_{S_L}S_L \frac{1}{2P^+V^-} \langle P, S_L |\ldots |P, S_L\rangle ,  \qquad \llangle \ldots \rrangle \equiv  s \Big\langle \ldots \Big\rangle ,
\end{split}
\end{equation}
one obtains the small-$x$ limit of Eq.~\eqref{eq:Gammaij},
\begin{align}\label{eq:GammaijL_polarized}
\Gamma_{WW}^{ij}(x,\un{k}) = & \ - \frac{1}{x}\frac{1}{g^2 \pi^3}\int d^2\xi d^2\zeta \, e^{-i\un{k}\cdot(\un{\xi}-\un{\zeta})}\left\langle \mathrm{tr}\left[V_{\un{\zeta}} \partial^i V_{\un{\zeta}}^{\dagger} V_{\un{\xi}}\partial^j V_{\un{\xi}}^{\dagger}\right]\right \rangle\\
& - \frac{i}{g^2 \pi^3} \int d^2\xi d^2\zeta \,  e^{-i\un{k}\cdot(\un{\xi}-\un{\zeta})} \llangle \mathrm{tr}\left[V_{\un{\zeta}} \partial^i V_{\un{\zeta}}^{\dagger} V_{\un{\xi}}^{j \, \textrm{G}[2]}V_{\un{\xi}}^{\dagger}\right]+\mathrm{tr}\left[ V_{\un{\zeta}}V^{i \, \textrm{G}[2]\dagger}_{\un{\zeta}} V_{\un{\xi}}\partial^j V^{\dagger}_{\un{\xi}}\right]\rrangle + {\cal O} (x) . \notag
\end{align}

Using 
\begin{equation}
 \Gamma^{ij}_{WW}(x, -\un{k}) = \Gamma^{ij}_{WW}(x, \un{k}) ,
\end{equation}
which follows from the lack of preferred transverse direction in a longitudinally polarized proton state, and employing the cyclic property of the traces in \eq{eq:GammaijL_polarized}, which follows from the properties of the eikonal Wilson lines explored in \cite{Mueller:2012bn, Kovchegov:2018znm}, one can see that the first term in Eq.~\eqref{eq:GammaijL_polarized} is real while the second term is purely imaginary. As a result, the small-$x$ limit of $g^{G \, WW}_{1L}(x, k_T^2)$ is determined by the second term while the small-$x$ limit of $h^{\perp G \, WW}_{1L}(x, k_T^2)$ is determined by the first term. To be specific, 
\begin{align}\label{eq:WW_G_final}
 g^{G \, WW}_{1L}(x, k_T^2) = -\frac{1}{2} i\epsilon^{ij} \Gamma_{WW}^{ij}(x, \un{k}) 
 = - \frac{1}{\as \, 4 \pi^4} \int d^2\xi d^2\zeta \, e^{-i\un{k}\cdot(\un{\xi}-\un{\zeta})} \, \epsilon^{ij} \, \mathrm{Re}\, \llangle \mathrm{tr}\left[V_{\un{\zeta}} \partial^i V_{\un{\zeta}}^{\dagger} V_{\un{\xi}}^{j\, \textrm{G}[2]}V_{\un{\xi}}^{\dagger}\right] \rrangle
\end{align}
and 
\begin{align}
 h^{\perp G \, WW}_{1L}(x, k_T^2) =& \frac{1}{2} \frac{(\epsilon^{li}k^j + \epsilon^{lj} k^i) k^l}{k_T^2}  \Gamma^{ij}_{WW}(x, k_T^2) \notag \\
 =&-\frac{1}{x} \frac{1}{\as \, 4 \pi^4} \frac{(\epsilon^{li} k^j + \epsilon^{lj} k^i) k^l}{2k_T^2}  \int d^2\xi d^2\zeta  e^{-i\un{k}\cdot(\un{\xi}-\un{\zeta})}\left\langle \tr \left[ V_{\un{\zeta}} \partial^i V_{\un{\zeta}}^{\dagger} V_{\un{\xi}}\partial^j V_{\un{\xi}}^{\dagger} \right] \right\rangle.\label{eq:DeltaH_final}
\end{align}
We see that, in the $A^-=0$ gauge we consider, the linearly polarized WW gluon TMD $h^{\perp G \, WW}_{1L}$ in the longitudinally polarized proton depends only on the gluon field $A^+$. The $1/x$ prefactor of \eq{eq:DeltaH_final} appears to indicate that this is an eikonal TMD. However, the eikonal gluon field $A^+$ is independent of the proton helicity and cannot give a non-zero $h^{\perp G \, WW}_{1L}$. Hence, $h^{\perp G \, WW}_{1L}$ appears to be zero at the eikonal order. In \cite{Li:2024fdb, Li:2024xra}, a sub-eikonal classical gluon field $A^+$ was constructed, which was non-linear in the sources of color charge (in the terminology of the McLerran--Venugopalan (MV) model \cite{McLerran:1993ni,McLerran:1993ka,McLerran:1994vd} modified for calculating helicity-dependent observables \cite{Cougoulic:2020tbc}). However, that $A^+$ field does not seem to couple to the operator on the right-hand side of \eq{eq:DeltaH_final} (when inserted in it once, along with multiple insertions of the eikonal $A^+$ field), again giving us zero $h^{\perp G \, WW}_{1L}$: this agrees with the expectation that $h^{\perp G \, WW}_{1L}$ is an eikonal TMD, which should receive no contribution from sub-eikonal fields. In general, the authors of the present paper could not find a realistic gluon field at the sub-eikonal order that couples to the longitudinal spin of the proton and gives a non-zero $h^{\perp G \, WW}_{1L}$. It is possible that $h^{\perp G \, WW}_{1L}$ is identically zero at sub-eikonal order, though a proof of such statement is left for future work\footnote{The potential contribution of $h^{\perp G}_{1L}$ to $J/\psi$ production in unpolarized electron scattering on longitudinally polarized proton was discussed in \cite{liu:2023unl}, see Eq.~(30) there. It may, therefore, be possible to measure $h^{\perp G \, WW}_{1L}$ TMD in such scattering processes.}. 

Below we will show that this linearly polarized gluon TMD does not contribute to the spin asymmetry $A_{LL}$ for the inclusive di-jet production in the back-to-back limit.

Defining
\begin{equation}\label{GWW_def0}
 G^{WW}_{10} (s) \equiv - \frac{1}{4 N_c}  \epsilon^{ij} \, \llangle  \mathrm{tr}\left[V_{{\un x}_0} \partial^i V_{{\un x}_0}^{\dagger} V_{{\un x}_1}^{j\, \textrm{G}[2]}V_{{\un x}_1}^{\dagger}\right]\rrangle (s) - \cc = - \frac{1}{2 N_c}  \epsilon^{ij} \, \mathrm{Re} \,\llangle  \mathrm{tr}\left[V_{{\un x}_0} \partial^i V_{{\un x}_0}^{\dagger} V_{{\un x}_1}^{j\, \textrm{G}[2]}V_{{\un x}_1}^{\dagger}\right]\rrangle (s)
\end{equation}
we rewrite \eq{eq:WW_G_final} as
\begin{align}\label{eq:WW_G_final2}
 g^{G \, WW}_{1L}(x, k_T^2) =
 \frac{N_c}{\as \, 2 \pi^4} \int d^2 x_1 d^2 x_0 \, e^{-i\un{k}\cdot \un{x}_{10}} \, G^{WW}_{10}  \left( s = \frac{Q^2}{x} \right) ,
\end{align}
where we have relabeled ${\un \xi} \to {\un x}_1$, ${\un \zeta} \to {\un x}_0$, with ${\un x}_{10} = \un x_1 - \un x_0$. The renormalization scale is labeled $Q^2$. To study the small-$x$ asymptotics of the WW gluon helicity TMD, one needs to identify observables in scattering processes that probe $G^{WW}_{10} (s)$ and derive the small-$x$ evolution equation for this quantity.


\subsection{Dipole gluon helicity TMD}

For the completeness of our discussion, we reproduce and slightly expand the derivation in \cite{Cougoulic:2022gbk} for the small-$x$ limit of the {\sl dipole} gluon helicity TMD. Using $\mathcal{U} = \mathcal{U}^{[+]}$ and $\mathcal{U}' = \mathcal{U}^{[-]}$ in Eq.~\eqref{eq:Gammaij}, one gets
\begin{align}\label{eq:Gamma_dip}
\Gamma^{ij}_{dip}(x, \un{k})  = &\frac{4}{xP^+ V^-} \frac{1}{(2\pi)^3} \frac{1}{2} \sum_{S_L}S_L \int d\xi^- d^2 \xi d\zeta^- d^2\zeta  e^{ixP^+(\xi^--\zeta^-)} e^{-i\un{k}\cdot(\un{\xi}-\un{\zeta})}  \\
&\qquad \times \left\langle P, S_L\left| \mathrm{tr}\left[F^{+i}(\zeta) \mathcal{U}^{[+]}[\zeta, \xi] F^{+j}(\xi) \mathcal{U}^{[-]}[\xi, \zeta]\right] \right|P, S_L\right\rangle_{\xi^+=\zeta^+=0} \notag \\
=&\frac{4}{xP^+ V^-} \frac{1}{(2\pi)^3} \frac{1}{2} \sum_{S_L}S_L \int d^2 \xi d^2\zeta \, e^{-i\un{k}\cdot(\un{\xi}-\un{\zeta})}\langle P, S_L| \mathrm{tr}\left[[L^i(x, \un{\zeta})]^{\dagger} L^j(x, \un{\xi})\right] |P, S_L\rangle   \notag \\
=& i\epsilon^{ij}  g_{1L}^{G \, dip}(x, k_T^2) + \frac{(\epsilon^{li} k^j + \epsilon^{lj} k^i)k^l}{k_T^2}  h_{1L}^{\perp G \, dip}(x, k_T^2) . \notag
\end{align}
We have defined (cf. \cite{Cougoulic:2022gbk})
\begin{align}
L^j(x, \un{\xi}) \equiv - \int\limits_{-\infty}^\infty d\xi^- e^{ixP^+ \xi^-} V_{\un{\xi}}[\infty, \xi^-] F^{+j}(\xi^-, \un{\xi}) V_{\un{\xi}}[\xi^-, -\infty]
=  - \frac{i}{g} \partial^j V_{\un{\xi}} - \frac{xs}{g} V_{\un{\xi}}^{j \, \textrm{G}[2]} + \mathcal{O}(x^2), 
\end{align}
also expanding it in the powers of $x$. For the correlator in \eq{eq:Gamma_dip} this expansion in $x$ gives
\begin{align}
\Gamma_{dip}^{ij}(x,\un{k}) = &\frac{1}{x} \frac{1}{g^2\pi^3} \int d^2\xi d^2\zeta e^{-i\un{k}\cdot(\un{\xi}- \un{\zeta})} \Big\langle \mathrm{tr}\left[\partial^i V_{\un{\zeta}}^{\dagger} \, \partial^j V_{\un{\xi}}  \right] \Big\rangle \\
&+\frac{i}{g^2\pi^3} \int d^2\xi d^2\zeta e^{-i\un{k}\cdot(\un{\xi}- \un{\zeta})} \llangle \mathrm{tr}\left[ V_{\un{\zeta}}^{i \, \textrm{G}[2] \dagger} \, \partial^j  V_{\un{\xi}} \right] - \mathrm{tr}\left[\partial^i V_{\un{\zeta}}^{\dagger} \, V_{\un{\xi}}^{j \, \textrm{G}[2]} \right]\rrangle  + \mathcal{O}(x). \notag
\end{align}
The dipole gluon helicity TMD can be expressed as
\begin{equation}\label{eq:dipole_G_final}
g_{1L}^{G \, dip}(x, k_T^2) = \frac{1}{\as 8 \pi^4} i\epsilon^{ij}k^j  \int d^2\xi d^2\zeta e^{-i\un{k}\cdot(\un{\xi}- \un{\zeta})} \llangle \mathrm{tr}\left[V_{\un{\xi}}^{i \, \textrm{G}[2]} V_{\un{\zeta}}^{\dagger}\right]\rrangle + \cc ,
\end{equation}
while the dipole linearly polarized gluon TMD in longitudinally polarized proton is
\begin{equation}\label{glue_dip_helicity}
h_{1L}^{\perp G \, dip}(x, k_T^2) =- \frac{1}{\as 8 \pi^4}\epsilon^{ij}k^j \int d^2\xi d^2\zeta e^{-i\un{k}\cdot(\un{\xi}- \un{\zeta})} \llangle \mathrm{tr}\left[V_{\un{\xi}}^{i \, \textrm{G}[2]} V_{\un{\zeta}}^{\dagger}\right]\rrangle + \cc .
\end{equation}
Defining \cite{Cougoulic:2022gbk, Kovchegov:2017lsr}
\begin{equation}\label{Gi_def}
G^i_{10} (s) \equiv \frac{1}{2N_c} \llangle \mathrm{tr}\left[V_{\un{x}_0}^{\dagger} \, V_{\un{x}_1}^{i \, \textrm{G}[2]} \right]\rrangle (s) + \cc ,
\end{equation}
we recast \eq{glue_dip_helicity} as
\begin{align}
    g_{1L}^{G \, dip}(x, k_T^2) = \frac{N_c}{\as \, 4 \pi^4} i\epsilon^{ij}k^j \int d^2 x_1 d^2 x_0 \, e^{-i\un{k}\cdot \un{x}_{10}} \, G_{10}^i \left( s = \frac{Q^2}{x} \right) .
\end{align}
We see that in order to understand the small-$x$ asymptotics of the dipole helicity TMD, one needs to study the small-$x$ evolution of the polarized Wilson line correlator (``polarized dipole amplitude") $G_{10}^i$, which has been extensively investigated in \cite{Kovchegov:2015pbl, Kovchegov:2018znm, Cougoulic:2022gbk, Borden:2024bxa}.  The polarized dipole amplitude $G_{10}^i$ also enters the total cross section in inclusive deep-inelastic electron-proton collisions \cite{Cougoulic:2022gbk} (and hence the structure function $g_1$) and particle (gluon) production cross section in the longitudinally polarized proton-proton collisions \cite{Kovchegov:2024aus}.

From Eqs.~\eqref{eq:WW_G_final} and \eqref{eq:dipole_G_final}, it is easy to verify that the two gluon helicity TMDs lead to the same gluon hPDF $\Delta G (x)$ when integrated over all $k_T$, that is 
\begin{equation}
\Delta G (x)  = \int d^2k\,  g_{1L}^{G \, WW}(x, k_T^2) = \int d^2k \, g_{1L}^{G \, dip}(x, k_T^2) =  \frac{1}{\as \, 2 \pi^2} \epsilon^{ij} \int d^2\xi \, \llangle \mathrm{tr}\left[ \partial^j V^{\dagger}_{\un{\xi}} \, V^{i \, \textrm{G}[2]}_{\un{\xi}} \right]\rrangle + \cc .
\end{equation}

\section{Longitudinal Double-Spin Asymmetry for Dijet Production}
\label{eq:computing_ALL_dijet}

In this Section, we show that double-spin asymmetry for inclusive dijet production in longitudinally polarized electron-proton collisions can probe the WW gluon helicity TMD. This is achieved by taking the back-to-back limit for the dijet production cross section.

\subsection{General setup}

For the longitudinally polarized deep inelastic electron-proton scattering, it is convenient to write the differential cross section in the polarization basis of the virtual photon, 
\begin{align}\label{DIS1}
E'  \, \frac{d \sigma}{d^3 k'} = \frac{2 M_p \alpha_{EM}^2}{s \, Q^4} \, L_{\mu\nu} \, W^{\mu\nu}  = \frac{2 M_p \alpha_{EM}^2}{s \, Q^4} \, \sum_{\lambda. \lambda' = 0, \pm 1} \, (-1)^{\lambda + \lambda'} \, L_{\lambda\lambda'} \, W^{\lambda\lambda'}.
\end{align}
Here $\lambda, \lambda' = \pm 1$ and $\lambda, \lambda' =0$ represent the transverse and the longitudinal polarizations of the virtual photon, respectively, while $\alpha_{EM} = e^2/(4 \pi)$ is the fine structure constant.  The leptonic tensor and the hadronic tensor in the polarization basis are
\begin{align}
L_{\lambda\lambda'} \equiv L^{\mu\nu} \, \epsilon_\mu^\lambda \, \epsilon_\nu^{\lambda' *}, \ \ \ W^{\lambda\lambda'} \equiv W^{\alpha\beta} \, \epsilon_\alpha^{\lambda *} \, \epsilon_\beta^{\lambda'} ,
\end{align}
with
\begin{align}
L^{\mu\nu} = 2 \, \left[ k^\mu k'^\nu + k^\nu k'^\mu - g^{\mu\nu} k \cdot k' + i \, \sigma_e \, \epsilon^{\mu\nu\rho\sigma} \, k_\rho \, k'_\sigma \right]
\end{align}
and 
\begin{align}
2 M_p \, W^{\alpha\beta} (P, q) = \frac{1}{2\pi} \, \sum_X \, \bra{P, S_L} j^\alpha (0)   \ket{X} \, \bra{X} j^\beta (0) \,  \ket{P, S_L} \, (2 \pi)^4 \, \delta^4 (P + q - p_X).
\end{align}
The incoming electron carries 4-momentum $k$ (with energy $E$) and polarization $\sigma_e$, the outgoing electron's momentum is $k'$ (with energy $E'$), the proton has mass $M_p$ and comes in with momentum $P$ and helicity $S_L$, and $j^{\alpha}(x) = \sum_f Z_f \bar{\psi}(x) \gamma^{\alpha} \psi(x)$ is the electromagnetic current operator for the quarks with fractional charge $Z_f$. We use $X$ to denote a particular final state with momentum $p_X$. For production processes, we may choose not to sum over all $X$, but only over a subset of $X$ included in the final state definition. In obtaining the expression \eqref{DIS1} in the polarization basis, we have used the numerator for the virtual photon propagator 
\begin{align}\label{metric}
g_{\mu\nu} = - \sum_{\lambda = \pm 1} \epsilon_{T \, \mu}^{\lambda *} \, \epsilon_{T \nu}^\lambda + \epsilon_{L \, \mu}^{*} \, \epsilon_{L \nu} + \frac{q_\mu \, q_\nu}{q^2} .
\end{align}
The last term in \eq{metric} vanishes when dotted into $L^{\mu\nu}$. We are left with the summation over transverse and longitudinal polarizations in \eq{DIS1}.

The center-of-mass energy squared for the electron and the proton is $s = (P+k)^2$.  
We ignore the mass of the electron. The momentum transfer is $q = k-k'$, and $q^2 = -Q^2$ represents the virtuality of the photon. For the leptonic tensor, one sums over the final spin states of the outgoing electron while keeping the initial spin state of the incoming electron $\sigma_e$ fixed. We follow the standard convention in the literature \cite{Bacchetta:2004jz, Braun:2005rg, Mantysaari:2020lhf, Bhattacharya:2022vvo,Bhattacharya:2024sno,Hatta:2016aoc}: the proton moves in the positive-$z$ direction while the electron propagates in the negative-$z$ direction. We will work in the ``dipole" ($q_T =0$) frame, where the momenta can be written as
\begin{subequations}
\begin{align}
& P^\mu \approx (P^+, 0^-, {\un 0}),  \\
& q^\mu = \left( - \frac{Q^2}{2 q^-}, q^-, {\un 0} \right), \\
& k^\mu = \left( \frac{k_\perp^2}{2 k^-} , k^-, {\un k} \right), \\
& k'^\mu = \left( \frac{k_\perp^2}{2 (k^- - q^-)} , k^- - q^-, {\un k} \right) 
\end{align}
\end{subequations}
with $P^+$ very large. Employing
\begin{align}
y = \frac{P \cdot q}{P \cdot k} = \frac{q^-}{k^-},
\end{align}
we see that conservation of the + momentum component in the electron--photon system gives
\begin{align}\label{relation1}
\frac{Q^2}{y^2} = \frac{k_\perp^2}{1-y}.
\end{align}
The virtual photon polarization vectors are 
\begin{align}
\epsilon_T^\lambda = (0,0, {\un \epsilon}_\lambda ) , \ \ \ \epsilon_L = \left( \frac{Q}{2 q^-}, \frac{q^-}{Q}, {\un 0} \right)
\end{align}
with ${\un \epsilon}_\lambda = - (1/\sqrt{2}) (-\lambda, i)$. 

A direct calculation employing \eq{relation1} yields \cite{Mantysaari:2020lhf, Kovchegov:2024wjs} (here $k_x = k_T \cos \phi$, $k_y = k_T \sin \phi$)
\begin{subequations}\label{Lij}
\begin{align}
& L_{TT} = \frac{Q^2}{y^2}  \, \delta_{\lambda \lambda'} \,  \left\{   \left[ 1 + (1-y)^2 \right] - \sigma_e \, \lambda \,  \left[ (1-y)^2 -1 \right] \right\}  -  Q^2 \, \delta_{\lambda, - \lambda'} \, \frac{2 (1-y)}{y^2} \, e^{- 2 i \lambda \, \phi}, \\
& L_{TL} = - e^{- i \lambda \, \phi} \, \sqrt{2 (1-y)} \, \frac{Q^2}{y^2} \, \left[ \lambda \,  (2-y) + \sigma_e \, y \right] = \left( L_{LT} \right)^* , \\
& L_{LL} = Q^2 \, \frac{4 (1-y)}{y^2} .
\end{align}
\end{subequations}

Using Eqs.~\eqref{Lij} in \eq{DIS1} we obtain (cf. Eq.~(3.20) in \cite{Arens:1996xw} and note that in that reference the virtual photon is moving in the positive $z$ direction, while here the photon is moving in the $-z$ direction, accounting for the sign difference in the definition of the azimuthal angles here and in Eq.~(2.1) of \cite{Arens:1996xw})
\begin{align}\label{DIS4}
& E'  \, \frac{d \sigma}{d^3 k'} =\frac{2 M_p \alpha_{EM}^2}{s \, Q^2 \, y^2} \, \Bigg\{ \left[ 1 + (1-y)^2 \right]  \, \sum_{\lambda = \pm 1} W^{\lambda \lambda}  - \sigma_e \,  \left[ (1-y)^2 -1 \right] \, \sum_{\lambda = \pm 1} \lambda \, W^{\lambda \lambda} - 4 (1-y) \, \cos (2 \phi) \, \mbox{Re} \, W^{1, - 1} \notag \\
& - 4 (1-y) \, \sin (2 \phi) \, \mbox{Im} \, W^{1, - 1} + 4 (1-y) \, W^{00} - 2 \, \sqrt{2 (1-y)} \, (2-y) \, \cos (\phi) \,  \sum_{\lambda = \pm 1} \lambda \, \mbox{Re} \,  W^{\lambda, 0}  \\
& - 2 \, \sqrt{2 (1-y)} \, (2-y) \, \sin (\phi) \,  \sum_{\lambda = \pm 1} \mbox{Im} \, W^{\lambda, 0} 
- 2 \, \sqrt{2 (1-y)} \, y \, \sigma_e \, \cos (\phi) \, \sum_{\lambda = \pm 1} \mbox{Re} \, W^{\lambda, 0}  \notag \\
& - 2 \, \sqrt{2 (1-y)} \, y \, \sigma_e \,  \sin (\phi) \,
 \sum_{\lambda = \pm 1} \lambda \, \mbox{Im} \, W^{\lambda, 0}  \Bigg\} . \notag
\end{align}
Note that $(W^{\lambda\lambda'})^* = W^{\lambda' \lambda}$. There are 9 different hadronic tensor structures multiplying different angle- and $\sigma_e$-dependent terms:
\begin{align}\label{9}
\sum_{\lambda = \pm 1} W^{\lambda \lambda}, \sum_{\lambda = \pm 1} \lambda \, W^{\lambda \lambda} ,  \mbox{Re} \, W^{1, - 1} , \mbox{Im} \, W^{1, - 1} , W^{00} , \sum_{\lambda = \pm 1} \lambda \, \mbox{Re} \,  W^{\lambda, 0} , \sum_{\lambda = \pm 1} \mbox{Im} \, W^{\lambda, 0}  , \sum_{\lambda = \pm 1} \mbox{Re} \, W^{\lambda, 0} , \sum_{\lambda = \pm 1} \lambda \, \mbox{Im} \, W^{\lambda, 0} .
\end{align}

For a longitudinally polarized proton, we are interested in calculating the double-spin asymmetries in the $q_T =0$ frame. Following the notation of \cite{Diehl:2005pc} we define them as 
\begin{align}\label{ALL_def}
 A_{LL}^{\gamma^*} = \frac{ d\sigma (++) - d\sigma(+-) }{d\sigma (++) + d\sigma(+-)} \equiv \frac{ d\sigma (++) - d\sigma(+-) }{2 \, d\sigma_{unpol}} \equiv   \frac{d \sigma^{DSA} }{d\sigma_{unpol}} ,
\end{align}
where
\begin{subequations} 
\begin{align}
& d \sigma^{DSA} = \frac{1}{4} \, \sum_{\sigma_e, S_L} \, \sigma_e \, S_L \, d \sigma (\sigma_e, S_L), \label{DSA_sigma} \\
& d\sigma_{unpol} = \frac{1}{4} \, \sum_{\sigma_e, S_L} \, d \sigma (\sigma_e, S_L) .
\end{align}
\end{subequations}
From Eqs.~\eqref{DIS4} and \eqref{DSA_sigma} we obtain the numerator of $A_{LL}$,
\begin{align}
\label{DSA1}
E'  \, \frac{d \sigma^{DSA}}{d^3 k'} =  \frac{2 M_p \alpha_{EM}^2}{s \, Q^2 \, y^2} \, \half \, \sum_{S_L} S_L \ &   \Bigg\{ \left[ 1-(1-y)^2 \right] \, \sum_{\lambda = \pm 1} \lambda \, W^{\lambda \lambda}  
- 2 \, \sqrt{2 (1-y)} \, y \,\cos (\phi) \, \sum_{\lambda = \pm 1} \mbox{Re} \, W^{\lambda, 0}  \notag \\
& - 2 \, \sqrt{2 (1-y)} \, y \,  \sin (\phi) \,
 \sum_{\lambda = \pm 1} \lambda \, \mbox{Im} \, W^{\lambda, 0} \Bigg\} . 
\end{align}
We see that DSA couples to 3 out of 9 structures in \eq{9}. The expression \eqref{DSA1} suggests that there are possible interference terms between the transversely polarized photon in the amplitude and the longitudinally polarized photon in the complex conjugate amplitude (and vice versa)
which can be measured if the azimuthal angle of the lepton is resolved. We will consider the cross section in \eq{DSA1} averaged over the azimuthal angles of the lepton and discard these interference terms. More specifically, we will concentrate on
\begin{align}\label{DSA22}
    \int\limits_0^{2 \pi} \frac{d \phi}{2 \pi} \, E'  \, \frac{d \sigma^{DSA}}{d^3 k'} =  \frac{2 M_p \alpha_{EM}^2}{s \, Q^2 \, y^2} \, \half \, \sum_{S_L} S_L \, \left[ 1-(1-y)^2 \right] \, \sum_{\lambda = \pm 1} \lambda \, W^{\lambda \lambda} .
\end{align}
Further, defining the virtual photon--proton scattering cross section by
\begin{align}
\sigma_{\lambda \lambda'}^{\gamma^* p} = \frac{4 \pi^2 \alpha_{EM} \, x}{Q^2} \, 2 M_p \, W^{\lambda \lambda'} ,
\end{align}
we rewrite \eq{DSA22} as
\begin{align}\label{DSA33}
    \int\limits_0^{2 \pi} \frac{d \phi}{2 \pi} \,  E'  \, \frac{d \sigma^{DSA}}{d^3 k'} =  \frac{ \alpha_{EM}}{4 \pi^2 \, Q^2 } \, \half \, \sum_{S_L} S_L \, (2-y) \, \sum_{\lambda = \pm 1} \lambda \, \sigma_{\lambda \lambda}^{\gamma^* p} 
\end{align}
since $Q^2 \approx  x ys$ if we neglect the proton mass. In the following, only the transversely polarized virtual photon is considered in calculating the double-spin asymmetry of inclusive dijet production (with only diagonal in polarization terms contributing).

\subsection{Inclusive di-jet production calculation}

\label{sec:dijet_production}

The inelastic di-jet production cross section is, in our infinite momentum frame,
\begin{align}\label{XS-2}
d\sigma^{\gamma^* p \to q {\bar q} X} = \frac{1}{8 q^0 E_P} \sum_n  \, \left\langle |M_{\gamma^* p \to q {\bar q} n}|^2 \right\rangle \, \frac{d^2 p_1 \, d p_1^-}{(2 \pi)^3 \, 2 p_1^-} \, \frac{d^2 p_2 \, d p_2^-}{(2 \pi)^3 \, 2 p_2^-} \, \prod_{i=1}^n  \frac{d^2 k_i  \, d k^{+}_i}{(2 \pi)^3 \, 2 k_i^{+}} \, (2 \pi)^4 \, \delta^4 \left( q+P - p_1 - p_2 - \sum_{j=1}^n k_j \right) 
\end{align}
with $p_1$ and $p_2$ the momenta of the two jets (a quark and an anti-quark, respectively, in our formalism), $k_i$ the momenta of the final-state particles other then the produced quark and anti-quark, $M$ the scattering amplitude, and the angle brackets denoting the sums/averaging over the final/initial-state quantum numbers. Here we assume that the two jets are produced in the current fragmentation region (see \fig{FIG:diagrams} below), such that their minus momentum components are large, $p_1^- , p_2^- \sim q^- \gg k_i^-, P^-$. (Note that $P^- \approx 0$ in our frame.) Additionally, one of the produced particles $k_i$ should carry most of the proton's plus momentum $P^+$: we choose to label this particle $n$, such that $k_n^+ \approx P^+$.\footnote{In the case when there are several such high-plus momentum particles, we will be left with the integrals over the plus momentum fractions of the proton's momentum carried by such particles, which will be a part of the saturation/color glass condensate (CGC) averaging.} We thus write 
\begin{align}\label{eq:p-_conservation}
& \int \prod_{i=1}^n \frac{d^2 k_i  \, d k^{+}_i}{(2 \pi)^3 \, 2 k_i^{+}} \, (2 \pi)^4 \, \delta^4 \left( q+P - p_1 - p_2 - \sum_{j=1}^n k_j \right) \notag\\
&  =  \int \prod_{i=1}^n \frac{d^2 k_i  \, d k^{+}_i}{(2 \pi)^3 \, 2 k_i^{+}} \, (2 \pi)^4 \, \delta (q^- - p_1^- - p_2^-) \, \delta (P^+ - k_n^+) \,   \delta^2 \left( {\un p}_1 + {\un p}_2 + \sum_{j=1}^{n-1} {\un k}_j + {\un k}_n  \right) \notag \\
& = 2 \pi \, \delta (q^- - p_1^- - p_2^-) \, \frac{1}{2 P^+} \, \int \prod_{i=1}^{n-1} \frac{d^2 k_i  \, d k^{+}_i}{(2 \pi)^3 \, 2 k_i^{+}}  .
\end{align}
Using this in \eq{XS-2} we get
\begin{align}\label{XS-1}
d\sigma^{\gamma^* p \to q {\bar q} X}  =  \frac{1}{4 \pi} \sum_n  \, \left\langle |A_{\gamma^* p \to q {\bar q} n}|^2 \right\rangle \, \frac{d^2 p_1 \, d^2 p_2}{(2 \pi)^4} \, \frac{d z}{z (1-z)} \, \int \prod_{i=1}^{n-1} \frac{d^2 k_i  \, d k^{+}_i}{(2 \pi)^3 \, 2 k_i^{+}} ,
\end{align}
where $A = M/(2s)$ and $z = p_1^- /q^-$ is the fraction of the photon's minus momentum carried by the produced quark.  Once again, we assume that $z = {\cal O} (1)$. In the saturation/color glass condensate (CGC) notation \cite{Gribov:1984tu, Iancu:2003xm, Weigert:2005us, JalilianMarian:2005jf, Gelis:2010nm, Albacete:2014fwa, Kovchegov:2012mbw, Morreale:2021pnn}, the remaining integrals over $k_i$ are included in the averaging (angle brackets):
\begin{align}\label{XS0}
d\sigma^{\gamma^* p \to q {\bar q} X}  =  \frac{1}{4 \pi}\, \left\langle |A_{\gamma^* p \to q {\bar q} X}|^2 \right\rangle \, \frac{d^2 p_1 \, d^2 p_2}{(2 \pi)^4} \, \frac{d z}{z (1-z)} .
\end{align}
Note that now, in \eq{XS0}, the angle brackets denote the saturation/CGC averaging as defined in \eq{CGC_ave}, and have a different meaning from the angle brackets in Eqs.~\eqref{XS-2} and \eqref{XS-1}.

At the sub-eikonal order, the diagrams with the $\gamma^* \to q {\bar q}$ splitting outside the shock wave contributing to the forward quark--anti-quark inclusive production cross section are shown in \fig{FIG:diagrams}. Using the rules of small-$x$ shock-wave calculations \cite{Iancu:2003xm, Weigert:2005us, JalilianMarian:2005jf, Gelis:2010nm, Albacete:2014fwa, Kovchegov:2012mbw, Morreale:2021pnn} while working in $A^- =0$ gauge we arrive at 
\begin{align}\label{XS1}
& z (1-z) \, \frac{d\sigma_{\lambda \lambda'}^{\gamma^* p \to q {\bar q} X ; \, \textrm{out}}}{d^2 p_1 \, d^2 p_2 \, d z} = \frac{1}{2 (2 \pi)^5} \,  \left\langle A_{\lambda'} \, A^*_\lambda \right\rangle \\ 
& = \frac{1}{2 (2 \pi)^5} \, \int d^2 x_1 \, d^2 x_{1'} \, d^2 x_2 \, d^2 x_{2'} \, d^2 x_0 \, e^{- i {\un p}_1 \cdot {\un x}_{11'} - i {\un p}_2 \cdot {\un x}_{22'}} \, \sum_{\sigma_1, \sigma_2, \sigma'_1, \sigma'_2, i } \notag \\ 
& \Bigg\{ \Psi_{\lambda', \sigma_1, \sigma_2; i, i}^{\gamma^* \to q {\bar q}} ({\un x}_{02}, z) \, \left[ \Psi_{\lambda, \sigma'_1, \sigma'_2; i, i}^{\gamma^* \to q {\bar q}} ({\un x}_{1'2'}, z) \right]^* \, \frac{1}{N_c} \left\langle  \tr \left[ \tord \left( V^\textrm{pol}_{{\un x}_1, {\un x}_0; \sigma'_1, \sigma_1} \, V_{{\un x}_2}^\dagger \right) \,  \atord  \left( V_{{\un x}_{2'}} \, V_{{\un x}_{1'}}^\dagger - 1 \right) \right]  \right\rangle (zs) \, \delta_{\sigma_2 , \sigma'_2} \notag \\
& -  \Psi_{\lambda', \sigma_1, \sigma_2; i, i}^{\gamma^* \to q {\bar q}} ({\un x}_{10}, z) \, \left[ \Psi_{\lambda, \sigma'_1, \sigma'_2; i, i}^{\gamma^* \to q {\bar q}} ({\un x}_{1'2'}, z) \right]^* \frac{1}{N_c} \left\langle  \tr \left[ \tord \left( V_{{\un x}_1} \,  V^{\textrm{pol} \, \dagger}_{{\un x}_2, {\un x}_0; - \sigma'_2, - \sigma_2} \right) \, \atord  \left( V_{{\un x}_{2'}} \, V_{{\un x}_{1'}}^\dagger - 1 \right) \right]  \right\rangle ((1-z) s)  \, \delta_{\sigma_1 , \sigma'_1} \notag \\
& +  \Psi_{\lambda', \sigma_1, \sigma_2; i, i}^{\gamma^* \to q {\bar q}} ({\un x}_{12}, z) \, \left[ \Psi_{\lambda, \sigma'_1, \sigma'_2; i, i}^{\gamma^* \to q {\bar q}} ({\un x}_{02'}, z) \right]^* \frac{1}{N_c} \left\langle \tr \left[  \tord \left(  V_{{\un x}_{1}} \, V_{{\un x}_{2}}^\dagger - 1 \right)  \,  \atord \left( V_{{\un x}_{2'}} \,  V^{\textrm{pol} \, \dagger}_{{\un x}_{1'}, {\un x}_0; \sigma_1, \sigma'_1}  \right) \right]  \right\rangle (zs) \, \delta_{\sigma_2 , \sigma'_2} \notag \\
& -  \Psi_{\lambda', \sigma_1, \sigma_2; i, i}^{\gamma^* \to q {\bar q}} ({\un x}_{12}, z) \, \left[ \Psi_{\lambda, \sigma'_1, \sigma'_2; i, i}^{\gamma^* \to q {\bar q}} ({\un x}_{1'0}, z) \right]^* \frac{1}{N_c} \left\langle  \tr \left[ \tord \left( V_{{\un x}_{1}} \, V_{{\un x}_{2}}^\dagger - 1 \right) \,  \atord \left( V^{\textrm{pol} }_{{\un x}_{2'}, {\un x}_0; -\sigma_2, - \sigma'_2} \, V_{{\un x}_{1'}}^\dagger \right) \right]  \right\rangle ((1-z) s) \, \delta_{\sigma_1 , \sigma'_1} \Bigg\} .\notag
\end{align}
Here positions of the quark and anti-quark lines are labeled by $\un x_1, \un x_2, \un x_{1'}, \un x_{2'}, \un x_0$, as shown in \fig{FIG:diagrams}, with $\un x_{ij} = \un x_i - \un x_j$ and $x_{ij} = |\un x_{ij}|$ (to be used later). The quark and anti-quark helicities are denoted by $\sigma_1, \sigma_2$ in the amplitude and by $\sigma'_1, \sigma'_2$ in the complex conjugate amplitude, while the virtual photon's transverse polarizations are $\lambda'$ in the amplitude and $\lambda$ in the complex conjugate amplitude. $\Psi_{\lambda', \sigma_1, \sigma_2; i, j}^{\gamma^* \to q {\bar q}}$ is the light-cone wave function of the $\gamma^* \to q {\bar q}$ splitting \cite{Lepage:1980fj, Brodsky:1997de} with $i, j$ the fundamental color indices of the quark and anti-quark. In addition, $\tord$ and $\atord$ denote the time ordering and anti-time ordering operations, respectively.

\begin{figure}[h]
\centering
\includegraphics[width= \linewidth]{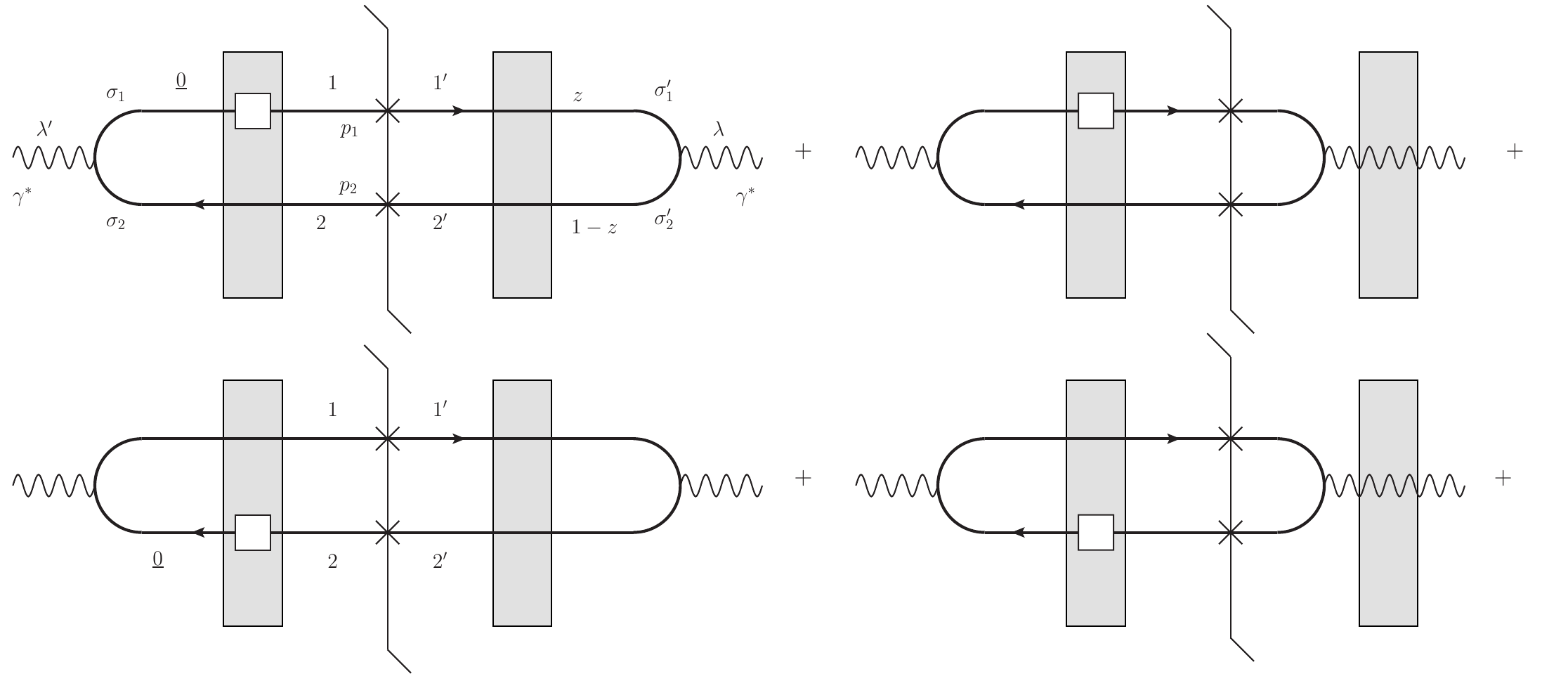}
\includegraphics[width= \linewidth]{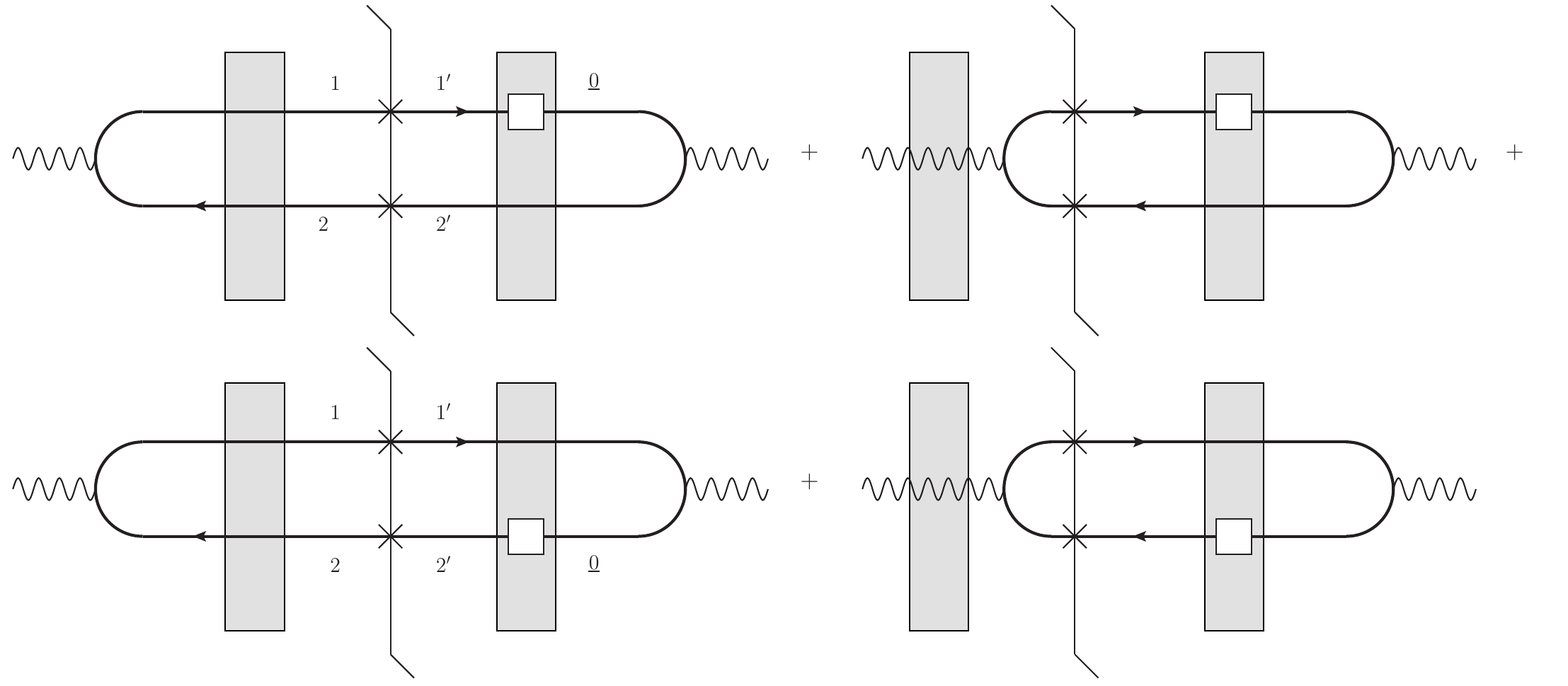}
\caption{Di-jet production diagrams contributing to the cross section at the sub-eikonal level with the $\gamma^* \to q {\bar q}$ splitting taking place outside the shock wave. The shaded rectangles denote the (proton) target shock wave, the crosses label the produced quark and anti-quark in the final state (denoted by the vertical cut), while the white square denotes the sub-eikonal interaction with the background field of the proton \cite{Kovchegov:2015pbl}.}
\label{FIG:diagrams}
\end{figure}

The interaction of the quark and anti-quark with the background field is given by its $S$-matrices, expanded in the inverse powers of energy to the sub-eikonal order. For the quark, the $S$-matrix is 
\begin{align}\label{Vxy_sub-eikonal}
V_{\un{x}, \un{y}; \sigma', \sigma}  = V_{\un x} \, \delta_{\sigma, \sigma'} \, \delta^2 ({\un x} - {\un y}) + V^{\textrm{pol}}_{\un{x}, \un{y}; \sigma', \sigma},
\end{align}
where the sub-eikonal part can be written in term of the so-called polarized Wilson lines \cite{Altinoluk:2014oxa,Balitsky:2015qba,Balitsky:2016dgz, Kovchegov:2017lsr, Kovchegov:2018znm, Chirilli:2018kkw, Jalilian-Marian:2018iui, Jalilian-Marian:2019kaf, Altinoluk:2020oyd, Kovchegov:2021iyc, Altinoluk:2021lvu, Kovchegov:2022kyy, Altinoluk:2022jkk, Altinoluk:2023qfr,Altinoluk:2023dww, Li:2023tlw}, 
\begin{align}\label{Vxy_sub-eikonal}
V_{\un{x}, \un{y}; \sigma', \sigma}^{\textrm{pol}} \equiv \sigma \, \delta_{\sigma, \sigma'} \, \left[ V_{\un x}^{\textrm{G} [1]} + V_{\un x}^{\textrm{q} [1]} \right] \, \delta^2 ({\un x} - {\un y}) + \delta_{\sigma, \sigma'} \, \left[ V_{{\ul x}, {\un y}}^{\textrm{G} [2]} + V_{{\ul x}}^{\textrm{q} [2]} \, \delta^2 ({\un x} - {\un y}) \right].
\end{align}
Here the fundamental polarized Wilson lines of the first and second type (denoted by [1] and [2] in the superscript, respectively) are
\begin{subequations}\label{VqG}
\begin{align}
& V_{\un x}^{\textrm{G} [1]}  = \frac{i \, g \, P^+}{s} \int\limits_{-\infty}^{\infty} d{x}^- V_{\un{x}} [ \infty, x^-] \, F^{12} (x^-, {\un x}) \, \, V_{\un{x}} [ x^-, -\infty]  , \label{VG1} \\
& V_{\un x}^{\textrm{q} [1]}  = \frac{g^2 P^+}{2 \, s} \int\limits_{-\infty}^{\infty} \!\! d{x}_1^- \! \int\limits_{x_1^-}^\infty d x_2^- V_{\un{x}} [ \infty, x_2^-] \, t^b \, \psi_{\beta} (x_2^-,\un{x}) \, U_{\un{x}}^{ba} [x_2^-, x_1^-] \, \left[ \gamma^+ \gamma^5 \right]_{\alpha \beta} \, \bar{\psi}_\alpha (x_1^-,\un{x}) \, t^a \, V_{\un{x}} [ x_1^-, -\infty] , \label{Vq1} \\
& V_{{\ul x}, {\un y}}^{\textrm{G} [2]}  = - \frac{i \, P^+}{s} \int\limits_{-\infty}^{\infty} d{z}^- d^2 z \ V_{\un{x}} [ \infty, z^-] \, \delta^2 (\un{x} - \un{z}) \, \cev{D}^i (z^-, {\un z}) \, D^i  (z^-, {\un z}) \, V_{\un{y}} [ z^-, -\infty] \, \delta^2 (\un{y} - \un{z}) , \label{VxyG2} \\
& V_{{\ul x}}^{\textrm{q} [2]} = - \frac{g^2 P^+}{2 \, s} \int\limits_{-\infty}^{\infty} \!\! d{x}_1^- \! \int\limits_{x_1^-}^\infty d x_2^- V_{\un{x}} [ \infty, x_2^-] \, t^b \, \psi_{\beta} (x_2^-,\un{x}) \, U_{\un{x}}^{ba} [x_2^-, x_1^-] \, \left[ \gamma^+ \right]_{\alpha \beta} \, \bar{\psi}_\alpha (x_1^-,\un{x}) \, t^a \, V_{\un{x}} [ x_1^-, -\infty] \label{Vq2},
\end{align}
\end{subequations}
with $\psi$ and $\bar \psi$ the quark and anti-quark background fields, respectively, and the adjoint light-cone Wilson line
\begin{align}\label{Vline}
U_{\un{x}} [x^-_f,x^-_i] = \mathcal{P} \exp \left[ ig \int\limits_{x^-_i}^{x^-_f} d{x}^- {\cal A}^+ (0^+, x^-, \un{x}) \right].
\end{align}
The anti-quark $S$-matrix is
\begin{align}\label{Vxy_sub-eikonal_anti}
{\overline V}_{\un{x}, \un{y}; \sigma', \sigma}  = & \ V_{\un x}^\dagger \, \delta_{\sigma, \sigma'} \, \delta^2 ({\un x} - {\un y}) + \sigma \, \delta_{\sigma, \sigma'} \, \left[ V_{\un x}^{\textrm{G} [1] \, \dagger} + V_{\un x}^{\textrm{q} [1] \, \dagger} \right] \, \delta^2 ({\un x} - {\un y}) - \delta_{\sigma, \sigma'} \, \left[ V_{{\ul x}, {\un y}}^{\textrm{G} [2] \, \dagger} + V_{{\ul x}}^{\textrm{q} [2] \, \dagger} \, \delta^2 ({\un x} - {\un y}) \right] \\
= & \ V_{\un x}^\dagger \, \delta_{\sigma, \sigma'} \, \delta^2 ({\un x} - {\un y}) - V^{\textrm{pol} \, \dagger}_{\un{x}, \un{y}; - \sigma', - \sigma} . \notag
\end{align}

The $\gamma^* \to q {\bar q}$ light-cone wave functions are (for massless quarks)
\begin{subequations}\label{LCwf}
\begin{align}
& \Psi_{\lambda = \pm 1, \sigma_1, \sigma_2; i, j}^{\gamma^* \to q {\bar q}} ({\un x}_{12}, z) = \frac{e Z_f}{2 \pi} \, \delta_{ij} \, \delta_{\sigma_1, - \sigma_2} \, z \, (1-z) \, (1-2 z + \sigma_1 \, \lambda) \, i Q \, \frac{{\un \epsilon}_\lambda \cdot {\un x}_{12}}{x_{12}} \, K_1 \left( x_{12} \, a_f \right)  \label{LCwfT}, \\
& \Psi_{\lambda = 0, \sigma_1, \sigma_2; i, j}^{\gamma^* \to q {\bar q}} ({\un x}_{12}, z) = - \frac{e Z_f}{2 \pi} \, \delta_{ij} \, \delta_{\sigma_1, - \sigma_2} \,  [z \, (1-z)]^{3/2} \, 2 Q \, K_0 \left( x_{12} \, a_f \right) .
\end{align}
\end{subequations}
Here $a_f = Q\sqrt{z(1-z)}$. These expressions can be compared to those in \cite{Kovchegov:2012mbw} and references therein: note that they have been calculated for the minus-(right-)moving virtual photon with the polarization 2-vector ${\un \epsilon}_\lambda = - (1/\sqrt{2}) (-\lambda, i)$ and for the ``anti-Brodsky-Lepage" spinors from \cite{Kovchegov:2018znm}.

The above derivations assume that there is no significant longitudinal (minus) momentum exchange when the quark-antiquark pair from the virtual photon splitting interacts with the proton, as indicated by the Dirac delta function $\delta(q^--p_1^--p_2^-)$ in Eq.~\eqref{eq:p-_conservation}. This is certainly true for the eikonal interactions. On the other hand, it is known \cite{Chirilli:2018kkw, Chirilli:2021lif, Altinoluk:2021lvu, Li:2023tlw} that at sub-eikonal order the $F^{+-}$ component of the background gluon field-strength tensor is responsible for longitudinal momentum exchange via what we label here as the polarized Wilson line of type-3, 
\begin{align}
V_{\un x}^{\textrm{G} [3]}  = \frac{i \, g \, P^+}{s} \int\limits_{-\infty}^{\infty} d{x}^- V_{\un{x}} [ \infty, x^-] \, F^{+-} (x^-, {\un x}) \, \, V_{\un{x}} [ x^-, -\infty]  . \label{VG3} 
\end{align}
The origin of this contribution, not accounted for in \cite{Kovchegov:2021iyc}, can be easily understood using the following line of argument. In the eikonal approximation, the $x^+$-dependence of the Wilson line on the $x^-$-light cone is often neglected and one puts $x^+ =0$ in most calculations (cf. Eqs.~\eqref{eq:V_def} and \eqref{Vline}). Let us, for a moment, restore this $x^+$ dependence, and define
\begin{equation}\label{eq:V_def2}
V_{\un{x}} (x^+) = \mathcal{P} \mathrm{exp}\left\{ig \int_{-\infty}^{\infty} d\xi^- A^+(x^+, \xi^-, \un{x})\right\} .
\end{equation}
The momentum space quark $S$-matrix is proportional to 
\begin{align}
    \int\limits_{-\infty}^\infty d x^+ \, e^{i (p_f^- - p_i^-) \, x^+} \, V_{\un{x}} (x^+)
\end{align}
with some initial and final momenta $p_i$ and $p_f$ for a quark scattering on the background-field shock wave. In the eikonal limit, if we replace $V_{\un{x}} (x^+) \approx V_{\un{x}} (0^+)$, we get
\begin{align}
    \int\limits_{-\infty}^\infty d x^+ \, e^{i (p_f^- - p_i^-) \, x^+} \, V_{\un{x}} (0^+) = 2 \pi \, \delta (p_f^- - p_i^-) \, V_{\un{x}} (0^+) , 
\end{align}
ensuring no minus-momentum transfer to the target.

At the sub-eikonal order we expand the eikonal Wilson line in the powers of $x^+$, obtaining
\begin{align}\label{V_expansion}
    & \int\limits_{-\infty}^\infty d x^+ \, e^{i (p_f^- - p_i^-) \, x^+} \, V_{\un{x}} (x^+) = \int\limits_{-\infty}^\infty d x^+ \, e^{i (p_f^- - p_i^-) \, x^+} \, \left[ V_{\un{x}} (0^+) + x^+ \, \pd^- V_{\un{x}} (0^+) + \ldots \right] \\
    & = 2 \pi \, \delta (p_f^- - p_i^-) \, V_{\un{x}} (0^+) +  2 \pi \, i \, \left[ \frac{\pd}{\pd p_f^-} \delta (p_f^- - p_i^-) \right] \, 2 \sqrt{p_f^- \, p_i^-} \, V_{\un x}^{\textrm{G} [3]} + \ldots \, . \notag
\end{align}
Here $V_{\un x}^{\textrm{G} [3]}$ is defined in \eq{VG3} with $s \approx 2 P^+ \sqrt{p_f^- \, p_i^-}$.  
We are also working in $A^- =0$ light-cone gauge, in which $F^{+-} = - \pd^- A^+$. Note that the derivative of the $\delta$-function in the last term in \eq{V_expansion} implies a non-zero minus momentum transfer to the proton target. (The overall minus component of momentum is, indeed, conserved in the scattering process.) Finally, the ellipses in \eq{V_expansion} denote the higher orders in the expansion, which are sub-sub-eikonal and beyond: they are not considered in our sub-eikonal calculation.

One may question whether the sub-eikonal operator $V_{\un{x}}^{\textrm{G}[3]}$ will contribute to the dijet production at hand and/or to the helicity evolution of \cite{Kovchegov:2015pbl,Kovchegov:2018znm,Cougoulic:2022gbk, Borden:2024bxa}. We have explicitly verified that the small-$x$ evolution of the ``polarized dipole scattering amplitude" constructed out of the $V_{\un{x}}^{\textrm{G}[3]}$ operator,
\begin{equation}\label{G3_def}
G^{[3]}_{10} (s) \equiv \frac{1}{2N_c} \llangle \mathrm{tr}\left[V_{\un{x}_0}^{\dagger} \, V_{\un{x}_1}^{\textrm{G}[3]} \right]\rrangle (s) + \cc ,
\end{equation} 
does not mix with the polarized dipole amplitudes employed in \cite{Kovchegov:2015pbl,Kovchegov:2018znm,Cougoulic:2022gbk, Borden:2024bxa} in the DLA, which resums powers of $\as \, \ln^2 (1/x)$ and $\as \, \ln (1/x) \, \ln (Q^2/\Lambda^2)$, with $\Lambda$ an infrared (IR) cutoff. The operators do mix in the single logarithmic approximation (SLA), which resums powers of $\as \, \ln (1/x)$, a sub-leading parameter in helicity evolution at small $x$. This is also visible from the evolution for $G_{10}^{WW}$ given in \eq{eq:final_result_WW} below and its DLA version \eqref{eq:WW_evol_DLA}: while $V_{\un{x}}^{\textrm{G}[3]}$ (or, more precisely, its adjoint version $U_{\un{x}}^{\textrm{G}[3]}$) appears in the former, it does not contribute to the latter. Moreover, the operator $V_{\un{x}}^{\textrm{G}[3]}$ does not couple to the longitudinal spin of the proton at the leading (Born) order of the calculations employed when constructing the initial conditions for small-$x$ DLA evolution, since it does not depend on the transverse gluon field $A^i$, which couples to the longitudinal proton spin at that order \cite{Kovchegov:2017lsr}.  We conclude that the DLA helicity evolution of \cite{Kovchegov:2015pbl,Kovchegov:2018znm,Cougoulic:2022gbk, Borden:2024bxa} is not affected by $V_{\un{x}}^{\textrm{G}[3]}$ and that one can safely ignore the contribution of $V_{\un{x}}^{\textrm{G}[3]}$ in the present DLA-level dijet production calculation. Even if the $V_{\un{x}}^{\textrm{G}[3]}$ contributes to the dijet production cross section, this contribution will be zero in the DLA and/or at the Born level of the initial conditions: it is, therefore, outside the precision of the present calculation.

\begin{figure}[h]
\centering
\includegraphics[width= 0.75\linewidth]{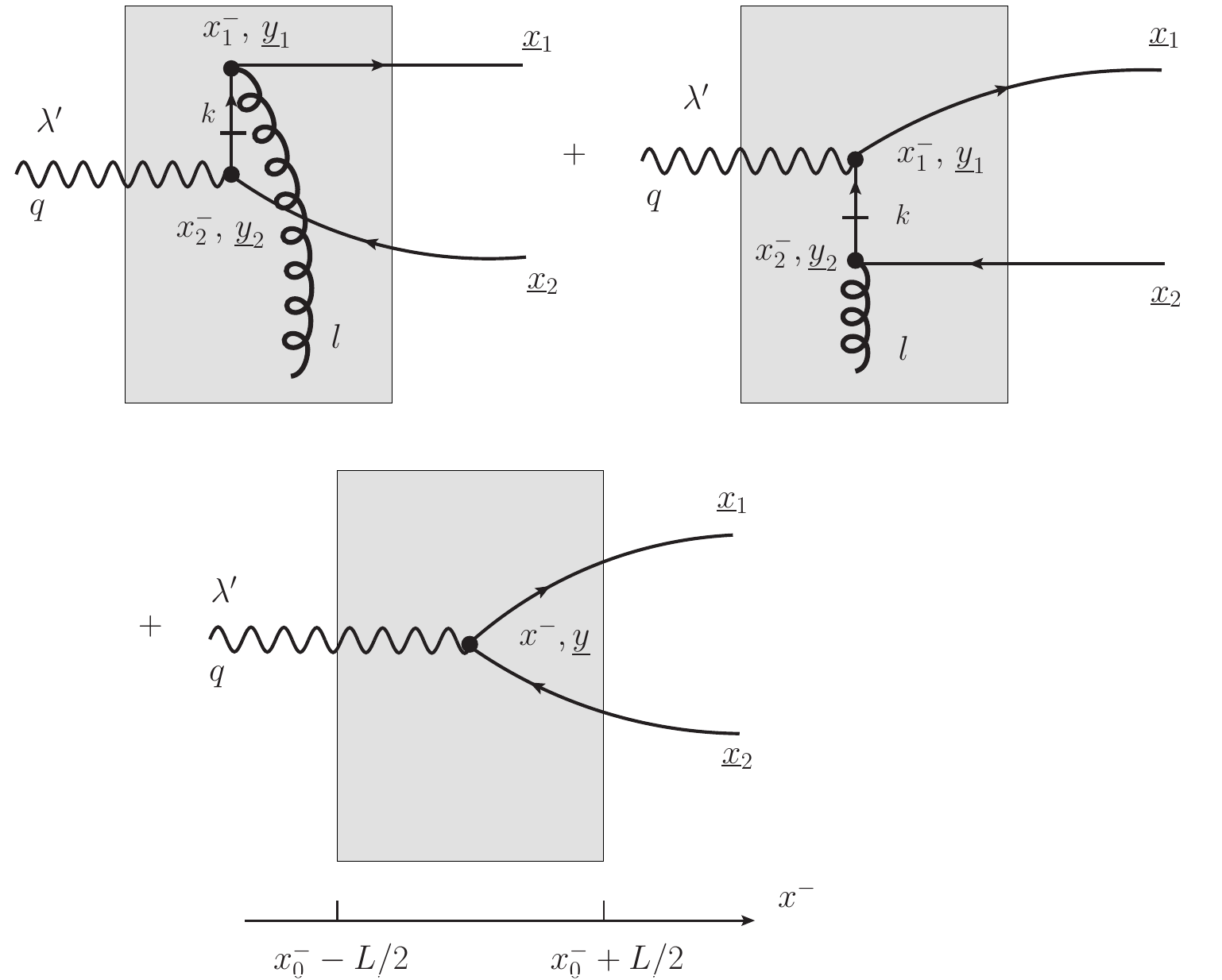}  
\caption{Di-jet production diagrams at the sub-eikonal level with the $\gamma^* \to q {\bar q}$ splitting taking place inside the shock wave. The vertical line with a notch across it denotes the instantaneous term for the quark and anti-quark propagators \cite{Lepage:1980fj, Brodsky:1997de}.}
\label{FIG:in_shock_wave}
\end{figure}

In addition to the diagrams in \fig{FIG:diagrams}, we need to calculate the contribution of the $\gamma^* \to q {\bar q}$ splitting happening inside the shock wave. The diagrams, at the amplitude level, are shown in \fig{FIG:in_shock_wave}, and also include the contributions of the instantaneous term in the light-cone perturbation theory terminology \cite{Lepage:1980fj, Brodsky:1997de}, denoted by a vertical line with a notch. The top two diagrams in \fig{FIG:in_shock_wave} give, for the transverse polarizations of the virtual photon considered here,
\begin{align}\label{XS_inst1}
& z (1-z) \, \frac{d\sigma_{\lambda \lambda'}^{\gamma^* p \to q {\bar q} X ; \, \textrm{inst} }}{d^2 p_1 \, d^2 p_2 \, d z} = \frac{1}{2 (2 \pi)^5} \, \int d^2 x_1 \, d^2 x_{1'} \, d^2 x_2 \, d^2 x_{2'} \, e^{- i {\un p}_1 \cdot {\un x}_{11'} - i {\un p}_2 \cdot {\un x}_{22'}} \, \sum_{\sigma_1, \sigma_2, i } \\ 
& \Bigg\{ \Psi_{\lambda', \sigma_1, \sigma_2}^{j , \, \gamma^* \to q {\bar q}} ({\un x}_{12}, z) \, \left[ \Psi_{\lambda, \sigma_1, \sigma_2; i, i}^{\gamma^* \to q {\bar q}} ({\un x}_{1'2'}, z) \right]^* \, \frac{1}{N_c} \left\langle  \tr \left[ \tord \left( {\cal O}_{{\un x}_1}^j (z s) \right) \,  \atord  \left( V_{{\un x}_{2'}} \, V_{{\un x}_{1'}}^\dagger - 1 \right) \right]  \right\rangle  \notag \\
& + \Psi_{\lambda', \sigma_2, \sigma_1}^{j , \, \gamma^* \to q {\bar q}} ({\un x}_{12}, z) \, \left[ \Psi_{\lambda, \sigma_1, \sigma_2; i, i}^{\gamma^* \to q {\bar q}} ({\un x}_{1'2'}, z) \right]^* \frac{1}{N_c} \left\langle  \tr \left[ \tord \left( {\cal O}_{{\un x}_1}^{j \, \dagger} ((1-z) s)  \right) \, \atord  \left( V_{{\un x}_{2'}} \, V_{{\un x}_{1'}}^\dagger - 1 \right) \right]  \right\rangle  \notag \\
& +  \Psi_{\lambda', \sigma_1, \sigma_2; i, i}^{\gamma^* \to q {\bar q}} ({\un x}_{12}, z) \, \left[ \Psi_{\lambda, \sigma_1, \sigma_2}^{j , \, \gamma^* \to q {\bar q}} ({\un x}_{1'2'}, z) \right]^* \frac{1}{N_c} \left\langle \tr \left[  \tord \left(  V_{{\un x}_{1}} \, V_{{\un x}_{2}}^\dagger - 1 \right)  \,  \atord \left( {\cal O}_{{\un x}_{1'}}^{j \, \dagger} (z s)  \right) \right]  \right\rangle  \notag \\
& + \Psi_{\lambda', \sigma_1, \sigma_2; i, i}^{\gamma^* \to q {\bar q}} ({\un x}_{12}, z) \, \left[ \Psi_{\lambda, \sigma_2, \sigma_1}^{j , \, \gamma^* \to q {\bar q}} ({\un x}_{1'2'}, z)  \right]^* \frac{1}{N_c} \left\langle  \tr \left[ \tord \left( V_{{\un x}_{1}} \, V_{{\un x}_{2}}^\dagger - 1 \right) \,  \atord \left( {\cal O}_{{\un x}_{1'}}^{j} ((1-z) s) \right) \right]  \right\rangle \Bigg\} ,\notag
\end{align}
where 
\begin{align}\label{eq:LCWF_inside}
\Psi_{\lambda', \sigma_1, \sigma_2}^{j , \, \gamma^* \to q {\bar q}} ({\un x}_{12}, z) = - 2 \, e Z_f \, \sqrt{z \, (1-z)} \, \delta_{\sigma_1, - \sigma_2} \, \delta_{\sigma_1 , \lambda'} \, \epsilon^j_{\lambda'} \, \delta^2 ({\un x}_{12})
\end{align}
for $\lambda' = \pm 1$ and 
\begin{align}
{\cal O}_{\un x}^j (s) = -  \frac{i g P^+}{s} \, \int\limits_{-\infty}^{\infty} d{x}^- \, V_{\un{x}} [ \infty, x^-] \, A^{j}  (x^-, {\un x}) \, V_{\un{x}} [ x^-, \infty]  .
\end{align}

A direct calculation of the third diagram in \fig{FIG:in_shock_wave} while neglecting the longitudinal phase, which, in this case, gives a sub-sub-eikonal correction, yields, at the amplitude level,
\begin{align}
& 2 \, e Z_f \, \sqrt{z \, (1-z)} \, \delta_{\sigma_1, - \sigma_2} \, \int\limits_{x_0^- -L/2}^{x_0^- + L/2} d x^- \,  \int d^2 x_1 \,  d^2 x_2 \,  e^{- i {\un p}_1 \cdot {\un x}_{1} - i {\un p}_2 \cdot {\un x}_{2}} \, V_{\un{x}_1} [ \infty, x^-] \, V_{\un{x}_2} [ x^-, \infty] \\ 
& \times \, \left[ \frac{1}{2 z q^-} \, \delta_{\lambda' \sigma_1} \, {\un \epsilon}_{\lambda'} \cdot {\un \nabla}_1 + \frac{1}{2 (1-z) q^-} \, \delta_{\lambda' \sigma_2} \, {\un \epsilon}_{\lambda'} \cdot {\un \nabla}_2  \right]  \, \delta^2 ({\un x}_{12}) , \notag 
\end{align}
where we introduced the position of the shock wave center $x_0^-$ explicitly, along with the shock wave width $L$. Writing
\begin{align}
{\un \nabla}_1 \, \delta^2 ({\un x}_{12}) = \thalf \, \left[ {\un \nabla}_1 - {\un \nabla}_2 \right] \, \delta^2 ({\un x}_{12})
\end{align}
and integrating by parts we obtain (while repeating the same trick for ${\un \nabla}_2$)
\begin{align}\label{d-d1}
& e Z_f \, \sqrt{z \, (1-z)} \, \delta_{\sigma_1, - \sigma_2} \, \epsilon^i_{\lambda'} \, \int d^2 x_1 \,  d^2 x_2 \,  e^{- i {\un p}_1 \cdot {\un x}_{1} - i {\un p}_2 \cdot {\un x}_{2}} \, \delta^2 ({\un x}_{12})  \\ 
& \times \, \left[ \frac{1}{2 z q^-} \, \delta_{\lambda' \sigma_1} -  \frac{1}{2 (1-z) q^-} \, \delta_{\lambda' \sigma_2}  \right] \, \int\limits_{x_0^- -L/2}^{x_0^- + L/2} d x^- \,  \left[ V_{\un{x}_1} [ \infty, x^-]  \,  \left( \cev{\pd}_1^i - \pd_1^i + i (p_1^i - p_2^i)  \right)  \, V_{\un{x}_1} [ x^-, \infty] \right] \notag 
\end{align}
with the partial derivatives acting now only on the Wilson lines. 

Simplifying 
\begin{subequations}
\begin{align}
& \int\limits_{x_0^- -L/2}^{x_0^- + L/2} d x^- \,  \left[ V_{\un{x}_1} [ \infty, x^-]  \,  \left( \cev{\pd}_1^i - \pd_1^i + i (p_1^i - p_2^i)  \right)  \, V_{\un{x}_1} [ x^-, \infty] \right] \\ 
& = - \int\limits_{-\infty}^{\infty} d x^- \, \left( V_{\un{x}_1} [ \infty, x^-]  \, \left( \pd_1^i - \cev{\pd}_1^i \right) \, V_{\un{x}_1} [ x^-, -\infty] \right) \, V_{{\un x}_1}^\dagger  - L \, V_{{\un x}_1} \left[ \pd_1^i - i (p_1^i - p_2^i) \right] \, V_{{\un x}_1}^\dagger + 2 x_0^- \, V_{{\un x}_1} \, \pd_1^i  \, V_{{\un x}_1}^\dagger \notag , \\
& \int\limits_{x_0^- -L/2}^{x_0^- + L/2} d x^- \,  \left[ V_{\un{x}_1} [ \infty, x^-]  \,  \left( \cev{\pd}_1^i - \pd_1^i + i (p_1^i - p_2^i)  \right)  \, V_{\un{x}_1} [ x^-, \infty] \right] \\ 
& = \int\limits^{-\infty}_{\infty} d x^- \, V_{{\un x}_1} \, \left( V_{\un{x}_1} [ - \infty, x^-]  \, \left( \pd_1^i - \cev{\pd}_1^i \right) \, V_{\un{x}_1} [ x^-, \infty] \right)  + L \, V_{{\un x}_1} \left[ {\cev \pd}_1^i + i (p_1^i - p_2^i) \right] \, V_{{\un x}_1}^\dagger  + 2 x_0^- \, V_{{\un x}_1} \, \pd_1^i  \, V_{{\un x}_1}^\dagger \notag ,
\end{align}
\end{subequations}
and dropping the terms proportional to $L$ and to $x_0^-$ we recast \eq{d-d1} as
\begin{align}\label{d-d2}
& \int d^2 x_1 \,  d^2 x_2 \,  e^{- i {\un p}_1 \cdot {\un x}_{1} - i {\un p}_2 \cdot {\un x}_{2}}  \, \left[ \frac{p^+_1}{2 z s} \, \Psi_{\lambda', \sigma_1, \sigma_2}^{i , \, \gamma^* \to q {\bar q}} ({\un x}_{12}, z) \, \int\limits_{-\infty}^{\infty} d x^- \, \left( V_{\un{x}_1} [ \infty, x^-]  \, \left( \pd_1^i - \cev{\pd}_1^i \right) \, V_{\un{x}_1} [ x^-, -\infty] \right) \, V_{{\un x}_1}^\dagger \right. \\ 
& \left.  + \frac{p_1^+}{2 (1-z) s} \, \Psi_{\lambda', \sigma_2, \sigma_1}^{i , \, \gamma^* \to q {\bar q}} ({\un x}_{12}, z)  \, \int\limits^{-\infty}_{\infty} d x^- \, V_{{\un x}_1} \, \left( V_{\un{x}_1} [ - \infty, x^-]  \, \left( \pd_1^i - \cev{\pd}_1^i \right) \, V_{\un{x}_1} [ x^-, \infty] \right) \right] . \notag 
\end{align}
We will return to the terms proportional to $L$ and $x_0^-$ shortly.

Adding to \eq{d-d2} the contributions to the sub-eikonal amplitude entering \eq{XS_inst1} we arrive at 
\begin{align}\label{d-d3}
& \int d^2 x_1 \,  d^2 x_2 \,  e^{- i {\un p}_1 \cdot {\un x}_{1} - i {\un p}_2 \cdot {\un x}_{2}}  \, \left[ \Psi_{\lambda', \sigma_1, \sigma_2}^{i , \, \gamma^* \to q {\bar q}} ({\un x}_{12}, z) \ \tord \left( V^{i \, \textrm{G} [2]}_{{\un x}_1}  \, V_{{\un x}_1}^\dagger \right) (zs)  +  \Psi_{\lambda', \sigma_2, \sigma_1}^{i , \, \gamma^* \to q {\bar q}} ({\un x}_{12}, z)  \ \tord \left( V_{{\un x}_1} \, V^{i \, \textrm{G} [2] \dagger}_{{\un x}_1} \right) ((1-z) s) \right] . 
\end{align}
We conclude that the final result for the $\gamma^* \to q {\bar q}$ splitting inside the shock wave is obtained from \eq{XS_inst1} by replacing 
\begin{align}
{\cal O}_{{\un x}_1}^j  \to V^{j \, \textrm{G} [2]}_{{\un x}_1}  \, V_{{\un x}_1}^\dagger  
\end{align} 
in the latter. Performing this replacement in \eq{XS_inst1} we arrive at the net contribution coming from the $\gamma^* \to q {\bar q}$ splitting being inside the shock wave,
\begin{align}\label{XS_inst2}
& z (1-z) \, \frac{d\sigma_{\lambda \lambda'}^{\gamma^* p \to q {\bar q} X ; \, \textrm{in} }}{d^2 p_1 \, d^2 p_2 \, d z} = \frac{1}{2 (2 \pi)^5} \, \int d^2 x_1 \, d^2 x_{1'} \, d^2 x_2 \, d^2 x_{2'} \, e^{- i {\un p}_1 \cdot {\un x}_{11'} - i {\un p}_2 \cdot {\un x}_{22'}} \, \sum_{\sigma_1, \sigma_2, i } \\ 
& \Bigg\{ \Psi_{\lambda', \sigma_1, \sigma_2}^{j , \, \gamma^* \to q {\bar q}} ({\un x}_{12}, z) \, \left[ \Psi_{\lambda, \sigma_1, \sigma_2; i, i}^{\gamma^* \to q {\bar q}} ({\un x}_{1'2'}, z) \right]^* \, \frac{1}{N_c} \left\langle  \tr \left[ \tord \left( V^{j \, \textrm{G} [2]}_{{\un x}_1}  \, V_{{\un x}_1}^\dagger \right)  \,  \atord  \left( V_{{\un x}_{2'}} \, V_{{\un x}_{1'}}^\dagger - 1 \right) \right]  \right\rangle (zs) \notag \\
& + \Psi_{\lambda', \sigma_2, \sigma_1}^{j , \, \gamma^* \to q {\bar q}} ({\un x}_{12}, z) \, \left[ \Psi_{\lambda, \sigma_1, \sigma_2; i, i}^{\gamma^* \to q {\bar q}} ({\un x}_{1'2'}, z) \right]^* \frac{1}{N_c} \left\langle  \tr \left[ \tord \left( V_{{\un x}_1} \, V^{j \, \textrm{G} [2] \dagger}_{{\un x}_1}  \right) \, \atord  \left( V_{{\un x}_{2'}} \, V_{{\un x}_{1'}}^\dagger - 1 \right) \right]  \right\rangle  ((1-z) s)   \notag \\
& +  \Psi_{\lambda', \sigma_1, \sigma_2; i, i}^{\gamma^* \to q {\bar q}} ({\un x}_{12}, z) \, \left[ \Psi_{\lambda, \sigma_1, \sigma_2}^{j , \, \gamma^* \to q {\bar q}} ({\un x}_{1'2'}, z) \right]^* \frac{1}{N_c} \left\langle \tr \left[  \tord \left(  V_{{\un x}_{1}} \, V_{{\un x}_{2}}^\dagger - 1 \right)  \,  \atord \left( V_{{\un x}_{1'}} \, V^{j \, \textrm{G} [2] \dagger}_{{\un x}_{1'}}  \right) \right]  \right\rangle (z s)  \notag \\
& + \Psi_{\lambda', \sigma_1, \sigma_2; i, i}^{\gamma^* \to q {\bar q}} ({\un x}_{12}, z) \, \left[ \Psi_{\lambda, \sigma_2, \sigma_1}^{j , \, \gamma^* \to q {\bar q}} ({\un x}_{1'2'}, z)  \right]^* \frac{1}{N_c} \left\langle  \tr \left[ \tord \left( V_{{\un x}_{1}} \, V_{{\un x}_{2}}^\dagger - 1 \right) \,  \atord \left( V^{j \, \textrm{G} [2]}_{{\un x}_{1'}}  \, V_{{\un x}_{1'}}^\dagger  \right) \right]  \right\rangle ((1-z) s) \Bigg\} . \notag
\end{align}
The final result for the forward inclusive dijet production is
\begin{align}\label{XS_tot}
z (1-z) \, \frac{d\sigma_{\lambda \lambda'}^{\gamma^* p \to q {\bar q} X }}{d^2 p_1 \, d^2 p_2 \, d z} = z (1-z) \, \frac{d\sigma_{\lambda \lambda'}^{\gamma^* p \to q {\bar q} X ; \, \textrm{out}}}{d^2 p_1 \, d^2 p_2 \, d z} + z (1-z) \, \frac{d\sigma_{\lambda \lambda'}^{\gamma^* p \to q {\bar q} X ; \, \textrm{in} }}{d^2 p_1 \, d^2 p_2 \, d z},
\end{align}
with the cross sections on the right given by Eqs.~\eqref{XS1} and \eqref{XS_inst2}.

\begin{figure}[h]
\centering
\includegraphics[width= 0.85\linewidth]{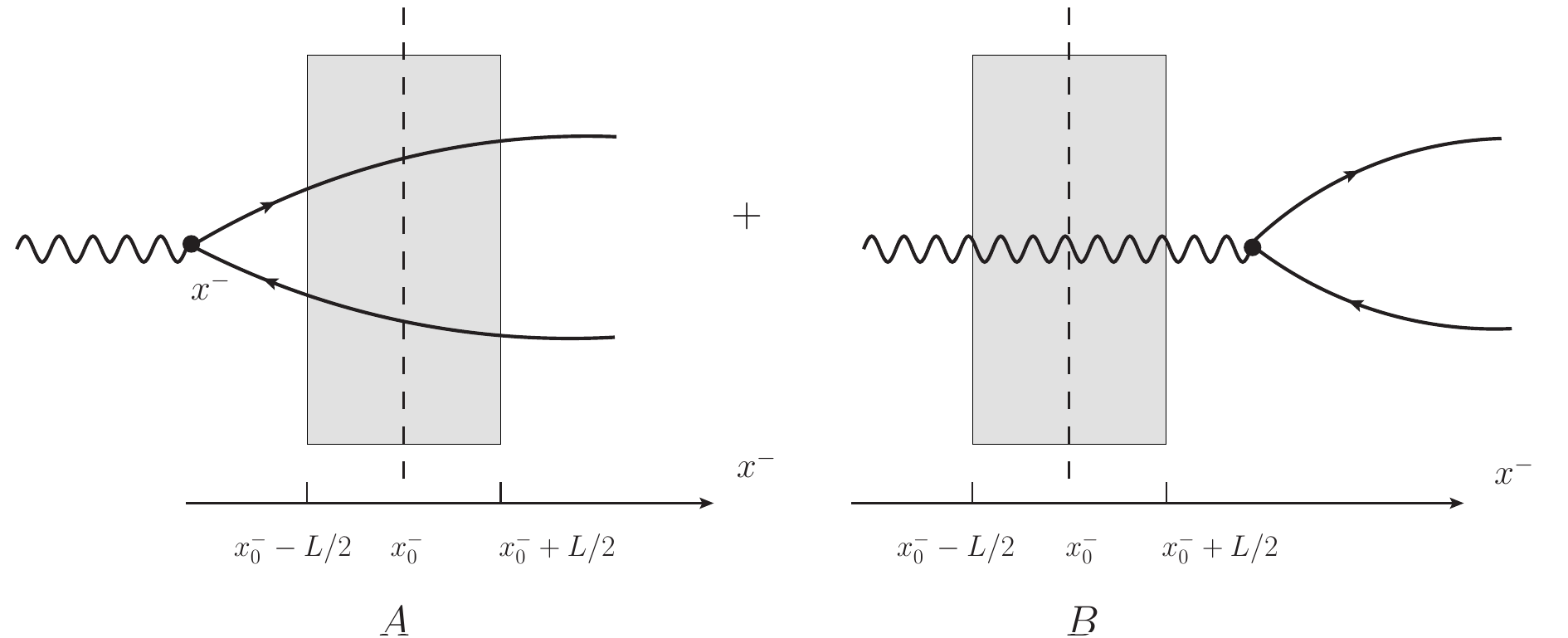}  
\caption{Diagrams illustrating that the eikonal contribution comes from the $\gamma^* \to q {\bar q}$ splitting happening outside the shock wave, either to the left (A) or to the right (B) of it.}
\label{FIG:remainder}
\end{figure}

In arriving at \eq{XS_inst2} we have neglected the terms proportional to $L$ and $x_0^-$, whose contribution to the sub-eikonal dijet production amplitude is
\begin{align}\label{remainder}
& \int d^2 x_1 \,  d^2 x_2 \,  e^{- i {\un p}_1 \cdot {\un x}_{1} - i {\un p}_2 \cdot {\un x}_{2}}  \, \left[ \frac{p^+_1}{2 z s} \, \Psi_{\lambda', \sigma_1, \sigma_2}^{i , \, \gamma^* \to q {\bar q}} ({\un x}_{12}, z) \, \left\{ L \, V_{{\un x}_1} \left[ \pd_1^i - i (p_1^i - p_2^i) \right] \, V_{{\un x}_1}^\dagger - 2 x_0^- \, V_{{\un x}_1} \, \pd_1^i  \, V_{{\un x}_1}^\dagger \right\} \right. \\ 
& \left.  + \frac{p_1^+}{2 (1-z) s} \, \Psi_{\lambda', \sigma_2, \sigma_1}^{i , \, \gamma^* \to q {\bar q}} ({\un x}_{12}, z)  \, \left\{ L \, V_{{\un x}_1} \left[ {\cev \pd}_1^i + i (p_1^i - p_2^i) \right] \, V_{{\un x}_1}^\dagger \right] + 2 x_0^- \, V_{{\un x}_1} \, \pd_1^i  \, V_{{\un x}_1}^\dagger \right\} \notag \\
& =  \int d^2 x_1 \,  d^2 x_2 \,  e^{- i {\un p}_1 \cdot {\un x}_{1} - i {\un p}_2 \cdot {\un x}_{2}}  \, \left[ \frac{p^+_1}{2 z s} \, \Psi_{\lambda', \sigma_1, \sigma_2}^{i , \, \gamma^* \to q {\bar q}} ({\un x}_{12}, z)  -  \frac{p_1^+}{2 (1-z) s} \, \Psi_{\lambda', \sigma_2, \sigma_1}^{i , \, \gamma^* \to q {\bar q}} ({\un x}_{12}, z)  \right] \notag \\ 
& \times \, \left\{ L \, V_{{\un x}_1} \left[ \pd_1^i - i (p_1^i - p_2^i) \right] \, V_{{\un x}_1}^\dagger - 2 x_0^- \, V_{{\un x}_1} \, \pd_1^i  \, V_{{\un x}_1}^\dagger \right\} \notag \\
& = \int d^2 x_1 \,  d^2 x_2 \,  e^{- i {\un p}_1 \cdot {\un x}_{1} - i {\un p}_2 \cdot {\un x}_{2}}  \, \left[ \frac{p^+_1}{2 z s} \, \Psi_{\lambda', \sigma_1, \sigma_2}^{i , \, \gamma^* \to q {\bar q}} ({\un x}_{12}, z)  -  \frac{p_1^+}{2 (1-z) s} \, \Psi_{\lambda', \sigma_2, \sigma_1}^{i , \, \gamma^* \to q {\bar q}} ({\un x}_{12}, z)  \right] \notag  \\ 
& \times \, \left\{ ( L - 2 x_0^- ) \, V_{{\un x}_1} \, \pd_1^i  \, V_{{\un x}_1}^\dagger \,  - i \, L \, (p_1^i - p_2^i)  \right\}  . \notag
\end{align}
These terms are canceled by removing the double-counting present in \eq{XS1}. Indeed, in arriving at \eq{XS1} we have integrated the $x^-$-position of the $\gamma^* \to q {\bar q}$ splitting in the eikonal production amplitude over $x^- \in [-\infty, 0]$ for the splitting happening before the shock wave interaction and over $x^- \in [0, \infty]$ for the splitting taking place after the interaction. While this is correct in the eikonal approximation, in which the shock wave can be assumed to be infinitely thin, when performing the calculation with sub-eikonal accuracy we need to treat the integration limits more carefully: the integration regions should be  $x^- \in [-\infty, (-L/2) + x_0^-]$ and $x^- \in [(L/2) + x_0^-, \infty]$. This is illustrated in \fig{FIG:remainder}, where the splitting vertex should be always outside the shock wave.

Therefore, we need to subtract the contributions with the $\gamma^* \to q {\bar q}$ splitting happening in the $x^- \in [(-L/2) + x_0^-, x^-_0]$ and $x^- \in [x^-_0, (L/2) + x_0^-]$ intervals. We further note that in \eq{XS1} the interaction with the shock wave is always given by infinite Wilson lines, even for $x^- \in [(-L/2) + x_0^-, x^-_0]$ and $x^- \in [x^-_0, (L/2) + x_0^-]$: therefore, we need to subtract a contribution with the complete infinite Wilson lines as well, even if $x^-$-position of the splitting is inside the shock wave. On the other hand, since in the subtraction the $x^-$-position of the splitting is located inside the shock wave, and is integrated over a finite $x^-$ range, instead of the semi-infinite one, the subtraction calculation leads to the wave function in \eq{eq:LCWF_inside}, akin to the last diagram in \fig{FIG:in_shock_wave}. 
To summarize, we need to subtract the contribution of diagram A from \fig{FIG:remainder}, with the splitting happening before the shock wave interaction (as far as the Wilson lines are concerned), but with the splitting vertex placed still in the shock wave, at $x^- \in [(-L/2) + x_0^-, x^-_0]$. We also need to subtract the contribution of diagram B from \fig{FIG:remainder} with the splitting occurring after the shock wave (for the Wilson lines, which are equal to 1 in this case), but with the splitting vertex at $x^- \in [x^-_0, (L/2) + x_0^-]$.

The sum of the subtractions for the diagrams A and B from \fig{FIG:remainder} yields
\begin{align}\label{remainder2}
& - \int d^2 x_1 \,  d^2 x_2 \,  e^{- i {\un p}_1 \cdot {\un x}_{1} - i {\un p}_2 \cdot {\un x}_{2}}  \, \left[ \frac{p^+_1}{2 z s} \, \Psi_{\lambda', \sigma_1, \sigma_2}^{i , \, \gamma^* \to q {\bar q}} ({\un x}_{12}, z)  -  \frac{p_1^+}{2 (1-z) s} \, \Psi_{\lambda', \sigma_2, \sigma_1}^{i , \, \gamma^* \to q {\bar q}} ({\un x}_{12}, z)  \right] \\
& \times \, \left\{ \int\limits_{(-L/2) + x_0^-}^{x^-_0} \!\!\!\!\! d x^- \,  \left[ V_{\un{x}_1}  \,  \left( \cev{\pd}_1^i - \pd_1^i + i (p_1^i - p_2^i)  \right)  \, V_{\un{x}_1}^\dagger  \right] +  \int\limits^{(L/2) + x_0^-}_{x^-_0} \!\!\!\!\! d x^- \,  \left[ 1  \,  \left( \cev{\pd}_1^i - \pd_1^i + i (p_1^i - p_2^i)  \right)  \, 1  \right] \right\} \notag \\
& = \int d^2 x_1 \,  d^2 x_2 \,  e^{- i {\un p}_1 \cdot {\un x}_{1} - i {\un p}_2 \cdot {\un x}_{2}}  \, \left[ \frac{p^+_1}{2 z s} \, \Psi_{\lambda', \sigma_1, \sigma_2}^{i , \, \gamma^* \to q {\bar q}} ({\un x}_{12}, z)  -  \frac{p_1^+}{2 (1-z) s} \, \Psi_{\lambda', \sigma_2, \sigma_1}^{i , \, \gamma^* \to q {\bar q}} ({\un x}_{12}, z)  \right] \,  L \, \left[ V_{\un{x}_1}  \, \pd_1^i \, V_{\un{x}_1}^\dagger - i \, (p_1^i - p_2^i)  \right] . \notag
\end{align}

In addition, the calculation leading to \eq{XS1} assumed $x_0^- =0$. For an infinitely thin shock wave located at $x_0^- \neq 0$, the calculation needs to be augmented by adding the integrals over $x^- \in [0, x_0^-]$ for the $\gamma^* \to q {\bar q}$ splitting to the left of the shock wave, and over $x^- \in [x_0^-, 0]$ for the splitting taking place to the right of the shock wave. Indeed, for the splitting happening before the shock wave, we need to integrate
\begin{align}
    \int\limits_{-\infty}^{x_0^-} d x^- \, e^{i x^- E}, 
\end{align}
where $E$ denotes the corresponding energy denominator. Expanding in the energy denominator to the linear (sub-eikonal) order, we obtain (while neglecting the regulator, for simplicity)
\begin{align}
    \int\limits_{-\infty}^{x_0^-} d x^- \, e^{i x^- E} = \frac{1}{i \, E} \, e^{i x_0^- E} = \frac{1}{i \, E} + x_0^- + \ldots =  \int\limits_{-\infty}^{0} d x^- \, e^{i x^- E} + \int\limits_{0}^{x^-_0} d x^- + \ldots \, .
\end{align}
The first term on the right is the integration that was done in obtaining \eq{XS1}. The second term is the correction of the type of the first term in the curly brackets in \eq{remainder2}: it accounts for the $\gamma^* \to q {\bar q}$ splitting vertex exactly, while neglecting the phase. Similar analysis can be easily carried out for the second term in the curly brackets of \eq{remainder2}.

Since we do not want to modify \eq{XS1}, we can simply subtract the above-mentioned corrections out of \eq{remainder2}. That is, we simply add the following term to \eq{remainder2} above: 
\begin{align}\label{remainder3}
& - \int d^2 x_1 \,  d^2 x_2 \,  e^{- i {\un p}_1 \cdot {\un x}_{1} - i {\un p}_2 \cdot {\un x}_{2}}  \, \left[ \frac{p^+_1}{2 z s} \, \Psi_{\lambda', \sigma_1, \sigma_2}^{i , \, \gamma^* \to q {\bar q}} ({\un x}_{12}, z)  -  \frac{p_1^+}{2 (1-z) s} \, \Psi_{\lambda', \sigma_2, \sigma_1}^{i , \, \gamma^* \to q {\bar q}} ({\un x}_{12}, z)  \right] \\
& \times \, \left\{ \int\limits_{x_0^-}^{0}  d x^- \,  \left[ V_{\un{x}_1}  \,  \left( \cev{\pd}_1^i - \pd_1^i + i (p_1^i - p_2^i)  \right)  \, V_{\un{x}_1}^\dagger  \right] +  \int\limits^{x_0^-}_{0} d x^- \,  \left[ 1  \,  \left( \cev{\pd}_1^i - \pd_1^i + i (p_1^i - p_2^i)  \right)  \, 1  \right] \right\} \notag \\
& = \int d^2 x_1 \,  d^2 x_2 \,  e^{- i {\un p}_1 \cdot {\un x}_{1} - i {\un p}_2 \cdot {\un x}_{2}}  \, \left[ \frac{p^+_1}{2 z s} \, \Psi_{\lambda', \sigma_1, \sigma_2}^{i , \, \gamma^* \to q {\bar q}} ({\un x}_{12}, z)  -  \frac{p_1^+}{2 (1-z) s} \, \Psi_{\lambda', \sigma_2, \sigma_1}^{i , \, \gamma^* \to q {\bar q}} ({\un x}_{12}, z)  \right] \,  (- 2 x_0^-) \, V_{\un{x}_1}  \, \pd_1^i \, V_{\un{x}_1}^\dagger . \notag
\end{align}
Subtracting Eqs.~\eqref{remainder2} and \eqref{remainder3} from \eq{remainder} we obtain zero, as desired. We have, therefore, justified neglecting the terms proportional to $L$ and $x_0^-$ in arriving at \eq{XS_inst2} above.

As detailed above, we are interested in the double-spin asymmetry in the longitudinally polarized $e+p$ scattering, with the numerator of the asymmetry given by \eq{DSA33}. Since our angle brackets \eqref{CGC_ave} already include the sum over the proton helicities weighed by $S_L$, we rewrite \eq{DSA33} as the numerator of $A_{LL}$ for our observable, inclusive dijet production,
\begin{align}\label{DSA34}
    \int\limits_0^{2 \pi} \frac{d \phi}{2 \pi} \,  E'  \, z (1-z) \, \frac{d \sigma^{DSA}}{d^3 k' \, d^2 p_1 \, d^2 p_2 \, d z} =  \frac{ \alpha_{EM}}{4 \pi^2 \, Q^2 } \,  \, (2-y) \, \sum_{\lambda = \pm 1} \lambda \, z (1-z) \, \frac{d\sigma_{\lambda \lambda}^{\gamma^* p \to q {\bar q} X }}{d^2 p_1 \, d^2 p_2 \, d z}.
\end{align}
Combining Eqs.~\eqref{XS1} and \eqref{XS_inst2} into \eq{XS_tot}, we write our final result for the part of the inclusive dijet production cross section dependent on the proton and virtual photon polarizations, 
\begin{tcolorbox}[colback=blue!10!white]
\begin{align}\label{XS_total}
& \sum_{\lambda = \pm 1} \lambda \, z (1-z) \, \frac{d\sigma_{\lambda \lambda}^{\gamma^* p \to q {\bar q} X }}{d^2 p_1 \, d^2 p_2 \, d z} = \frac{1}{2 (2 \pi)^5} \, \int d^2 x_1 \, d^2 x_{1'} \, d^2 x_2 \, d^2 x_{2'} \, e^{- i {\un p}_1 \cdot {\un x}_{11'} - i {\un p}_2 \cdot {\un x}_{22'}} \, \sum_{\lambda = \pm 1} \lambda \, \Bigg\{ \int d^2 x_0 \sum_{\sigma_1, \sigma_2, \sigma'_1, \sigma'_2, i } \notag  \\ 
& \left[ \Psi_{\lambda, \sigma_1, \sigma_2; i, i}^{\gamma^* \to q {\bar q}} ({\un x}_{02}, z) \, \left[ \Psi_{\lambda, \sigma'_1, \sigma'_2; i, i}^{\gamma^* \to q {\bar q}} ({\un x}_{1'2'}, z) \right]^* \, \frac{1}{N_c} \left\langle  \tr \left[ \tord \left( V^\textrm{pol}_{{\un x}_1, {\un x}_0; \sigma'_1, \sigma_1} \, V_{{\un x}_2}^\dagger \right) \,  \atord  \left( V_{{\un x}_{2'}} \, V_{{\un x}_{1'}}^\dagger - 1 \right) \right]  \right\rangle (zs) \, \delta_{\sigma_2 , \sigma'_2} \right. \notag \\
& -  \Psi_{\lambda, \sigma_1, \sigma_2; i, i}^{\gamma^* \to q {\bar q}} ({\un x}_{10}, z) \, \left[ \Psi_{\lambda, \sigma'_1, \sigma'_2; i, i}^{\gamma^* \to q {\bar q}} ({\un x}_{1'2'}, z) \right]^* \frac{1}{N_c} \left\langle  \tr \left[ \tord \left( V_{{\un x}_1} \,  V^{\textrm{pol} \, \dagger}_{{\un x}_2, {\un x}_0; - \sigma'_2, - \sigma_2} \right) \, \atord  \left( V_{{\un x}_{2'}} \, V_{{\un x}_{1'}}^\dagger - 1 \right) \right]  \right\rangle ((1-z) s)  \, \delta_{\sigma_1 , \sigma'_1} \notag \\
& +  \Psi_{\lambda, \sigma_1, \sigma_2; i, i}^{\gamma^* \to q {\bar q}} ({\un x}_{12}, z) \, \left[ \Psi_{\lambda, \sigma'_1, \sigma'_2; i, i}^{\gamma^* \to q {\bar q}} ({\un x}_{02'}, z) \right]^* \frac{1}{N_c} \left\langle \tr \left[  \tord \left(  V_{{\un x}_{1}} \, V_{{\un x}_{2}}^\dagger - 1 \right)  \,  \atord \left( V_{{\un x}_{2'}} \,  V^{\textrm{pol} \, \dagger}_{{\un x}_{1'}, {\un x}_0; \sigma_1, \sigma'_1}  \right) \right]  \right\rangle (zs) \, \delta_{\sigma_2 , \sigma'_2} \notag \\
& \left. -  \Psi_{\lambda, \sigma_1, \sigma_2; i, i}^{\gamma^* \to q {\bar q}} ({\un x}_{12}, z) \, \left[ \Psi_{\lambda, \sigma'_1, \sigma'_2; i, i}^{\gamma^* \to q {\bar q}} ({\un x}_{1'0}, z) \right]^* \frac{1}{N_c} \left\langle  \tr \left[ \tord \left( V_{{\un x}_{1}} \, V_{{\un x}_{2}}^\dagger - 1 \right) \,  \atord \left( V^{\textrm{pol} }_{{\un x}_{2'}, {\un x}_0; -\sigma_2, - \sigma'_2} \, V_{{\un x}_{1'}}^\dagger \right) \right]  \right\rangle ((1-z) s) \, \delta_{\sigma_1 , \sigma'_1} \right] \notag \\
 & + \sum_{\sigma_1, \sigma_2, i } \, \left[ \Psi_{\lambda, \sigma_1, \sigma_2}^{j , \, \gamma^* \to q {\bar q}} ({\un x}_{12}, z) \, \left[ \Psi_{\lambda, \sigma_1, \sigma_2; i, i}^{\gamma^* \to q {\bar q}} ({\un x}_{1'2'}, z) \right]^* \, \frac{1}{N_c} \left\langle  \tr \left[ \tord \left( V^{j \, \textrm{G} [2]}_{{\un x}_1}  \, V_{{\un x}_1}^\dagger \right)  \,  \atord  \left( V_{{\un x}_{2'}} \, V_{{\un x}_{1'}}^\dagger - 1 \right) \right]  \right\rangle (zs) \right. \notag \\
& + \Psi_{\lambda, \sigma_2, \sigma_1}^{j , \, \gamma^* \to q {\bar q}} ({\un x}_{12}, z) \, \left[ \Psi_{\lambda, \sigma_1, \sigma_2; i, i}^{\gamma^* \to q {\bar q}} ({\un x}_{1'2'}, z) \right]^* \frac{1}{N_c} \left\langle  \tr \left[ \tord \left( V_{{\un x}_1} \, V^{j \, \textrm{G} [2] \dagger}_{{\un x}_1}  \right) \, \atord  \left( V_{{\un x}_{2'}} \, V_{{\un x}_{1'}}^\dagger - 1 \right) \right]  \right\rangle  ((1-z) s)   \notag \\
& +  \Psi_{\lambda, \sigma_1, \sigma_2; i, i}^{\gamma^* \to q {\bar q}} ({\un x}_{12}, z) \, \left[ \Psi_{\lambda, \sigma_1, \sigma_2}^{j , \, \gamma^* \to q {\bar q}} ({\un x}_{1'2'}, z) \right]^* \frac{1}{N_c} \left\langle \tr \left[  \tord \left(  V_{{\un x}_{1}} \, V_{{\un x}_{2}}^\dagger - 1 \right)  \,  \atord \left( V_{{\un x}_{1'}} \, V^{j \, \textrm{G} [2] \dagger}_{{\un x}_{1'}}  \right) \right]  \right\rangle (z s)  \notag \\
& \left. + \Psi_{\lambda, \sigma_1, \sigma_2; i, i}^{\gamma^* \to q {\bar q}} ({\un x}_{12}, z) \, \left[ \Psi_{\lambda, \sigma_2, \sigma_1}^{j , \, \gamma^* \to q {\bar q}} ({\un x}_{1'2'}, z)  \right]^* \frac{1}{N_c} \left\langle  \tr \left[ \tord \left( V_{{\un x}_{1}} \, V_{{\un x}_{2}}^\dagger - 1 \right) \,  \atord \left( V^{j \, \textrm{G} [2]}_{{\un x}_{1'}}  \, V_{{\un x}_{1'}}^\dagger  \right) \right]  \right\rangle ((1-z) s) \right]  \Bigg\} . 
\end{align}
\end{tcolorbox}
\eq{XS_total} is a general formal expression. It employs the virtual photon wavefunctions given in Eqs.~\eqref{LCwfT} and \eqref{eq:LCWF_inside} and the polarized Wilson lines given in Eqs.~\eqref{Vxy_sub-eikonal} and \eqref{VjG2}. The cross section is certainly very interesting in the general kinematic regions as it can probe various gluon and quark helicity-related TMDs at small $x$. In the following, we would like to focus on the back-to-back (correlation) limit in which the magnitude of the transverse momentum for the quark-antiquark dijet is much larger than the transverse momentum exchange with the proton. In this particular kinematic region, the cross section will be shown to uniquely probe the WW gluon helicity TMD at small $x$.


\subsection{Back-to-back expansion}
\label{sec:expansion}

Let us simplify \eq{XS_total} in the limit when the two jets are back-to-back. Defining
\begin{align}
{\un \Delta} \equiv {\un p}_1 + {\un p}_2 , \ \ \ {\un p} \equiv (1-z) \, {\un p}_1 - z \, {\un p}_2 ,
\end{align}
such that 
\begin{align}\label{p1p2}
{\un p}_1 = {\un p} + z \, {\un \Delta} , \ \ \  {\un p}_2 = - {\un p} + (1-z) \, {\un \Delta} ,
\end{align}
we can recast the Fourier phase as 
\begin{align}
e^{- i {\un p}_1 \cdot {\un x}_{11'} - i {\un p}_2 \cdot {\un x}_{22'}} = e^{- i \, {\un p} \cdot ({\un x}_{12} - {\un x}_{1'2'}) - i \, {\un \Delta} \cdot ({\un b} - {\un b}')} ,  
\end{align}
where 
\begin{align}
{\un b} \equiv z \, {\un x}_1 + (1-z) \, {\un x}_2, \ \ \ {\un b}' \equiv z \, {\un x}_{1'} + (1-z) \, {\un x}_{2'}.
\end{align}
This allows us to expand the cross section in powers of  $\Delta_\perp/p_{\perp}$ to the lowest non-trivial order. We would like to work in the regime where $p_\perp \sim Q \gg \Lambda_{QCD} \sim \Delta_\perp \sim 1/R_p$ with $R_p$ the proton radius. If the saturation scale is important, we would then work in the regime where  $p_\perp \sim Q \gg Q_s \gg \Delta_\perp \sim 1/R_p$. We will be only interested in the leading term in the expansion.

The polarized Wilson lines $V^{\mathrm{pol}}_{\un{x}, \un{y},\sigma', \sigma}$ in \eq{XS_total} contains both type-1 and type-2 subeikonal operators as can be seen from the definition in \eq{Vxy_sub-eikonal}. 
First we consider the contribution of type-1 sub-eikonal operators in Eq.~\eqref{XS_total}:
\begin{align}
& \sum_{\lambda = \pm 1} \lambda \, z (1-z) \, \frac{d\sigma_{\lambda \lambda}^{[1] \, \gamma^* p \to q {\bar q} X }}{d^2 p \, d^2 \Delta \, d z} \approx \frac{1}{2 (2 \pi)^5} \, \int d^2 b \, d^2 b' \, d^2 x_{12} \, d^2 x_{1'2'} \, e^{- i \, {\un p} \cdot ({\un x}_{12} - {\un x}_{1'2'}) - i \, {\un \Delta} \cdot ({\un b} - {\un b}')}  \notag \\ 
& \times \, \sum_{\lambda = \pm 1, \sigma_1, \sigma_2, i } \, \lambda \, \Psi_{\lambda, \sigma_1, \sigma_2; i, i}^{\gamma^* \to q {\bar q}} ({\un x}_{12}, z) \, \left[ \Psi_{\lambda, \sigma_1, \sigma_2; i, i}^{\gamma^* \to q {\bar q}} ({\un x}_{1'2'}, z) \right]^* \, \sigma_1 \, \Bigg\{ \frac{1}{N_c} \left\langle  \tr \left[ \tord \left( V^{\textrm{pol} \, [1]}_{{\un x}_1} \, V_{{\un x}_2}^\dagger \right) \,    \atord \left( V_{{\un x}_{2'}} \, V_{{\un x}_{1'}}^\dagger - 1 \right) \right]  \right\rangle (zs)  \notag \\
& - \frac{1}{N_c} \left\langle  \tr \left[ \tord \left( V_{{\un x}_1} \,  V^{\textrm{pol} \, [1] \, \dagger}_{{\un x}_2} \right) \,  \atord \left( V_{{\un x}_{2'}} \, V_{{\un x}_{1'}}^\dagger - 1 \right) \right]  \right\rangle ((1-z) s)    +  \frac{1}{N_c} \left\langle \tr \left[\tord \left(  V_{{\un x}_{1}} \, V_{{\un x}_{2}}^\dagger - 1 \right)  \, \atord \left( V_{{\un x}_{2'}} \,  V^{\textrm{pol} \, [1] \, \dagger}_{{\un x}_{1'}}  \right) \right]  \right\rangle  (zs)\notag \\
& - \frac{1}{N_c} \left\langle  \tr \left[\tord \left( V_{{\un x}_{1}} \, V_{{\un x}_{2}}^\dagger - 1 \right) \, \atord \left( V^{\textrm{pol} \, [1]}_{{\un x}_{2'}} \, V_{{\un x}_{1'}}^\dagger \right) \right]  \right\rangle ((1-z) s) \Bigg\} \label{eq:type_1_Xsec} .
\end{align}
Here $V^{\textrm{pol} \, [1]}_{\un x} = V_{\un x}^{\textrm{G} [1]} + V_{\un x}^{\textrm{q} [1]}$ \cite{Kovchegov:2018znm, Cougoulic:2022gbk}. We need to expand the operators in the integrand assuming that $x_{12} , x_{1'2'} \sim 1/p_\perp \ll b_\perp, b'_\perp \sim R_p \sim 1/\Lambda_{QCD}$.  
At the lowest non-trivial order, using 
\begin{align}\label{eq:x1_x2_in_b}
{\un x}_1 = {\un b} + (1-z) \, {\un x}_{12}, \ \ \ {\un x}_2 = {\un b} - z \, {\un x}_{12},
\end{align}
we get
\begin{align}\label{eq:eikonal_exp}
V_{{\un x}_{1}} \, V_{{\un x}_{2}}^\dagger - 1 = x_{12}^i \, V_{\un b} \, \left( \pd^i \, V_{\un b}^\dagger \right) + {\cal O} (x_{12}^2) 
\end{align}
and 
\begin{subequations}
\begin{align}
&V^{\mathrm{pol} \, [1]}_{{\un x}_1} \, V_{{\un x}_2}^{\dagger}= V^{\mathrm{pol} \, [1]}_{{\un b}} \, V_{{\un b}}^\dagger + \mathcal{O}(x_{12}),\\
&V_{{\un x}_1} \,  V^{\textrm{pol} \, [1] \, \dagger}_{{\un x}_2} = V_{{\un b}} \,  V^{\textrm{pol} \, [1] \, \dagger}_{{\un b}} + \mathcal{O}(x_{12}).
\end{align}
\end{subequations}
The factor resulting from summing over polarizations and colors is
\begin{align}
&\sum_{\lambda = \pm 1, \sigma_1, \sigma_2, i } \, \lambda \, \Psi_{\lambda, \sigma_1, \sigma_2; i, i}^{\gamma^* \to q {\bar q}} ({\un x}_{12}, z) \, \left[ \Psi_{\lambda, \sigma_1, \sigma_2; i, i}^{\gamma^* \to q {\bar q}} ({\un x}_{1'2'}, z) \right]^* \, \sigma_1 \\ 
& = N_c \, \frac{\alpha_{EM} \, Z_f^2}{\pi} \, 4(1-2z) [z(1-z)]^2 \, Q^2 \, \frac{\un{x}_{12} \cdot \un{x}_{1'2'}}{x_{12} x_{1'2'}} K_1(x_{12} a_f) K_1(x_{1'2'}a_f). \notag
\end{align}
In the back-to-back limit, \eq{eq:type_1_Xsec} becomes
\begin{align}\label{DSA10}
& \sum_{\lambda = \pm 1} \lambda \, z (1-z) \, \frac{d\sigma_{\lambda \lambda}^{[1] \, \gamma^* p \to q {\bar q} X }}{d^2 p \, d^2 \Delta \, d z}
\approx \frac{\alpha_{EM} Z_f^2}{(2 \pi)^6 \, s}  4 \, (1-2z) z(1-z) \int d^2 b \, d^2 b' \, d^2 x_{12} \, d^2 x_{1'2'} \, e^{- i \, {\un p} \cdot ({\un x}_{12} - {\un x}_{1'2'}) - i \, {\un \Delta} \cdot ({\un b} - {\un b}')}  \notag \\ 
& \times \, a_f^2 \, \frac{\un{x}_{12} \cdot \un{x}_{1'2'}}{x_{12} x_{1'2'}} K_1(x_{12} a_f) K_1(x_{1'2'}a_f) \, \Bigg\{ \frac{x_{2'1'}^i}{z} \llangle  \tr \left[ \tord \left( V^{\textrm{pol} \, [1]}_{{\un b}} \, V_{{\un b}}^\dagger \right) \, \atord  \left( V_{{\un b}'} \, \left( \pd^i \, V_{{\un b}'}^\dagger \right) \right) \right]  \rrangle (zs) \\
& - \frac{x_{2'1'}^i}{1-z} \llangle  \tr \left[  \tord \left( V_{{\un b}} \,  V^{\textrm{pol} \, [1] \, \dagger}_{{\un b}} \right) \, \atord  \left( V_{{\un b}'} \, \left( \pd^i \, V_{{\un b}'}^\dagger \right) \right) \right]  \rrangle ((1-z)s)   +  \frac{x_{12}^i}{z} \llangle \tr \left[  \tord  \left(  V_{\un b} \, \left( \pd^i \, V_{\un b}^\dagger \right) \right)  \,  \atord \left( V_{{\un b}'} \,  V^{\textrm{pol} \, [1] \, \dagger}_{{\un b}'}  \right) \right]  \rrangle (zs) \notag \\
& - \frac{x_{12}^i}{1-z} \llangle  \tr \left[\tord \left( V_{\un b} \, \left( \pd^i \, V_{\un b}^\dagger \right) \right) \, \atord\left( V^{\textrm{pol} \, [1]}_{{\un b}'} \, V_{{\un b}'}^\dagger \right) \right]  \rrangle ((1-z)s) \Bigg\}  \notag ,
\end{align}
where we have switched to the double-bracket notation from \eq{CGC_ave}.
Using the Fourier transformations 
\begin{subequations}\label{FT}
\begin{align}
& \int d^2 x \, e^{- i\un{p}\cdot\un{x}} \, \frac{x^l}{x_\perp} \, a_f \, K_1(x_\perp \, a_f) = (- 2\pi i) \, \frac{p^l}{p_\perp^2+a_f^2},\\
&\int d^2x \, e^{-i\un{p}\cdot\un{x}} \, \frac{x^i \, x^l}{x_\perp} \, a_f \, K_1( x_\perp \, a_f) = (2\pi) \left[\frac{\delta^{il}}{p_\perp^2+a_f^2} - \frac{2p^ip^l}{(p_\perp^2+a_f^2)^2}\right] , \label{FT2}
\end{align}
\end{subequations}
Equation \eqref{DSA10} can be rewritten as 
\begin{align}
& \sum_{\lambda = \pm 1} \lambda \, z (1-z) \, \frac{d\sigma_{\lambda \lambda}^{[1] \, \gamma^* p \to q {\bar q} X }}{d^2 p \, d^2 \Delta \, d z}
\approx \frac{\alpha_{EM} Z_f^2}{(2 \pi)^4 \, s}  \, 4 (1-2z) z(1-z) \, \frac{p^i(a_f^2-p_\perp^2)}{(p_\perp^2+a_f^2)^3} \, i \int d^2 b \, d^2 b' \, e^{ - i \, {\un \Delta} \cdot ({\un b} - {\un b}')} \label{DSA10pspace} \\ 
& \times  \Bigg\{ \frac{1}{z} \llangle  \tr \left[ \tord \left( V^{\textrm{pol} \, [1]}_{{\un b}} \, V_{{\un b}}^\dagger \right) \, \atord  \left( V_{{\un b}'} \, \left( \pd^i \, V_{{\un b}'}^\dagger \right) \right) \right]  \rrangle (zs)+  \frac{1}{z} \llangle \tr \left[  \tord  \left(  V_{\un b} \, \left( \pd^i \, V_{\un b}^\dagger \right) \right)  \, \atord  \left( V_{{\un b}'} \,  V^{\textrm{pol} \, [1] \, \dagger}_{{\un b}'}  \right) \right]  \rrangle (zs)\notag \\
& \ \ \ \ \ - \frac{1}{1-z} \llangle  \tr \left[  \tord \left( V_{{\un b}} \,  V^{\textrm{pol} \, [1] \, \dagger}_{{\un b}} \right) \, \atord \left( V_{{\un b}'} \, \left( \pd^i \, V_{{\un b}'}^\dagger \right) \right) \right]  \rrangle ((1-z)s) \notag \\ 
& \ \ \ \ \ - \frac{1}{1-z} \llangle  \tr \left[ \tord \left( V_{\un b} \, \left( \pd^i \, V_{\un b}^\dagger \right) \right) \, \atord \left( V^{\textrm{pol} \, [1]}_{{\un b}'} \, V_{{\un b}'}^\dagger \right) \right]  \rrangle ((1-z)s) \Bigg\}  \notag .
\end{align}
The result is subleading in powers of $\Delta_{\perp}/p_T$ because the integrals over $b$ and $b'$ in it are of the type 
\begin{equation}\label{eq:relate_twist_3}
i \int d^2 b \, d^2 b' \, e^{- i \, {\un \Delta} \cdot ({\un b} - {\un b}')} \, \left[ \llangle  \tr \left[  \tord \left( V^{\textrm{pol} \, [1]}_{{\un b}} \, V_{{\un b}}^\dagger \right) \,  \atord  \left( V_{{\un b}'} \, \left( \pd^i \, V_{{\un b}'}^\dagger \right) \right) \right]  \rrangle + \llangle \tr \left[  \tord  \left(  V_{\un b} \, \left( \pd^i \, V_{\un b}^\dagger \right) \right)  \, \atord  \left( V_{{\un b}'} \,  V^{\textrm{pol} \, [1] \, \dagger}_{{\un b}'}  \right) \right]  \rrangle \right].
\end{equation}
If we put ${\un \Delta} =0$ in \eq{eq:relate_twist_3}, we would be left with an expression dependent on the transverse index $i$, but, after the integrations over $\un b$ and $\un b'$ are carried out, without a transverse vector to ``carry" this index. (Note that the proton is longitudinally polarized.) Therefore, \eq{eq:relate_twist_3} is zero for ${\un \Delta} =0$. For ${\un \Delta} \neq 0$, \eq{eq:relate_twist_3} is proportional to $\Delta^i$ and/or $\epsilon^{ij} \, \Delta^j$ multiplied by some function of $\Delta_\perp$ and $R_p$, which is regular at $\Delta_\perp =0$ due to confinement. Therefore, \eq{DSA10pspace} goes to zero as $\sim \Delta_\perp /p_\perp^3$ when $\Delta_\perp \to 0$, and is outside of the precision of our approximation.
In Appendix \ref{sec:twist-3TMD}, we show that, when $V_{\un{b}}^{\mathrm{pol}[1]} = V_{\un{b}}^{\textrm{G}[1]}$ (when only the gluon sub-eikonal contribution is kept), \eq{eq:relate_twist_3} becomes equal to $8\pi^4 \as \Delta^i  h^{\perp\, WW}_{3L}(x, \Delta^2_{\perp})$, that is, it is related to the twist-3 WW gluon helicity-flip TMD $h^{\perp\, WW}_{3L}$. Another interesting feature of \eq{DSA10} is that it vanishes when $z=1/2$.

To summarize, since we are only interested in the zeroth order result in the expansion of $\Delta_{\perp}/p_T$, we ignore contributions from the type-1 sub-eikonal operators.

Next, we continue to compute the contributions from type-2 sub-eikonal operators. The contribution of the $V_{{\ul x}}^{\textrm{q} [2]}$ operator can also be shown to be of the order $\sim \Delta_\perp /p_\perp^3$,  
similar to that of type-1 operator $V_{\un{x}}^{\mathrm{pol} \, [1]}$ analyzed above. We therefore discard these terms at leading order in the back-to-back expansion.  We are left with the contribution from the operator $V_{{\ul x}_1, {\un x}_2}^{\textrm{G} [2]}$. Using the definition in \eq{VxyG2}, one can show the following expressions holds for arbitrary functions $f(\un{x}_1)$ and $g(\un{x}_0)$ (cf. \cite{Kovchegov:2024wjs}):
\begin{align}\label{eq:VxyG2_VxiG[2]}
\int d^2x_1 d^2x_0 f(\un{x}_1) g(\un{x}_0)\,  V^{\textrm{G}[2]}_{\un{x}_1, \un{x}_0} =& - \frac{i P^+}{s} \int d^2x_1 f(\un{x}_1) g(\un{x}_1)  \int_{-\infty}^{\infty} dz^- \left[ V_{\un{x}_1}[\infty, z^-] \overleftarrow{D}_{\un{x}_1}^i\overrightarrow{D}_{\un{x}_1}^iV_{\un{x}_1}[z^-, -\infty] \right] \\
& - i\int d^2x_1 \left[ g(\un{x}_1) \, \partial_{\un 1}^i f(\un{x}_1) - f(\un{x}_1) \, \partial_{\un 1}^i g(\un{x}_1) \right] \, V^{i \, \textrm{G}[2]}_{\un{x}_1} \notag \\
&+\frac{i P^+}{2 \, s} \int d^2x_1 \left[] g(\un{x}_1) \, \partial_{\un 1}^2 f(\un{x}_1)  + f(\un{x}_1) \, \partial_{\un 1}^2g(\un{x}_1)\right] \, \int_{-\infty}^{\infty} dz^- \, V_{\un{x}_1} \notag . 
\end{align}
Similar to the above, the contribution of the first term on the right-hand side of \eq{eq:VxyG2_VxiG[2]} to the dijet cross section can be shown to start only at the linear order in $\Delta_{\perp}/p_T$ in the back-to-back limit. In fact, if one further requires that the dijet cross section is symmetric under $\un{p}_1\rightarrow -\un{p}_1$ and $\un{p}_2\rightarrow -\un{p}_2$, even the linear order (in $\Delta_\perp$) contributions from the first term on the right of \eq{eq:VxyG2_VxiG[2]} vanish in the back-to-back limit. Moreover, one can ignore contribution from the third term on the right of \eq{eq:VxyG2_VxiG[2]} as it is insensitive to the proton helicity state and thus will not contribute to the double-spin asymmetry. Therefore, for practical analysis, one can use the following prescription to simplify the terms containing $V^{\textrm{G}[2]}_{\un{x}, \un{y}}$ \cite{Kovchegov:2023yzd, Kovchegov:2024wjs} :
\begin{subequations}\label{G2_eqns}
 \begin{align}
 & \int d^2 x_1 \, d^2 x_0 \, f ({\un x}_1) \, g ({\un x}_0) \, V^{\textrm{G} [2]}_{{\un x}_1, {\un x}_0} \to - i \, \int d^2 x_1 \, \left[ g ({\un x}_1) \, \pd_{\un 1}^i f ({\un x}_1) - f ({\un x}_1) \,  \pd_{\un 1}^i g ({\un x}_1) \right] \, V_{{\un x}_1}^{i \, \textrm{G} [2]} , \\
 & \int d^2 x_1 \, d^2 x_0 \, f ({\un x}_1) \, g ({\un x}_0) \, \left( V^{\textrm{G} [2]}_{{\un x}_1, {\un x}_0} \right)^\dagger \to i \, \int d^2 x_1 \, \left[ g ({\un x}_1) \, \pd_{\un 1}^i f ({\un x}_1) - f ({\un x}_1) \,  \pd_{\un 1}^i g ({\un x}_1) \right] \, \, \left(  V_{{\un x}_1}^{i \, \textrm{G} [2]}  \right)^\dagger .
 \end{align}
\end{subequations}
Employing Eqs.~\eqref{G2_eqns} we obtain the contribution to the dijet cross section coming from type-2 subeikonal operators,
\begin{align}\label{eq:[2]_before_exp}
& \sum_{\lambda = \pm 1} \lambda \, z (1-z) \, \frac{d\sigma_{\lambda \lambda}^{[2] \, \gamma^* p \to q {\bar q} X }}{d^2 p \, d^2 \Delta \, d z} \approx \frac{1}{2 (2 \pi)^5} \, \int d^2 b \, d^2 b' \, d^2 x_{12} \, d^2 x_{1'2'} \, e^{- i \, {\un p} \cdot ({\un x}_{12} - {\un x}_{1'2'}) - i \, {\un \Delta} \cdot ({\un b} - {\un b}')} \, \sum_{\lambda = \pm 1} \lambda \,  \sum_{\sigma_1, \sigma_2, i } \\ 
& \times \Bigg\{ \left[ (i \, \pd_1^j + p^j) \Psi_{\lambda, \sigma_1, \sigma_2; i, i}^{\gamma^* \to q {\bar q}} ({\un x}_{12}, z) \right] \, \left[ \Psi_{\lambda, \sigma_1, \sigma_2; i, i}^{\gamma^* \to q {\bar q}} ({\un x}_{1'2'}, z) \right]^* \, \frac{1}{N_c} \left\langle  \tr \left[\tord \left( V^{j \, \textrm{G}  [2]}_{{\un x}_1} \, V_{{\un x}_2}^\dagger \right) \, \atord\left( V_{{\un x}_{2'}} \, V_{{\un x}_{1'}}^\dagger - 1 \right) \right]  \right\rangle (zs) \notag \\
& - \left[ (i \, \pd_1^j +  p^j)  \Psi_{\lambda, \sigma_1, \sigma_2; i, i}^{\gamma^* \to q {\bar q}} ({\un x}_{12}, z) \right] \, \left[ \Psi_{\lambda, \sigma_1, \sigma_2; i, i}^{\gamma^* \to q {\bar q}} ({\un x}_{1'2'}, z) \right]^* \frac{1}{N_c} \left\langle  \tr \left[\tord \left( V_{{\un x}_1} \,  V^{j \, \textrm{G}  [2] \, \dagger}_{{\un x}_2} \right) \,\atord \left( V_{{\un x}_{2'}} \, V_{{\un x}_{1'}}^\dagger - 1 \right) \right]  \right\rangle ((1-z) s) \notag \\
& +  \Psi_{\lambda, \sigma_1, \sigma_2; i, i}^{\gamma^* \to q {\bar q}} ({\un x}_{12}, z) \, \left[ (i \, \pd_{1'}^j + p^j) \Psi_{\lambda, \sigma_1, \sigma_2; i, i}^{\gamma^* \to q {\bar q}} ({\un x}_{1'2'}, z) \right]^* \frac{1}{N_c} \left\langle \tr \left[\tord \left(  V_{{\un x}_{1}} \, V_{{\un x}_{2}}^\dagger - 1 \right)  \, \atord \left( V_{{\un x}_{2'}} \,  V^{j \, \textrm{G} [2] \, \dagger}_{{\un x}_{1'}}  \right) \right]  \right\rangle (zs) \notag \\
& -  \Psi_{\lambda, \sigma_1, \sigma_2; i, i}^{\gamma^* \to q {\bar q}} ({\un x}_{12}, z) \, \left[ (i \, \pd_{1'}^j + p^j)  \Psi_{\lambda, \sigma_1, \sigma_2; i, i}^{\gamma^* \to q {\bar q}} ({\un x}_{1'2'}, z) \right]^* \frac{1}{N_c} \left\langle  \tr \left[\tord \left( V_{{\un x}_{1}} \, V_{{\un x}_{2}}^\dagger - 1 \right) \,\atord \left( V^{j \, \textrm{G}  [2]}_{{\un x}_{2'}} \, V_{{\un x}_{1'}}^\dagger \right) \right]  \right\rangle ((1-z) s)  \notag \\
 & + \Psi_{\lambda, \sigma_1, \sigma_2}^{j , \, \gamma^* \to q {\bar q}} ({\un x}_{12}, z) \, \left[ \Psi_{\lambda, \sigma_1, \sigma_2; i, i}^{\gamma^* \to q {\bar q}} ({\un x}_{1'2'}, z) \right]^* \, \frac{1}{N_c} \left\langle  \tr \left[\tord \left( V^{j \, \textrm{G} [2]}_{{\un x}_1}  \, V_{{\un x}_1}^\dagger \right)  \, \atord \left( V_{{\un x}_{2'}} \, V_{{\un x}_{1'}}^\dagger - 1 \right) \right]  \right\rangle (zs)  \notag \\
& + \Psi_{\lambda, \sigma_2, \sigma_1}^{j , \, \gamma^* \to q {\bar q}} ({\un x}_{12}, z) \, \left[ \Psi_{\lambda, \sigma_1, \sigma_2; i, i}^{\gamma^* \to q {\bar q}} ({\un x}_{1'2'}, z) \right]^* \frac{1}{N_c} \left\langle  \tr \left[ \tord \left( V_{{\un x}_1} \, V^{j \, \textrm{G} [2] \dagger}_{{\un x}_1}  \right) \, \atord \left( V_{{\un x}_{2'}} \, V_{{\un x}_{1'}}^\dagger - 1 \right) \right]  \right\rangle  ((1-z) s)   \notag \\
& +  \Psi_{\lambda, \sigma_1, \sigma_2; i, i}^{\gamma^* \to q {\bar q}} ({\un x}_{12}, z) \, \left[ \Psi_{\lambda, \sigma_1, \sigma_2}^{j , \, \gamma^* \to q {\bar q}} ({\un x}_{1'2'}, z) \right]^* \frac{1}{N_c} \left\langle \tr \left[ \tord \left(  V_{{\un x}_{1}} \, V_{{\un x}_{2}}^\dagger - 1 \right)  \,\atord \left( V_{{\un x}_{1'}} \, V^{j \, \textrm{G} [2] \dagger}_{{\un x}_{1'}}  \right) \right]  \right\rangle (z s)  \notag \\
& + \Psi_{\lambda, \sigma_1, \sigma_2; i, i}^{\gamma^* \to q {\bar q}} ({\un x}_{12}, z) \, \left[ \Psi_{\lambda, \sigma_2, \sigma_1}^{j , \, \gamma^* \to q {\bar q}} ({\un x}_{1'2'}, z)  \right]^* \frac{1}{N_c} \left\langle  \tr \left[ \tord  \left( V_{{\un x}_{1}} \, V_{{\un x}_{2}}^\dagger - 1 \right) \,\atord \left( V^{j \, \textrm{G} [2]}_{{\un x}_{1'}}  \, V_{{\un x}_{1'}}^\dagger  \right) \right]  \right\rangle ((1-z) s)   \Bigg\} . \notag
\end{align}
To perform the expansion around the back-to-back limit in powers of $x_{12}/R_p$ and $x_{1'2'}/R_p$, one needs Eqs.~\eqref{eq:x1_x2_in_b}, \eqref{eq:eikonal_exp} and 
\begin{subequations}
\begin{align}
&V^{j \, \textrm{G}  [2]}_{{\un x}_1} \, V_{{\un x}_2}^\dagger = V^{j \, \textrm{G}  [2]}_{{\un b}} \, V_{{\un b}}^\dagger + \mathcal{O}(x_{12}) , \\ 
&V_{{\un x}_1} \,  V^{j \, \textrm{G}  [2] \, \dagger}_{{\un x}_2} = V_{{\un b}} \,  V^{j \, \textrm{G}  [2] \, \dagger}_{{\un b}}  + \mathcal{O}(x_{12}) .
\end{align}
\end{subequations}

After the expansion, one can replace $p^j \to i {\cev \nabla}_1^j = - i {\cev \pd}_1^j$  and integrate by parts, which is equivalent to the replacing $p^j \to i \, \pd_1^j$ in the first two terms and $p^j \to i \, \pd_{1'}^j$ in the third and the fourth terms in the curly brackets. Furthermore, we also need the identity
\begin{align}
V^{j \, \textrm{G}  [2]}_{{\un b}} \, V_{{\un b}}^\dagger = - V_{{\un b}} \,  V^{j \, \textrm{G}  [2] \, \dagger}_{{\un b}}.
\end{align}
With all these intermediate steps of simplification, and switching to the double angle brackets, we rewrite \eq{eq:[2]_before_exp} as
\begin{align}\label{eq:before_compute_wff}
& \sum_{\lambda = \pm 1} \lambda \, z (1-z) \, \frac{d\sigma_{\lambda \lambda}^{\gamma^* p \to q {\bar q} X }}{d^2 p \, d^2 \Delta \, d z} \approx \frac{1}{2 (2 \pi)^5 \, s} \, \int d^2 b \, d^2 b' \, d^2 x_{12} \, d^2 x_{1'2'} \, e^{- i \, {\un p} \cdot ({\un x}_{12} - {\un x}_{1'2'}) - i \, {\un \Delta} \cdot ({\un b} - {\un b}')} \, \sum_{\lambda = \pm 1} \lambda \,  \sum_{\sigma_1, \sigma_2, i } \\ 
& \times \Bigg\{ \left[ \frac{2 i}{z (1-z)} \, \pd_1^j  \Psi_{\lambda, \sigma_1, \sigma_2; i, i}^{\gamma^* \to q {\bar q}} ({\un x}_{12}, z) + \frac{1}{z} \Psi_{\lambda, \sigma_1, \sigma_2}^{j , \, \gamma^* \to q {\bar q}} ({\un x}_{12}, z) - \frac{1}{1-z} \Psi_{\lambda, \sigma_2, \sigma_1}^{j , \, \gamma^* \to q {\bar q}} ({\un x}_{12}, z)  \right] \, \left[  x_{2'1'}^m \, \Psi_{\lambda, \sigma_1, \sigma_2; i, i}^{\gamma^* \to q {\bar q}} ({\un x}_{1'2'}, z) \right]^* \notag \\ 
& \times \, \frac{1}{N_c} \llangle  \tr \left[ \tord  \left( V^{j \, \textrm{G}  [2]}_{{\un b}} \, V_{{\un b}}^\dagger \right) \, \atord \left( V_{{\un b}'} \, \left( \pd^m \, V_{{\un b}'}^\dagger \right) \right) \right]  \rrangle (s)   \notag \\
& +  x^m_{12} \, \Psi_{\lambda, \sigma_1, \sigma_2; i, i}^{\gamma^* \to q {\bar q}} ({\un x}_{12}, z) \, \left[ \frac{2 i}{z (1-z)} \, \pd_{1'}^j \, \Psi_{\lambda, \sigma_1, \sigma_2; i, i}^{\gamma^* \to q {\bar q}} ({\un x}_{1'2'}, z) +  \frac{1}{z} \Psi_{\lambda, \sigma_1, \sigma_2}^{j , \, \gamma^* \to q {\bar q}} ({\un x}_{1'2'}, z)  - \frac{1}{1-z} \Psi_{\lambda, \sigma_2, \sigma_1}^{j , \, \gamma^* \to q {\bar q}} ({\un x}_{1'2'}, z) \right]^* \notag \\
& \times \, \frac{1}{N_c} \llangle \tr \left[ \tord \left(  V_{\un b} \, \left( \pd^m \, V_{\un b}^\dagger \right) \right)  \, \atord \left( V_{{\un b}'} \,  V^{j \, \textrm{G} [2] \, \dagger}_{{\un b}'}  \right) \right]  \rrangle (s) \Bigg\} . \notag
\end{align}
We have also dropped the factors of $z$ and $1-z$ in the arguments of the matrix elements, since we are interested in the regime where $z \sim 1-z \sim {\cal O} (1)$.

The factors involving the virtual photon wave functions can be further computed with the help of the identity
\begin{align}
{\un \epsilon}_\lambda \cdot {\un v} \ {\un \epsilon}^*_\lambda \cdot {\un w} = \frac{1}{2} \, {\un v} \cdot {\un w} + \frac{i}{2} \, \lambda \, {\un v} \times {\un w} .
\end{align}
Note that ${\un \epsilon}_\lambda = - (1/\sqrt{2}) (-\lambda, i)$. Using the expressions in \eq{LCwfT} and \eq{eq:LCWF_inside}, one obtains
\begin{align}\label{wave_functions_overlap}
& \sum_{\lambda = \pm 1} \lambda \,  \sum_{\sigma_1, \sigma_2, i } \left[ \frac{2 i}{z (1-z)} \, \pd_1^j  \Psi_{\lambda, \sigma_1, \sigma_2; i, i}^{\gamma^* \to q {\bar q}} ({\un x}_{12}, z) + \frac{1}{z} \Psi_{\lambda, \sigma_1, \sigma_2}^{j , \, \gamma^* \to q {\bar q}} ({\un x}_{12}, z) - \frac{1}{1-z} \Psi_{\lambda, \sigma_2, \sigma_1}^{j , \, \gamma^* \to q {\bar q}} ({\un x}_{12}, z)  \right] \, \left[ \Psi_{\lambda, \sigma_1, \sigma_2; i, i}^{\gamma^* \to q {\bar q}} ({\un x}_{1'2'}, z) \right]^*   \\
& = N_c \, \left( \frac{e Z_f}{2 \pi} \right)^2 \, 8 \, \left[ z^2 + (1-z)^2 \right] \, \epsilon^{kl} \, \frac{x_{1'2'}^l}{x_{1'2'}} \, a_f \, K_1 \left( x_{1'2'} \, a_f \right) \left[ \frac{\delta^{jk} x_{12}^2 - 2 \, x_{12}^j x_{12}^k}{x_{12}^3}  \, a_f \, K_1 \left( x_{12} \, a_f \right) - \frac{x_{12}^j x_{12}^k}{x_{12}^2} \, a_f^2\, K_0 \left( x_{12} \, a_f \right)  \right]. \notag
\end{align}
To be explicit, one has
\begin{align}\label{eq:cancel_outside}
& \sum_{\lambda = \pm 1} \lambda \,  \sum_{\sigma_1, \sigma_2, i } \left[ \frac{2 i}{z (1-z)} \, \pd_1^j  \Psi_{\lambda, \sigma_1, \sigma_2; i, i}^{\gamma^* \to q {\bar q}} ({\un x}_{12}, z) \right] \left[ \Psi_{\lambda, \sigma_1, \sigma_2; i, i}^{\gamma^* \to q {\bar q}} ({\un x}_{1'2'}, z) \right]^*  \notag  = N_c \, \left( \frac{e Z_f}{2 \pi} \right)^2 \, 8 \, \left[ z^2 + (1-z)^2 \right] \,  \\
&\times\epsilon^{kl} \, \frac{x_{1'2'}^l}{x_{1'2'}} \, a_f \, K_1 \left( x_{1'2'} \, a_f \right)\left[ \frac{\delta^{jk} x_{12}^2 - 2 \, x_{12}^j x_{12}^k}{x_{12}^3}  \, a_f \, K_1 \left( x_{12} \, a_f \right) - \frac{x_{12}^j x_{12}^k}{x_{12}^2} \, a_f^2\, K_0 \left( x_{12} \, a_f \right) + \delta^{jk} \pi \delta^2(\un{x}_{12})  \right]
\end{align}
and 
\begin{align}\label{eq:cancel_inside}
& \sum_{\lambda = \pm 1} \lambda \,  \sum_{\sigma_1, \sigma_2, i } \left[\frac{1}{z} \Psi_{\lambda, \sigma_1, \sigma_2}^{j , \, \gamma^* \to q {\bar q}} ({\un x}_{12}, z) - \frac{1}{1-z} \Psi_{\lambda, \sigma_2, \sigma_1}^{j , \, \gamma^* \to q {\bar q}} ({\un x}_{12}, z)  \right] \, \left[ \Psi_{\lambda, \sigma_1, \sigma_2; i, i}^{\gamma^* \to q {\bar q}} ({\un x}_{1'2'}, z) \right]^*  \notag \\
& = -N_c \, \left( \frac{e Z_f}{2 \pi} \right)^2 \, 8 \, \left[ z^2 + (1-z)^2 \right] \, \epsilon^{jl} \, \frac{x_{1'2'}^l}{x_{1'2'}} \, a_f \, K_1 \left( x_{1'2'} \, a_f \right) \pi \delta^2(\un{x}_{12}).
\end{align}
The terms in Eqs.~\eqref{eq:cancel_outside} and \eqref{eq:cancel_inside} containing $\delta^2(\un{x}_{12})$ cancel in their sum, \eq{wave_functions_overlap}. It is interesting to see that in the back-to-back limit, the contribution to the cross section coming from the virtual photon splitting inside the shockwave \eqref{eq:cancel_inside} cancels out a part of the contribution from the splitting occurring outside the shockwave \eqref{eq:cancel_outside}.

Equation \eq{eq:before_compute_wff} now becomes 
\begin{align}\label{DSA103}
 &\sum_{\lambda = \pm 1} \lambda \, z (1-z) \, \frac{d\sigma_{\lambda \lambda}^{\gamma^* p \to q {\bar q} X }}{d^2 p \, d^2 \Delta \, d z} \notag \\
 \approx & - \frac{16}{2 (2 \pi)^5 \, s} \, N_c \, \left( \frac{e Z_f}{2 \pi} \right)^2 \, \left[ z^2 + (1-z)^2 \right] \, \int d^2 b \, d^2 b' \, d^2 x_{12} \, d^2 x_{1'2'} \, e^{- i \, {\un p} \cdot ({\un x}_{12} - {\un x}_{1'2'}) - i \, {\un \Delta} \cdot ({\un b} - {\un b}')} \notag \\ 
& \times \, x_{1'2'}^m  \, \epsilon^{kl} \, \frac{x_{1'2'}^l}{x_{1'2'}} \, a_f \, K_1 \left( x_{1'2'} \,a_f \right) \,  \left[ \frac{\delta^{jk} x_{12}^2 - 2 \, x_{12}^j x_{12}^k}{x_{12}^3}  \, a_f \, K_1 \left( x_{12} \, a_f \right)  \right.\left. - \frac{x_{12}^j x_{12}^k}{x_{12}^2} \, a_f^2 \, K_0 \left( x_{12} \, a_f \right)  \right] \,\notag\\
&\times \frac{1}{N_c} \mbox{Re} \, \llangle  \tr \left[ \tord \left( V^{j \, \textrm{G}  [2]}_{{\un b}} \, V_{{\un b}}^\dagger \right) \,  \atord \left( V_{{\un b}'} \, \left( \pd^m \, V_{{\un b}'}^\dagger \right) \right) \right]  \rrangle (s) . 
\end{align}
Carrying out the Fourier transformations with respect to $\un{x}_{12}$ and $\un{x}_{1'2'}$ by employing Eqs.~\eqref{FT} along with 
\begin{align}
& \int  d^2 x_{12} \, e^{- i \, {\un p} \cdot {\un x}_{12}} \left[ \frac{\delta^{jk} x_{12}^2 - 2 \, x_{12}^j x_{12}^k}{x_{12}^3}  \, a_f \, K_1 \left( x_{12} \, a_f \right) - \frac{x_{12}^j x_{12}^k}{x_{12}^2} \, a_f^2 \, K_0 \left( x_{12} \, a_f \right)  \right] \notag \\
& = \int  d^2 x_{12} \, e^{- i \, {\un p} \cdot {\un x}_{12}} \left[ - \pd_1^j \left( \frac{x_{12}^k}{x_{12}} \, a_f \, K_1 \left( x_{12} \, a_f \right) \right) - \pi \, \delta^{jk} \, \delta^2 ({\un x}_{12})    \right] = 2 \pi \frac{p^j p^k}{p_T^2 + a_f^2} - \pi \, \delta^{jk}
\end{align}
and using the identity 
\begin{align}
- \epsilon^{mk} p^j p^k + \epsilon^{jk} p^m p^k = p_T^2 \, \epsilon^{jm}
\end{align}
we rewrite \eq{DSA103} as
\begin{align}\label{DSA105}
 \sum_{\lambda = \pm 1} \lambda \, z (1-z) \, \frac{d\sigma_{\lambda \lambda}^{\gamma^* p \to q {\bar q} X }}{d^2 p \, d^2 \Delta \, d z} \approx & - \frac{8}{2 (2 \pi)^5 \, s} \, N_c \, \left( e Z_f \right)^2 \, \left[ z^2 + (1-z)^2 \right] \, \int d^2 b \, d^2 b' \, e^{- i \, {\un \Delta} \cdot ({\un b} - {\un b}')} \\ 
& \times \, \frac{p_T^2 - a_f^2}{(p_T^2 + a_f^2)^2} \, \epsilon^{jm} \, \frac{1}{N_c} \mbox{Re} \, \llangle  \tr \left[\tord \left( V^{j \, \textrm{G}  [2]}_{{\un b}} \, V_{{\un b}}^\dagger \right) \,  \atord \left( V_{{\un b}'} \, \left( \pd^m \, V_{{\un b}'}^\dagger \right) \right) \right]  \rrangle (s) . \notag 
\end{align}
Recall that the discussion in Section~\ref{sec:WW_TMD} resulted in the small-$x$ WW gluon helicity TMD given by \eq{eq:WW_G_final}, which we state here once again, 
\begin{align}\label{glue_hel_TMD46}
& g_{1L}^{G \, WW} (x, k_T^2) =  - \frac{1}{\as  \, 4 \pi^4} \,  \int d^2 x_1 \, d^2 x_0 \, e^{- i {\un k} \cdot {\un x}_{10}} \,  \epsilon^{ij} \,  \mbox{Re} \, \llangle \tr \Big[  V_{{\un x}_0} \, \left( \partial^i V_{{\un x}_0}^\dagger \right) \, V_{{\un x}_1}^{j \textrm{G} [2]} \, V_{{\un x}_1}^\dagger \Big] 
\rrangle . 
\end{align}
Comparing this with \eq{DSA105}, we note that, per \cite{Mueller:2012bn,Kovchegov:2018znm},  
\begin{align}
 & \llangle  \tr \left[ \tord \left( V^{j \, \textrm{G}  [2]}_{{\un b}} \, V_{{\un b}}^\dagger \right) \,   \atord \left( V_{{\un b}'} \, \left( \pd^m \, V_{{\un b}'}^\dagger \right) \right) \right]  \rrangle =  \llangle  \tr \left[ \tord \left( V^{j \, \textrm{G}  [2]}_{{\un b}} \, V_{{\un b}}^\dagger \, V_{{\un b}'} \, \left( \pd^m \, V_{{\un b}'}^\dagger \right) \right) \right]  \rrangle \notag \\
 & = \llangle  \tr \left[ \tord \left( V_{{\un b}'} \, \left( \pd^m \, V_{{\un b}'}^\dagger \right)  \, V^{j \, \textrm{G}  [2]}_{{\un b}} \, V_{{\un b}}^\dagger \right) \right]  \rrangle = \llangle  \tr \left[ V_{{\un b}'} \, \left( \pd^m \, V_{{\un b}'}^\dagger \right)  \, V^{j \, \textrm{G}  [2]}_{{\un b}} \, V_{{\un b}}^\dagger \right]\rrangle.
\end{align}
We employ this observation while combining Eqs.~\eqref{DSA105} and \eqref{glue_hel_TMD46} to write
\begin{align}\label{DSA106}
\sum_{\lambda = \pm 1} \lambda \, z (1-z) \, \frac{d\sigma_{\lambda \lambda}^{\gamma^* p \to q {\bar q} X }}{d^2 p \, d^2 \Delta \, d z} \approx - \frac{2 \, \as \, \alpha_{EM} \, Z_f^2}{s} \, \left[ z^2 + (1-z)^2 \right] \, \frac{p_T^2 - a_f^2}{(p_T^2 + a_f^2)^2} \  g_{1L}^{G \, WW} \left( x \approx \frac{p_T^2}{s}, \Delta_T^2 \right) .
\end{align}
Eq.~\eqref{DSA106} is the main result of this Section. When combined with \eq{DSA34} (and summed over the flavors $f$), it shows that the longitudinal double-spin asymmetry for the quark-antiquark dijet production cross section (averaged over the electron's azimuthal angles) in the back-to-back limit ($p_\perp \sim Q \gg \Delta_\perp$) uniquely probes the WW gluon helicity distribution $g_{1L}^{G \, WW} (x, k_T^2)$.  The $\epsilon^{jm}$ factor in \eq{DSA105} is crucial to uniquely project out the WW gluon helicity distribution. It is clear that the photon splittings both inside and outside the shockwave are needed to correctly generate the factor $\epsilon^{jm}$. In this sense, the part of the process where the virtual photon splits into a quark-antiquark pair inside the shockwave must be considered. The leading-order result \eqref{DSA106}, however, does not contain linearly polarized gluon distribution in the longitudinally polarized proton $ h_{1L}^{\perp G \, WW}(x, k_T^2)$. This is quite different from the dijet production cross section in the unpolarized electron-proton collisions: in that case, the cross section in the back-to-back limit not only probes the (unpolarized) WW gluon distribution, but it also depends on the linearly polarized gluon distribution \cite{Metz:2011wb,Dominguez:2011br, Dumitru:2015gaa, Dumitru:2018kuw}.

\section{small-$x$ Evolution Equation for WW Gluon Helicity Distribution}
\label{sec:deriving_helicity_evolution}

In this Section, we derive the small-$x$ evolution equation governing the small-$x$ behavior of the WW gluon helicity TMD. We use the same operator method that has been used in deriving the small-$x$ helicity evolution equation for dipole gluon helicity TMD \cite{Kovchegov:2018znm, Cougoulic:2022gbk, Borden:2024bxa}. 

In Appendix~\ref{appendix:reproduce_unpolWW_evol}, using this operator method, we reproduce the small-$x$ evolution equation for unpolarized WW gluon TMD that was already obtained in \cite{Dominguez:2011gc}. Unlike the approach in \cite{Dominguez:2011gc} where the evolution equation was derived from the JIMWLK evolution equation for the color quadrupole (cf. \cite{Jalilian-Marian:2004vhw}), the operator treatment in Appendix~\ref{appendix:reproduce_unpolWW_evol} gives a somewhat more physical derivation in which the corresponding Feynman diagrams are explicitly computed. 

To determine the small-$x$ asymptotics for the WW gluon helicity TMD, we employ the definition \eqref{GWW_def0}, which we restate here for convenience, 
\begin{equation}\label{eq:WW_helicity_operator}
G^{WW}_{10} (s) \equiv \frac{1}{2N_c}\epsilon^{ji} \, \textrm{Re} \llangle \mathrm{tr}\left[V_{\un{x}_0}\partial^i V_{\un{x}_0}^{\dagger} V^{j \, \textrm{G}[2]}_{\un{x}_1} V^{\dagger}_{\un{x}_1}\right]\rrangle (s),
\end{equation}
along with
\begin{align}\label{gww_GW}
    g_{1L}^{G \, WW} (x, k_T^2) =  \frac{N_c}{\as  \, 2 \pi^4} \,  \int d^2 x_1 \, d^2 x_0 \, e^{- i {\un k} \cdot {\un x}_{10}} \, G^{WW}_{10} \left(s = \frac{Q^2}{x}\right) .
\end{align}
Our goal is to construct the small-$x$ evolution equation for $G^{WW}_{10} (s)$. Knowing the high-energy asymptotics of $G^{WW}_{10} (s)$ readily gives one the small-$x$ asymptotics of $g_{1L}^{G \, WW} (x, k_T^2)$ via \eq{gww_GW}.

\begin{figure}[h]
    \centering
    \includegraphics[width=0.8\textwidth]{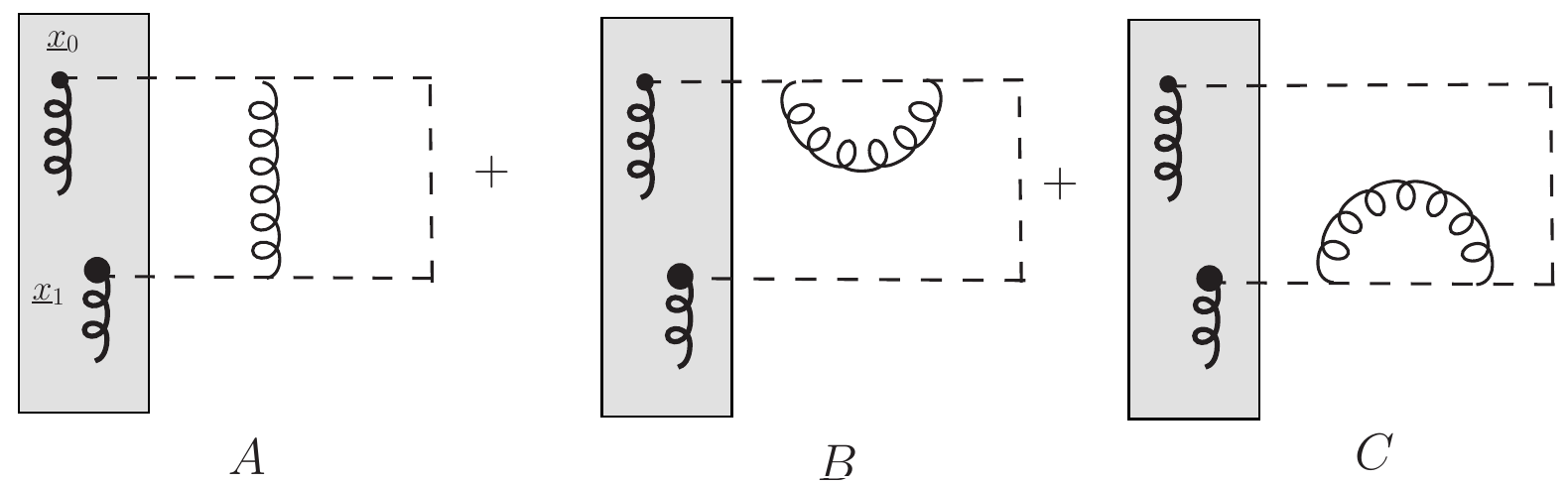}
    \caption{Diagrams representing one step of evolution for WW gluon distribution resulting from virtual gluon corrections. Dashed lines denote the Wilson line staple in the adjoint representation. The two background gluon fields stay inside the shockwave.}
	\label{fig:helicity_WW_diagrams_1}
\end{figure}

To facilitate construction of an evolution equation for $G^{WW}_{10} (s)$, we can recast it as
\begin{equation}\label{eq:WW_helicity_operator_adj}
G^{WW}_{10} (s)  = \frac{-g^2P^+}{(2s) \, (4 N_c)}\epsilon^{ji}\int_{-\infty}^{\infty} dx_0^- dx_1^-\, \llangle U_{\un{x}_0}^{ba}[\infty, x_0^-]\, E^i_{1,a}(x_0^-, \un{x}_0)\,  U^{bc}_{\un{x}_1}[\infty, x_1^-]\, E^j_{2,c}(x_1^-, \un{x}_1) \rrangle (s) + \cc \, .
\end{equation}
We have used  
\begin{equation}
V_{\un{x}_0}\partial^i V_{\un{x}_0}^{\dagger}
=-ig\int_{-\infty}^{\infty} dx_0^-\,  E^i_{1,a}(x_0^-, \un{x}_0) \, U_{\un{x}_0}^{\dagger ab}[\infty, x_0^-]t^b 
\end{equation}
and 
\begin{equation}\label{eq:VjG[2]_adjoint}
V^{j \, \textrm{G}[2]}_{\un{x}_1}V_{\un{x}_1}^{\dagger} = \frac{-igP^+}{s} \int_{-\infty}^{\infty} dx_1^- \, E^j_{2,c}(x_1^-, \un{x}_1) \, U^{\dagger cd}_{\un{x}_1}[\infty, x_1^-] t^d
\end{equation}
with  
\begin{equation}
E^j_{2, a}(x_1^-, \un{x}_1) \equiv x_1^- \partial^j A^{a \, +} (x_1^-, \un{x}_1) + A^{a \, j} (x_1^-, \un{x}_1),
\end{equation}
\begin{equation}
E^i_{1, a}(x_0^-, \un{x}_0) \equiv \partial^i A^{a \, +} (x_0^-, \un{x}_0).
\end{equation}
We will use the LCOT method \cite{Kovchegov:2018znm, Cougoulic:2022gbk} in alignment with Wilson's renormalization group approach \cite{Peskin:1995ev} to derive the small-$x$ evolution equation for $G^{WW}_{10} (s)$ in Eq.~\eqref{eq:WW_helicity_operator_adj}. The contributing diagrams are given in Figs.~\ref{fig:helicity_WW_diagrams_1}, ~\ref{fig:helicity_WW_diagrams_2} and ~\ref{fig:helicity_WW_diagrams_3}, in which the dashed lines denote the (future-pointing) adjoint Wilson line staple. These diagrams are organized according to the number of the ``end point" operators $E_1^i$ and $E_2^j$ located inside the shockwave. For example, in Fig.~\ref{fig:helicity_WW_diagrams_1}, both $E_1^i$ and $E_2^j$ are located inside the shockwave. In \fig{fig:helicity_WW_diagrams_2}, one of the end point operators is inside the shockwave while the other becomes a quantum fluctuating field and is located to the left of (before) the shockwave. In Fig.~\ref{fig:helicity_WW_diagrams_3}, neither of the two end point operators is inside the shockwave. They both become quantum fluctuating fields that need to be integrated out. 

We note that in addition to the diagrams F and I shown in \fig{fig:helicity_WW_diagrams_2}, there are diagrams with the ``end point" vertex to the right (after) the shock wave, with the gluon line (representing the quantum field) connecting to either of the two Wilson lines to the right of the  shock wave. Such diagrams are zero by color algebra, and are not shown here.

\subsection{Diagrams $A$, $B$ and $C$}

Diagrams $A$, $B$, and $C$ are the virtual diagrams similar to those contributing to the BK/JIMWLK equation. Their expressions are well-known and are straightforward to compute. We have 
\begin{equation}
A+ B+ C = -\frac{\alpha_s N_c}{2\pi^2}\int\limits^z_{\Lambda^2/s}  \frac{dz'}{z'} \int d^2x_2\,  \mathcal{K}_0(\un{x}_{20}, \un{x}_{21})\, 
\frac{1}{4 N_c} \, \epsilon^{ji} \llangle \mathrm{tr}\left[V_{\un{x}_0}\partial^iV_{\un{x}_0}^{\dagger} V^{j \, \textrm{G}[2]}_{\un{x}_1}V^{\dagger}_{\un{x}_1}\right] + \cc \rrangle (z' s).
\end{equation}
Here the kernel function is defined by
\begin{equation}\label{eq:KN_K0}
   \mathcal{K}_0(\un{x}_{20}, \un{x}_{21}) \equiv \frac{x_{10}^2}{x_{20}^2\,x_{21}^2}
\end{equation}
while $z'$ is the longitudinal momentum fraction of the projectiles momentum carried by the emitted gluon, which is bounded from above by the longitudinal momentum fraction $z$ of the ``parent" operator. As before, ${\un x}_{ij} = {\un x}_i - {\un x}_j$, with $x_{ij} = |\un x_{ij}|$.

\subsection{Diagrams $D, E, F, G, H, I$}

\begin{figure}
    \centering
    \includegraphics[width=0.9\textwidth]{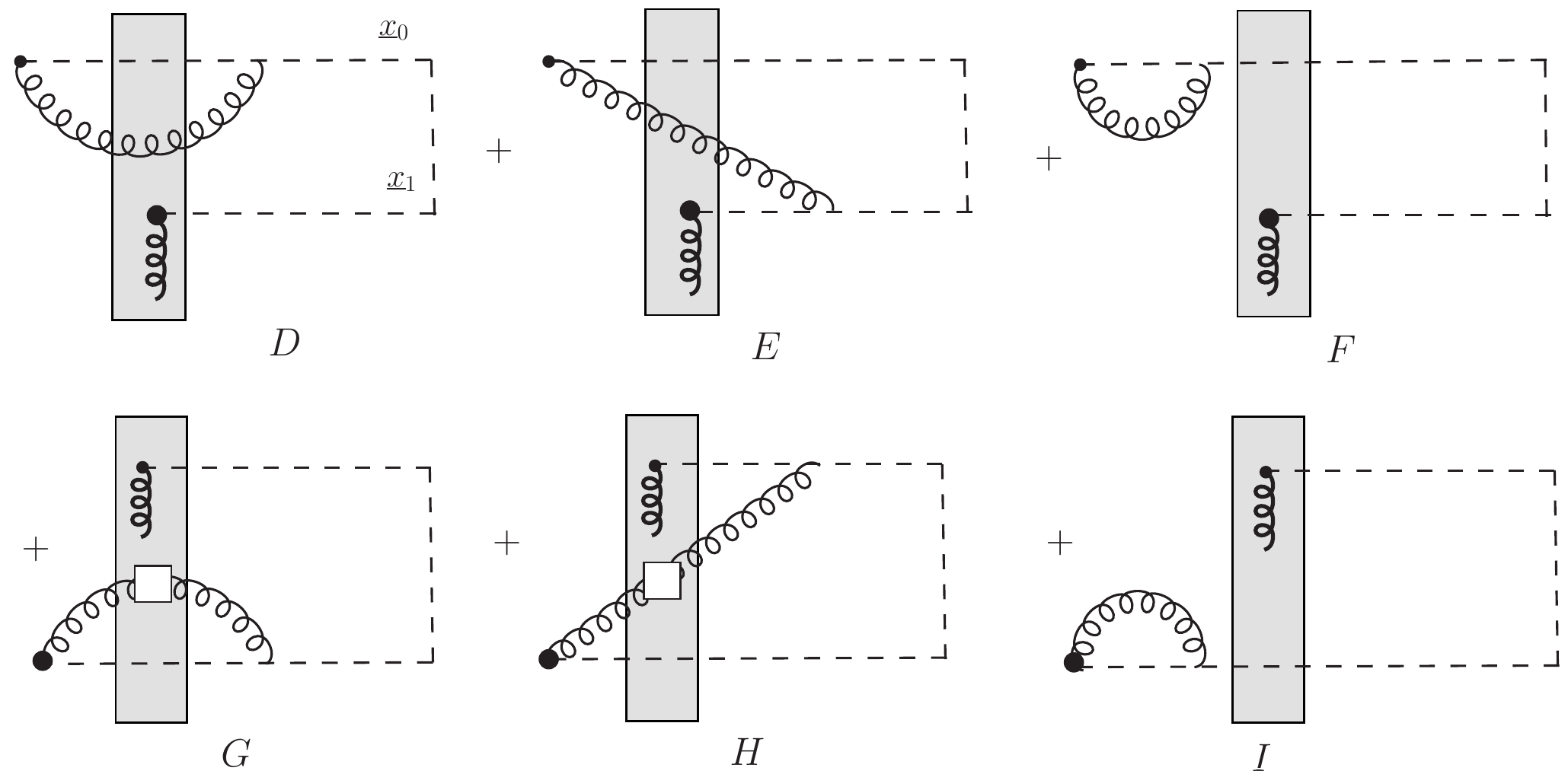}
    \caption{Diagrams for one step of evolution for WW gluon distribution. Dashed lines denote the Wilson line staple in the adjoint representation. One of the two background gluon fields remains inside the shockwave, while the other is outside of the shockwave.}
	\label{fig:helicity_WW_diagrams_2}
\end{figure}

Moving on to Fig.~\ref{fig:helicity_WW_diagrams_2}, we note that the diagrams $F$ and $I$ vanish due to the vanishing color factors. We will now compute the diagrams $D$ and $E$ and then analyze the diagrams $G$ and $H$.

For diagram $E$, the field $E^j_2$ at transverse position $\un{x}_1$ stays inside the shockwave and is treated as background fields. On the other hand, the chromo-electric field $E^i_1$ at transverse position $\un{x}_0$ is located before the shockwave and is treated as a quantum fluctuating field. The other quantum fluctuating field comes from the eikonal Wilson line at transverse position $\un{x}_1$ after the shockwave. One needs to average over these two quantum fluctuating fields, which can be done by employing the following eikonal background-field gluon propagator 
\begin{align}\label{eik}
\int\limits_{-\infty}^0 dx_{2'}^- \,  
\int\limits_0^\infty dx_2^- \, 
\contraction[2ex]
{}
{a}{^{+ \, a}
(x_{2'}^- , \ul{x}_1) \Big] \:}
{a}
\: 
a^{+ \, a} (x_{2'}^- , \ul{x}_1) \:
a^{\, + \, b} (x_2^- , \ul{x}_0)   = - \frac{1}{4 \pi^3} \, \int\limits^z_{\Lambda^2/s} \frac{d z'}{z'} \, \int d^2 x_2 \, U^{ba}_{{\un x}_2} \, \frac{{\un x}_{21} \cdot {\un x}_{20}}{x_{21}^2 \, x_{20}^2} .
\end{align}
This gives, after using Eq.~\eqref{eq:VjG[2]_adjoint}, 
\begin{equation}\label{eq:E_expr}
E = -\frac{\alpha_s}{\pi^2}  \int\limits^z_{\Lambda^2/s} \frac{d z'}{z'} \, \int d^2x_2\,  \mathcal{E}^j (\un{x}_{20}, \un{x}_{21})\, \frac{1}{4 N_c} \, \llangle \mathrm{Tr}\left[U^{\dagger}_{\un{x}_2} T^a U_{\un{x}_0}\right]\mathrm{tr}\left[V^{j \, \textrm{G} [2]}_{\un{x}_1}V^{\dagger}_{\un{x}_1} t^a \right] + \cc \rrangle (z' s), 
\end{equation}
where we have defined the kernel function
\begin{align}\label{eq:kernel_E2i}
\mathcal{E}^j (\un{x}_{20}, \un{x}_{21}) \equiv \epsilon^{ji}\partial^i_{\un{x}_0} \left(\frac{\un{x}_{20}\cdot\un{x}_{21}}{x_{20}^2\, x_{21}^2}\right) =\frac{x_{10}^2}{x_{20}^2\, x_{21}^2} \left(-\frac{\epsilon^{ji}x_{10}^i}{x_{10}^2} + \frac{\epsilon^{ji}x_{20}^i}{x_{20}^2}\right) - \frac{\epsilon^{ji}x_{20}^i}{x_{20}^4} + \frac{\epsilon^{ji}x_{21}^i}{x_{21}^2} \, \pi \delta^2(\un{x}_{20}).
\end{align}

Diagram $D$ can be calculated similarly. In fact, its expression can be obtained from \eq{eq:E_expr} by replacing $\mathcal{E}^j(\un{x}_{20}, \un{x}_{21}) \to \mathcal{E}^j(\un{x}_{20}, \un{x}_{20})$ and changing the overall sign. Combining diagrams $D$ and $E$, one gets
\begin{equation}\label{eq:helicity_D+E}
D+E  =  -\frac{\alpha_s}{\pi^2}  \int\limits^z_{\Lambda^2/s} \frac{d z'}{z'} \, \int d^2x_2\,\, \mathcal{K}^j_0 (\un{x}_{20}, \un{x}_{21})\, \frac{1}{4 N_c} \, \llangle \mathrm{Tr}\left[U^{\dagger}_{\un{x}_2} T^a U_{\un{x}_0}\right]\mathrm{tr}\left[V^{j \, \textrm{G}[2]}_{\un{x}_1}V^{\dagger}_{\un{x}_1} t^a\right] + \cc \rrangle (z' s)
\end{equation}
with the combined kernel function
\begin{equation}\label{eq:KN_K0j}
\mathcal{K}_0^j(\un{x}_{20}, \un{x}_{21}) \equiv \frac{x_{10}^2}{x_{20}^2\, x_{21}^2} \, \epsilon^{ji} \, \left(\frac{ x_{20}^i}{x_{20}^2} -\frac{x_{10}^i}{x_{10}^2} \right) .
\end{equation}
The term containing $\delta^2(\un{x}_{20})$ from \eq{eq:kernel_E2i} vanishes due to the vanishing color factor in the operator multiplying it. \\

In diagram $H$, the eikonal background field $E_1^i$ at transverse position $\un{x}_0$ stays inside the shockwave while the sub-eikonal-order field $E_2^j$ at transverse position $\un{x}_1$ becomes a quantum fluctuating field and is located before the shockwave. The other quantum fluctuating field comes from the eikonal Wilson line at $\un{x}_0$ after the shockwave. The relevant sub-eikonal gluon propagator in the background field is
\begin{align}\label{Vi5}
& \int\limits_{-\infty}^0 dx_{2'}^- \,  
\int\limits_0^\infty dx_2^- \, \Big[ x_{2'}^- {\pd}^i a^{+ \, a} (x_{2'}^- , \ul{x}_1)
\contraction[2ex]
{}
{+}{a^{i \, a}
(x_{2'}^- , \ul{x}_1) \Big] \:}
{a}
\: 
+ a^{i \, a} (x_{2'}^- , \ul{x}_1) \Big] \:
a^{+ \, b} (x_2^- , \ul{x}_0)  \\ 
& = \frac{1}{(2 \pi)^3} \int\limits_{0}^{q^-} d k^- \Bigg\{ \int d^2 x_2 \left[ \frac{\epsilon^{ij} x_{20}^j}{x_{20}^2} - 2 x_{21}^i \frac{{\un x}_{21} \times {\un x}_{20}}{x_{21}^2 \, x_{20}^2} \right] \left( U_{{\ul x}_2}^{\textrm{pol} [1]} \right)^{b a}  - i  \int d^2 x_2 d^2 x_{2'} \left[ \frac{ x_{20}^i}{x_{20}^2} - 2 x_{2'1}^i \frac{{\un x}_{2'1} \cdot {\un x}_{20}}{x_{2'1}^2 \, x_{20}^2} \right] \left( U_{{\ul x}_2, {\un x}_{2'}}^{\textrm{pol} [2]} \right)^{b a} \notag \\
& + \int d^2 x_2 \, \frac{x_{20}^j}{x_{20}^2} \, \left( \delta^{ij} - 2 \, \frac{x_{21}^i \, x_{21}^j}{x_{21}^2} \right) \, \left( U_{\un x_2}^{\textrm{G} [3]} \right)^{ba} \Bigg\}, \notag
\end{align}
where the $U$'s are the adjoint counter-parts of the fundamental polarized Wilson lines defined in Eqs.~\eqref{Vxy_sub-eikonal}, \eqref{VqG}, and \eqref{VG3} (see \cite{Cougoulic:2022gbk} for this propagator, excluding the $U_{\un x_2}^{\textrm{G} [3]}$ term). Here $U_{{\ul x}_2, {\un x}_{2'}}^{\textrm{pol} [2]} = U_{{\ul x}_2, {\un x}_{2'}}^{\textrm{G} [2]}  + U_{{\ul x}_2}^{\textrm{q} [2]} \, \delta^2 ({\ul x}_2 - {\un x}_{2'})$ \cite{Cougoulic:2022gbk}.  
In \eq{Vi5}, $k^-$ is the minus momentum of the gluon, whose integration range is capped by the minus momentum of the probe, $q^-$, as a stand-in for the more detailed limits of integration introduced in the next expression.

Employing the propagator \eqref{Vi5}, we obtain the following expression for the diagram $H$, 
\begin{align}\label{eq:H_expr}
H 
=-\frac{\alpha_s}{4\pi^2} \int\limits^z_{\Lambda^2/s} \frac{d z'}{z'} \,  \int d^2x_2\, \frac{1}{4 N_c} \ &  \Bigg\{\mathcal{K}^i_1(\un{x}_{20}, \un{x}_{21}) \, \llangle \mathrm{tr}\left[V_{\un{x}_0}\partial^iV_{\un{x}_0}^{\dagger}t^a \right] \, \mathrm{Tr}\left[U_{\un{x}_2}^{\mathrm{pol}[1]}U_{\un{x}_1}^{\dagger} T^a \right] \rrangle (z' s) \\
&+ \mathcal{K}_2^{i,j}(\un{x}_{20}, \un{x}_{21})\, \llangle \mathrm{tr}\left[V_{\un{x}_0}\partial^iV_{\un{x}_0}^{\dagger}t^a \right] \, \mathrm{Tr}\left[U_{\un{x}_2}^{j \, \textrm{G}[2]}U_{\un{x}_1}^{\dagger} T^a \right] \rrangle (z' s) \notag \\
& - \mathcal{K}^i_2(\un{x}_{20}, \un{x}_{21} )\, \llangle \mathrm{tr}\left[V_{\un{x}_0}\partial^iV_{\un{x}_0}^{\dagger}t^a \right] \, \mathrm{Tr}\left[U_{\un{x}_2}^{\textrm{G}[3]}U_{\un{x}_1}^{\dagger} T^a \right] \rrangle (z' s)  + \cc \Bigg\}, \notag
\end{align}
where we have introduced the kernel functions 
\begin{subequations}\label{eq:KN_K1i_K2i}
\begin{align}
&\mathcal{K}_1^i(\un{x}_{20}, \un{x}_{21}) \equiv  -\epsilon^{ji}\,\frac{x_{20}^l}{x_{20}^2} \left[\delta^{jm} - 2\, \frac{x_{21}^jx_{21}^m}{x_{21}^2}\right]\epsilon^{lm},\\
&\mathcal{K}_2^i(\un{x}_{20}, \un{x}_{21}) \equiv  - \epsilon^{ji}\, \frac{x_{20}^l}{x_{20}^2} \left[\delta^{jm} - 2 \, \frac{x_{21}^jx_{21}^m}{x_{21}^2}\right]\delta^{lm}, \\
\label{eq:KN_K2in}
&\mathcal{K}_2^{i,j}(\un{x}_{20}, \un{x}_{21}) \equiv - (\partial^j_{\un{x}_0} - \partial^j_{\un{x}_1}) \mathcal{K}_2^i(\un{x}_{20}, \un{x}_{21}).
\end{align}
\end{subequations}
In arriving at \eq{eq:H_expr}, we have discarded contributions from terms containing $U^{\textrm{q} [2]}_{\un{x}_2}$: these terms are imaginary and cancel out when adding the complex conjugate counterpart. Similar cancellation happens for the term involving $U^{\textrm{G}[2]}_{\un{x}_2, \un{x}_{2'}}$. Using \eq{eq:VxyG2_VxiG[2]}, one can show that the only surviving piece is proportinal to $U^{j \, \textrm{G}[2]}_{\un{x}_2}$.

The expression for the diagram $G$ can be obtained from \eq{eq:H_expr} by replacing $\un{x}_{20}$ in the kernel functions $\mathcal{K}_1^i(\un{x}_{20}, \un{x}_{21})$,  $\mathcal{K}_2^{i,n}(\un{x}_{20}, \un{x}_{21})$, $\mathcal{K}_2^i(\un{x}_{20}, \un{x}_{21})$ with $\un{x}_{21}$ and changing the overall sign. Summing up diagram $H$ and diagram $G$, one gets
\begin{align}\label{eq:helicity_H+G}
H+G 
=& - \frac{\alpha_s}{4\pi^2} \int\limits^z_{\Lambda^2/s} \frac{d z'}{z'} \, \int d^2x_2\, \frac{1}{4 N_c} \, \Bigg\{ \left[ \mathcal{K}^i_1(\un{x}_{20}, \un{x}_{21}) - \mathcal{K}^i_1(\un{x}_{21}, \un{x}_{21})\right] \, \llangle \mathrm{tr}\left[V_{\un{x}_0}\partial^iV_{\un{x}_0}^{\dagger}t^a \right] \, \mathrm{Tr}\left[U_{\un{x}_2}^{\mathrm{pol}[1]}U_{\un{x}_1}^{\dagger} T^a \right] \rrangle (z's)  \notag \\
& + \left[ \mathcal{K}_2^{i,j}(\un{x}_{20}, \un{x}_{21}) - \mathcal{K}_2^{i,j}(\un{x}_{21}, \un{x}_{21})\right] \, \llangle \mathrm{tr}\left[V_{\un{x}_0}\partial^iV_{\un{x}_0}^{\dagger}t^a \right] \, \mathrm{Tr}\left[U_{\un{x}_2}^{j \, \textrm{G}[2]}U_{\un{x}_0}^{\dagger} T^a \right] \rrangle (z' s) \notag\\
&- \left[ \mathcal{K}^i_2(\un{x}_{20}, \un{x}_{21} ) - \mathcal{K}^i_2(\un{x}_{21}, \un{x}_{21} )\right] \, \llangle \mathrm{tr}\left[V_{\un{x}_0}\partial^iV_{\un{x}_0}^{\dagger}t^a \right] \, \mathrm{Tr}\left[U_{\un{x}_2}^{\textrm{G}[3]}U_{\un{x}_1}^{\dagger} T^a \right] \rrangle (z' s) + \cc \Bigg\}.
\end{align}
To be more precise, we note that
\begin{align}
    \mathcal{K}_2^{i,j}(\un{x}_{21}, \un{x}_{21}) = \lim_{{\un x}_0 \to {\un x}_1} \, \mathcal{K}_2^{i,j}(\un{x}_{20}, \un{x}_{21}) .
\end{align}

\subsection{Diagrams $J, K, L, M$}

\begin{figure}[h]
    \centering
    \includegraphics[width=0.9\textwidth]{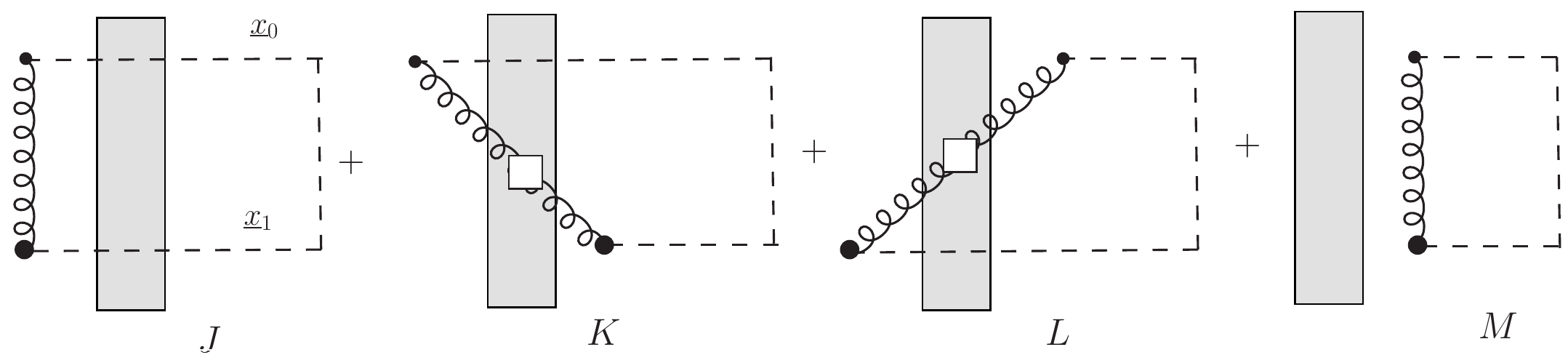}
    \caption{Diagrams for one step evolution of WW gluon distribution. Dashed lines denote the Wilson line staple in the adjoint representation. Both background gluon fields are located outside of the shockwave.}
	\label{fig:helicity_WW_diagrams_3}
\end{figure}

We proceed by considering the diagrams in \fig{fig:helicity_WW_diagrams_3}. In that figure, the diagrams $J$ and $M$ vanish because they contain the factor $\epsilon^{ji} k^j k^i=0$ with $k$ the gluon momentum.

To calculate the diagrams $K$ and $L$ we will need the propagator \eqref{Vi5} along with its counterpart with different $x^-$-orderings of the endpoints,
\begin{align}\label{Vi6}
& \int\limits_{-\infty}^0 dx_{2'}^- \,  
\int\limits_0^\infty dx_2^- \, \Big[ x_2^- {\pd}^i a^{+ \, b} (x_2^- , \ul{x}_1)
\contraction[2ex]
{}
{+}{a^{i \, b}
(x_2^- , \ul{x}_1) \Big] \:}
{a}
\: 
+ a^{i \, b} (x_2^- , \ul{x}_1) \Big] \:
a^{+ \, a} (x_{2'}^- , \ul{x}_0) \\ & = \frac{1}{(2 \pi)^3} \int\limits_0^{p_2^-} d k^- \Bigg\{ \int d^2 x_2 \left[ \frac{\epsilon^{ij} x_{20}^j}{x_{20}^2} - 2 x_{21}^i \frac{{\un x}_{21} \times {\un x}_{20}}{x_{21}^2 \, x_{20}^2} \right] \left( U_{{\ul x}_2}^{\textrm{pol} [1]} \right)^{b a} + i  \int d^2 x_2 d^2 x_{2'} \left[ \frac{ x_{2'0}^i}{x_{2'0}^2} - 2 x_{21}^i \frac{{\un x}_{21} \cdot {\un x}_{2'0}}{x_{21}^2 \, x_{2'0}^2} \right] \left( U_{{\ul x}_2, {\un x}_{2'}}^{\textrm{pol} [2]} \right)^{b a} \notag \\
& + \int d^2 x_2 \, \frac{x_{20}^j}{x_{20}^2} \, \left( \delta^{ij} - 2 \, \frac{x_{21}^i \, x_{21}^j}{x_{21}^2} \right) \, \left( U_{\un x_2}^{\textrm{G} [3]} \right)^{ba} \Bigg\} . \notag
\end{align}
(See \cite{Cougoulic:2022gbk} again for the derivation of all but the $U_{\un x_2}^{\textrm{G} [3]}$ terms.)

In the diagram $K$, both fields at transverse positions $\un{x}_0$ and $\un{x}_1$ are quantum fluctuating fields. The field $E^i_1$ at $\un{x}_0$ is located before the shockwave, while the field $E^j_2$ at $\un{x}_1$ is located after the shockwave. Similar to the above calculations, employing \eq{eq:VxyG2_VxiG[2]}, we obtain, for the diagram $K$,
\begin{align}
K = -\frac{\alpha_s}{8\pi^2} \int\limits^z_{\Lambda^2/s} \frac{d z'}{z'} \,  \int d^2x_2\, \frac{1}{4 N_c} \, \Bigg\{ & \, \mathcal{J}_1(\un{x}_{20}, \un{x}_{21}) \, \llangle \mathrm{Tr}\left[U^{\mathrm{pol}[1]}_{\un{x}_2} U^{\dagger}_{\un{x}_0}\right] \rrangle (z' s) +  \mathcal{J}_2^{i}(\un{x}_{20}, \un{x}_{21}) \, \llangle \mathrm{Tr}\left[U^{i \, \textrm{G} [2]}_{\un{x}_2}U^{\dagger}_{\un{x}_0}\right] \rrangle (z' s) \notag  \\ 
& - \mathcal{J}_2(\un{x}_{20}, \un{x}_{21}) \, \llangle \mathrm{Tr}\left[U^{\textrm{G}[3]}_{\un{x}_2}U^{\dagger}_{\un{x}_0}\right] \rrangle (z' s) + \cc \Bigg\}, 
\end{align} 
where we have introduced the kernel functions 
\begin{subequations}\label{Js}
\begin{align}
&\mathcal{J}_1(\un{x}_{20}, \un{x}_{21})
= \partial^i_{\un{x}_0} \mathcal{K}^i_1(\un{x}_{20}, \un{x}_{21}), \, \label{eq:KN_J1}\\
&\mathcal{J}_2(\un{x}_{20}, \un{x}_{21}) = \partial^i_{\un{x}_0}\mathcal{K}^i_2(\un{x}_{20}, \un{x}_{21}),\, \label{eq:KN_J2}\\
&\mathcal{J}_2^{i}(\un{x}_{20}, \un{x}_{21})= -(\partial_{\un{x}_{0}}^i - \partial^i_{\un{x}_{1}}) \, \mathcal{J}_2(\un{x}_{20}, \un{x}_{21}).\, \label{eq:KN_J2n}
\end{align}
\end{subequations}

We continue by computing the diagram $L$ in which the quantum fluctuating field at the transverse position $\un{x}_0$ is located after the shockwave while the quantum fluctuating field at $\un{x}_1$ is located after the shockwave. We obtain 
\begin{align}
L = -\frac{\alpha_s}{8\pi^2} \int\limits^z_{\Lambda^2/s} \frac{d z'}{z'} \,  \int d^2x_2\, \frac{1}{4 N_c} \, \Bigg\{ & \, \mathcal{J}_1(\un{x}_{20}, \un{x}_{21}) \, \llangle \mathrm{Tr}\left[U^{\mathrm{pol}[1]}_{\un{x}_2} U^{\dagger}_{\un{x}_1}\right] \rrangle (z' s)  +  \mathcal{J}_2^{i}(\un{x}_{20}, \un{x}_{21}) \, \llangle \mathrm{Tr}\left[U^{i \, \textrm{G} [2]}_{\un{x}_2}U^{\dagger}_{\un{x}_1}\right] \rrangle (z' s) \notag \\
& - \mathcal{J}_2(\un{x}_{20}, \un{x}_{21}) \, \llangle \mathrm{Tr}\left[U^{\textrm{G}[3]}_{\un{x}_2}U^{\dagger}_{\un{x}_1}\right] \rrangle (z' s)  + \cc \Bigg\}.
\end{align} 
Combining diagram $K$ and diagram $L$, one gets
\begin{align}\label{eq:helicity_K+L_adjoint}
K+L =  -\frac{\alpha_s}{8\pi^2} \int\limits^z_{\Lambda^2/s} \frac{d z'}{z'} \,  \int d^2x_2\, \frac{1}{4 N_c} \,  \Bigg\{ & \,\mathcal{J}_1(\un{x}_{20}, \un{x}_{21}) \, \llangle \mathrm{Tr}\left[U^{\mathrm{pol}[1]}_{\un{x}_2} U^{\dagger}_{\un{x}_0}\right]+\mathrm{Tr}\left[U^{\mathrm{pol}[1]}_{\un{x}_2} U^{\dagger}_{\un{x}_1}\right]\rrangle (z's)  \\
& +  \mathcal{J}_2^{i}(\un{x}_{20}, \un{x}_{21}) \, \llangle\mathrm{Tr}\left[U^{i \,, \textrm{G} [2]}_{\un{x}_2}U^{\dagger}_{\un{x}_0}\right]+\mathrm{Tr}\left[U^{i \, \textrm{G} [2]}_{\un{x}_2}U^{\dagger}_{\un{x}_1}\right] \rrangle (z's) \notag \\ 
& - \mathcal{J}_2(\un{x}_{20}, \un{x}_{21}) \, \llangle \mathrm{Tr}\left[U^{\textrm{G}[3]}_{\un{x}_2}U^{\dagger}_{\un{x}_0}\right]+\mathrm{Tr}\left[U^{\textrm{G}[3]}_{\un{x}_2}U^{\dagger}_{\un{x}_1}\right] \rrangle (z's) + \cc \Bigg\}. \notag
\end{align}

\subsection{Small-$x$ evolution equation in DLA}
\label{sec:DLA_evol}

Combining the results of the above calculations of the diagrams in Figs.~\ref{fig:helicity_WW_diagrams_1}, ~\ref{fig:helicity_WW_diagrams_2} and ~\ref{fig:helicity_WW_diagrams_3} we arrive at the following expression for the snall-$x$ evolution of $G^{WW}_{10}$:
\begin{tcolorbox}[colback=gray!10!white]
\begin{align}\label{eq:final_result_WW}
& G^{WW}_{10} (zs) = G^{WW \, (0)}_{10} (z s) \\
 &-\frac{\alpha_s N_c}{2\pi^2}\int^z_{\Lambda^2/s}\frac{dz'}{z'} \int d^2x_2\,  \mathcal{K}_0(\un{x}_{20}, \un{x}_{21})\, \frac{1}{4 N_c}\epsilon^{ji}\llangle \mathrm{tr}\left[V_{\un{x}_0}\partial^iV_{\un{x}_0}^{\dagger} V^{j \, \textrm{G}[2]}_{\un{x}_1}V^{\dagger}_{\un{x}_1}\right] +\mbox{\mbox{c.c.}}\rrangle (z's) \notag \\ 
&-\frac{\alpha_sN_c}{\pi^2}  \int^z_{\Lambda^2/s}\frac{dz'}{z'} \int d^2x_2\,\, \mathcal{K}^i_0 (\un{x}_{20}, \un{x}_{21})\, \frac{1}{4 N_c^2}\llangle \mathrm{Tr}\left[U^{\dagger}_{\un{x}_2} T^a U_{\un{x}_0}\right]\mathrm{tr}\left[V^{i \, \textrm{G}[2]}_{\un{x}_1}V^{\dagger}_{\un{x}_1} t^a\right]+\mbox{\mbox{c.c.}}\rrangle (z's) \notag \\ 
&-\frac{\alpha_sN_c}{4\pi^2}\int^z_{\Lambda^2/s}\frac{dz'}{z'}\int d^2x_2\, \Bigg\{\left(\mathcal{K}^i_1(\un{x}_{20}, \un{x}_{21}) - \mathcal{K}^i_1(\un{x}_{21}, \un{x}_{21})\right) \, \frac{1}{4 N_c^2}\llangle \mathrm{tr}\left[V_{\un{x}_0}\partial^iV_{\un{x}_0}^{\dagger}t^a\right] \, \mathrm{Tr}\left[U_{\un{x}_2}^{\mathrm{pol}[1]}U_{\un{x}_1}^{\dagger} T^a \right]\rrangle (z's) \notag \\ 
&\qquad\qquad + \left(\mathcal{K}_2^{i,j}(\un{x}_{20}, \un{x}_{21}) - \mathcal{K}_2^{i,j}(\un{x}_{21}, \un{x}_{21})\right)\, \frac{1}{4 N_c^2}\llangle \mathrm{tr}\left[V_{\un{x}_0}\partial^iV_{\un{x}_0}^{\dagger}t^a\right] \, \mathrm{Tr}\left[U_{\un{x}_2}^{j \, \textrm{G}[2]}U_{\un{x}_1}^{\dagger} T^a\right] \rrangle (z's) \notag \\ 
&\qquad \qquad- \left( \mathcal{K}^i_2(\un{x}_{20}, \un{x}_{21} ) - \mathcal{K}^i_2(\un{x}_{21}, \un{x}_{21} )\right)\frac{1}{4 N_c^2}\llangle \mathrm{tr}\left[V_{\un{x}_0}\partial^iV_{\un{x}_0}^{\dagger}t^a\right] \, \mathrm{Tr}\left[U_{\un{x}_2}^{\textrm{G}[3]}U_{\un{x}_1}^{\dagger} T^a \right]\rrangle (z's) + \mbox{\mbox{c.c.}}\Bigg\} \notag \\ 
& -\frac{\alpha_sN_c}{8\pi^2}\int^z_{\Lambda^2/s}\frac{dz'}{z'} \int d^2x_2\, \Bigg\{\mathcal{J}_1(\un{x}_{20}, \un{x}_{21}) \, \frac{1}{4 N_c^2}\llangle\mathrm{Tr}\left[U^{\mathrm{pol}[1]}_{\un{x}_2} U^{\dagger}_{\un{x}_0}\right]+\mathrm{Tr}\left[U^{\mathrm{pol}[1]}_{\un{x}_2} U^{\dagger}_{\un{x}_1}\right]\rrangle (z's) \, \notag \\ 
&\qquad \qquad  +  \mathcal{J}_2^{i}(\un{x}_{20}, \un{x}_{21})\, \frac{1}{4 N_c^2}\llangle\mathrm{Tr}\left[U^{i \, \textrm{G}[2]}_{\un{x}_2}U^{\dagger}_{\un{x}_0}\right]+\mathrm{Tr}\left[U^{i \, \textrm{G}[2]}_{\un{x}_2}U^{\dagger}_{\un{x}_1}\right] \rrangle (z's) \notag \\ 
&\qquad \qquad - \mathcal{J}_2(\un{x}_{20}, \un{x}_{21})\, \frac{1}{4 N_c^2}\llangle\mathrm{Tr}\left[U^{\textrm{G}[3]}_{\un{x}_2}U^{\dagger}_{\un{x}_0}\right]+\mathrm{Tr}\left[U^{\textrm{G}[3]}_{\un{x}_2}U^{\dagger}_{\un{x}_1}\right]\rrangle(z's)+\mbox{\mbox{c.c.}}\Bigg\}. \notag 
\end{align}
\end{tcolorbox}
Here $G^{WW \, (0)}_{10} (z s)$ represents the inhomogeneous term, given by the initial conditions for the evolution. The expressions of the kernel functions are given in Eqs.~\eqref{eq:KN_K0}, ~\eqref{eq:KN_K0j},~\eqref{eq:KN_K1i_K2i}, and \eqref{Js}.  
The integration over the transverse position $\un{x}_2$ should be understood as constrained by the $x^-$-direction life-time ordering, which typically leads to $\theta(zx_{10}^2 - z'x_{21}^2)$ multiplying the kernel, along with the constraint from the ultraviolet (UV) region, $\mathrm{min}\{x_{20}^2, x_{21}^2\} >1/z's$. 

However, before specifying the integration limits, we need to clarify the approximation under which \eq{eq:final_result_WW} was derived. Similar to the unpolarized small-$x$ evolution \cite{Balitsky:1995ub,Balitsky:1998ya,Kovchegov:1999yj,Kovchegov:1999ua,Jalilian-Marian:1997dw,Jalilian-Marian:1997gr,Weigert:2000gi,Iancu:2001ad,Iancu:2000hn,Ferreiro:2001qy}, this equation resums logarithms of energy (multiplied by $\as$) resulting from the longitudinal integrals over $z'$. Since the leading-order evolution for helicity observables is double-logarithmic, \eq{eq:final_result_WW} should contain this DLA contribution, along with the part of the single-logarithmic contribution labeled SLA$_L$ in \cite{Kovchegov:2021lvz}. Here the subscript $L$ indicates that the single logarithm of energy in the SLA contribution arises from the longitudinal ($z'$) integral. Since the single logarithm of energy may also arise from the transverse position integrals (SLA$_T$), both in the UV \cite{Kovchegov:2021lvz} and in the IR, none of which is included in our calculation, \eq{eq:final_result_WW} does not contain the complete DLA+SLA evolution. Rather, it contains only the DLA+SLA$_L$ piece. Therefore, it should be understood as an incomplete result at the SLA order, out of which we can extract the complete DLA contribution. This is what we will do next, while employing the large-$N_c$ limit.

The DLA term extraction is easier to perform for the impact parameter-integrated quantities. We, therefore, define
\begin{equation}\label{eq:integrate_impact_parameter}
\int d^2\left(\frac{\un{x}_0+\un{x}_1}{2}\right) \, G^{WW}_{10} (zs) \equiv G^{WW}(x_{10}^2, zs), \ \ \ \int d^2\left(\frac{\un{x}_0+\un{x}_1}{2}\right) \, G^{WW \, (0)}_{10} (zs) \equiv G^{WW \, (0)}(x_{10}^2, zs).
\end{equation}  
Further, we will employ the polarized dipole amplitude defined in \eq{Gi_def}: when integrated over all impact parameters it can be decomposed as \cite{Kovchegov:2017lsr, Cougoulic:2022gbk}
\begin{equation}\label{decomp21}
\int d^2\left(\frac{\un{x}_0+\un{x}_1}{2}\right)G^i_{10}( zs) = x_{10}^i \, G_1(x_{10}^2, zs) + \epsilon^{ij} x_{10}^j \, G_2(x_{10}^2, zs).
\end{equation} 
In addition, we define
\begin{equation}
G_{10}(zs) \equiv \frac{1}{2N_c} \llangle \mathrm{tr}\left[V_{\un{x}_0}^{\dagger} V_{\un{x}_1}^{\textrm{G}[1]}\right] + \mathrm{tr}\left[V_{\un{x}_1}^{\textrm{G}[1]\dagger} V_{\un{x}_0}\right]\rrangle(zs)
\end{equation}
and employ the definition \eqref{G3_def} above. The unpolarized quark dipole $S$-matrix is
\begin{equation}
S_{10}(zs) = \frac{1}{2 N_c} \left\langle \mathrm{tr}\left[V_{\un{x}_0}V_{\un{x}_1}^{\dagger}\right] + \mathrm{tr}\left[V_{\un{x}_1}V_{\un{x}_0}^{\dagger}\right] \right\rangle (zs) \approx \frac{1}{N_c} \left\langle \mathrm{tr}\left[V_{\un{x}_0}V_{\un{x}_1}^{\dagger}\right] \right\rangle (zs) \approx \frac{1}{N_c} \left\langle \mathrm{tr}\left[V_{\un{x}_1}V_{\un{x}_0}^{\dagger}\right] \right\rangle (zs).
\end{equation}
where we neglect the odderon contribution as a higher-order correction in $\as$.

The Wilson line structures in Eq.~\eqref{eq:final_result_WW} can be further simplified to be expressed in terms of (polarized) Wilson lines in the fundamental representation. We will work in the large-$N_c$ limit and ignore all contributions from quark loops so that $U^{\mathrm{pol}[1]}_{\un{x}_2} \approx U^{\textrm{G}[1]}_{\un{x}_2}$.

The Wilson line structure in 
line 3 of \eq{eq:final_result_WW} can be rewritten as
\begin{equation}\label{eq:new_structure_1}
\frac{1}{N_c^2}\llangle\mathrm{Tr}\left[U^{\dagger}_{\un{x}_2} T^a U_{\un{x}_0}\right]\mathrm{tr}\left[V^{j \, \textrm{G}[2]}_{\un{x}_1}V^{\dagger}_{\un{x}_1} t^a \right]\rrangle\\
= \frac{1}{2}\frac{1}{N_c}\llangle \mathrm{tr}\left[V^{j \,\textrm{G}[2]}_{\un{x}_1}V^{\dagger}_{\un{x}_1} V_{\un{x}_0}V_{\un{x}_2}^{\dagger}\right]\rrangle \, S_{02}(zs) + \mbox{c.c.} .
\end{equation}
We have used the large-$N_c$ approximation so that averaging over color-singlet objects factorizes out.
The Wilson line structures in lines 4, 5, and 6 of \eq{eq:final_result_WW}  
can be rewritten at large $N_c$ as
\begin{subequations}\label{largeNc_identities}
\begin{align}
&\frac{1}{N_c^2}\llangle \mathrm{tr}\left[V_{\un{x}_0}\partial^iV_{\un{x}_0}^{\dagger}t^a\right] \mathrm{Tr}\left[U_{\un{x}_2}^{\textrm{G}[1]}U_{\un{x}_1}^{\dagger} T^a\right]\rrangle\notag\\
= &\frac{1}{N_c}\llangle \mathrm{tr}\left[V_{\un{x}_0}\partial^iV_{\un{x}_0}^{\dagger}V_{\un{x}_2}^{\textrm{G}[1]} V_{\un{x}_1}^{\dagger}\right]\rrangle \, S_{21}(zs)+ \frac{1}{N_c} \left\langle \mathrm{tr}\left[V_{\un{x}_0}\partial^iV_{\un{x}_0}^{\dagger} V_{\un{x}_2}V_{\un{x}_1}^{\dagger}\right] \right\rangle \, G_{21}(zs) + \mbox{c.c.} , \label{eq:new_structure_2}\\
&\frac{1}{N_c^2}\llangle \mathrm{tr}\left[V_{\un{x}_0}\partial^iV_{\un{x}_0}^{\dagger}t^a\right] \mathrm{Tr}\left[U_{\un{x}_2}^{j \, \textrm{G}[2]}U_{\un{x}_1}^{\dagger} T^a\right]\rrangle \notag\\
= &\frac{1}{2} \frac{1}{N_c}\llangle \mathrm{tr}\left[V_{\un{x}_0}\partial^iV_{\un{x}_0}^{\dagger}V_{\un{x}_2}^{j \, \textrm{G}[2]} V_{\un{x}_1}^{\dagger}\right]\rrangle \, S_{21}( zs)+ \frac{1}{2}\frac{1}{N_c} \left\langle\mathrm{tr}\left[V_{\un{x}_0}\partial^iV_{\un{x}_0}^{\dagger} V_{\un{x}_2}V_{\un{x}_1}^{\dagger}\right] \right\rangle \, G_{21}^j ( zs)+\mbox{c.c.} , \label{eq:new_structure_3}\\
&\frac{1}{N_c^2}\llangle \mathrm{tr}\left[V_{\un{x}_0}\partial^iV_{\un{x}_0}^{\dagger}t^a\right] \mathrm{Tr}\left[U_{\un{x}_2}^{\textrm{G}[3]}U_{\un{x}_1}^{\dagger} T^a\right]\rrangle\notag\\
= &\frac{1}{2}\frac{1}{N_c}\llangle \mathrm{tr}\left[V_{\un{x}_0}\partial^iV_{\un{x}_0}^{\dagger}V_{\un{x}_2}^{\textrm{G}[3]} V_{\un{x}_1}^{\dagger}\right]\rrangle \, S_{21}(zs)+ \frac{1}{2}\frac{1}{N_c}\left\langle \mathrm{tr}\left[V_{\un{x}_0}\partial^iV_{\un{x}_0}^{\dagger} V_{\un{x}_2}V_{\un{x}_1}^{\dagger}\right]\right\rangle \, G^{[3]}_{21}(zs) + \mbox{c.c.} . \label{eq:new_structure_4}
\end{align}
\end{subequations}
In arriving at Eqs.~\eqref{largeNc_identities} we have assumed that the eikonal correlator $\left\langle \mathrm{tr}\left[V_{\un{x}_0}\partial^iV_{\un{x}_0}^{\dagger} V_{\un{x}_2}V_{\un{x}_1}^{\dagger}\right]\right\rangle$ is real.

The Wilson line structures in the last three lines of \eq{eq:final_result_WW} can be recast at large $N_c$ as
\begin{subequations}
\begin{align}
&\frac{1}{N^2_c}\llangle\mathrm{Tr}\left[U^{\textrm{G}[1]}_{\un{x}_2} U^{\dagger}_{\un{x}_0}\right]\rrangle 
=4\, G_{20}(zs)\, S_{02}(zs), \qquad\frac{1}{N_c^2}\llangle \mathrm{Tr}\left[U^{\textrm{G}[1]}_{\un{x}_2} U^{\dagger}_{\un{x}_1}\right] \rrangle
=4\, G_{21}( zs)\, S_{12}( zs), \label{eq:KL_structure_1}\\
&\frac{1}{N_c^2}\llangle\mathrm{Tr}\left[U^{i \, \textrm{G}[2]}_{\un{x}_2}U^{\dagger}_{\un{x}_0}\right] \rrangle
=2\, G^i_{20}(zs)\, S_{02}(zs), \qquad \frac{1}{N_c^2}\llangle \mathrm{Tr}\left[U^{i \, \textrm{G}[2]}_{\un{x}_2}U^{\dagger}_{\un{x}_1}\right] \rrangle= 
2\, G_{21}^i (zs)\, S_{12}(zs), \label{eq:KL_structure_2}\\
&\frac{1}{N^2_c}\llangle\mathrm{Tr}\left[U^{\textrm{G}[3]}_{\un{x}_2} U^{\dagger}_{\un{x}_0}\right]\rrangle 
=2\, G^{[3]}_{20}(zs)\, S_{02}(zs), \qquad\frac{1}{N_c^2}\llangle \mathrm{Tr}\left[U^{\textrm{G}[3]}_{\un{x}_2} U^{\dagger}_{\un{x}_1}\right] \rrangle
=2\, G^{[3]}_{21}( zs)\, S_{12}( zs).\label{eq:KL_structure_3}
\end{align} 
\end{subequations}

From Eqs.~\eqref{eq:new_structure_1}, ~\eqref{eq:new_structure_2}, ~\eqref{eq:new_structure_3} and ~\eqref{eq:new_structure_4}, one can see that apart from the polarized dipole amplitudes  $G^i_{20}(zs)$ and $G_{20}(zs)$, whose evolutions equations have been derived and studied in \cite{Kovchegov:2015pbl, Kovchegov:2017lsr, Kovchegov:2018znm, Cougoulic:2022gbk}, the evolution \eqref{eq:final_result_WW} for
$G^{WW}_{10} (zs)$ appears to contain new polarized Wilson line correlators that involve three different transverse coordinates, 
\begin{align}
&\frac{1}{N_c}\llangle \mathrm{tr}\left[V^{j \,\textrm{G}[2]}_{\un{x}_1}V^{\dagger}_{\un{x}_1} V_{\un{x}_0}V_{\un{x}_2}^{\dagger}\right]\rrangle , \qquad \frac{1}{N_c}\llangle \mathrm{tr}\left[V_{\un{x}_0}\partial^iV_{\un{x}_0}^{\dagger}V_{\un{x}_2}^{\textrm{G}[1]} V_{\un{x}_1}^{\dagger}\right]\rrangle ,\\
&\frac{1}{N_c}\llangle \mathrm{tr}\left[V_{\un{x}_0}\partial^iV_{\un{x}_0}^{\dagger}V_{\un{x}_2}^{j \, \textrm{G}[2]} V_{\un{x}_1}^{\dagger}\right]\rrangle, \quad \frac{1}{N_c}\llangle \mathrm{tr}\left[V_{\un{x}_0}\partial^iV_{\un{x}_0}^{\dagger}V_{\un{x}_2}^{\textrm{G}[3]} V_{\un{x}_1}^{\dagger}\right]\rrangle.
\end{align}
These are not the polarized dipole amplitudes and not sub-eikonal quadrupoles either. Due to these new objects, the small-$x$ evolution equation in Eq.~\eqref{eq:final_result_WW} is not closed in the large-$N_c$ limit at the DLA+SLA level of accuracy. This is very similar to the observation for small-$x$ evolution of unpolarized WW gluon distribution \cite{Dominguez:2011gc, Dominguez:2011br}. However, as we will show in the following, in the DLA, these new objects drop out of the evolution equation and the final evolution equation does close in the large-$N_c$ limit in DLA.

We now analyze whether and how each term in Eq.~\eqref{eq:final_result_WW} contributes in the DLA. There are three kinematic regions of interest: $\un{x}_2\rightarrow \un{x}_0$, $\un{x}_2\rightarrow \un{x}_1$ and ${x}_{21}\sim {x}_{20} \gg {x}_{10}$. The first two kinematic regions may generate UV-divergent logarithmic terms while the third kinematic region is relevant to the IR-divergent logarithmic terms. In the DLA, we also approximate the eikonal Wilson line correlator $S_{20}(zs) \approx S_{21}(zs) \approx 1$, which reflects the linearized (dilute) regime and the fact that the unpolarized evolution \cite{Balitsky:1995ub,Balitsky:1998ya,Kovchegov:1999yj,Kovchegov:1999ua,Jalilian-Marian:1997dw,Jalilian-Marian:1997gr,Weigert:2000gi,Iancu:2001ad,Iancu:2000hn,Ferreiro:2001qy} is single-logarithmic. We will further integrate over the impact parameters on both sides of the evolution equation while employing \eq{eq:integrate_impact_parameter}.

When carrying out the integral over $\un x_2$ in the second line of \eq{eq:final_result_WW},
one finds that there is no IR-divergent logarithmic term. However, there are UV-divergent logarithmic terms when $\un{x}_{2}\rightarrow \un{x}_0$ or $\un{x}_2\rightarrow \un{x}_1$. Therefore, under DLA, the dominant contribution coming from the virtual diagrams of \fig{fig:helicity_WW_diagrams_1} is 
\begin{align}\label{eq:UV_log_cancel_1}
-\frac{\alpha_s N_c}{2 \pi }\int\limits_{1/sx_{10}^2}^{z}\frac{dz'}{z'} \, \left[ \int\limits^{x_{10}^2}_{1/(z' s)} \frac{d x_{21}^2}{x_{21}^2} + \int\limits^{x_{10}^2}_{1/(z' s)} \frac{d x_{20}^2}{x_{20}^2} \right] \, G^{WW} (x_{10}^2, z's)
\to -\frac{\alpha_s N_c}{\pi }\int\limits_{1/sx_{10}^2}^{z}\frac{dz'}{z'} \int\limits_{1/z's}^{x_{10}^2} \frac{dx_{21}^2}{x_{21}^2} \, \Gamma^{WW} (x_{10}^2, x_{21}^2, z's).
\end{align}
According to \cite{Kovchegov:2015pbl, Kovchegov:2016zex, Kovchegov:2018znm, Cougoulic:2022gbk}, the Wilson line correlator in the virtual diagrams should be treated as the neighbor correlator $\Gamma^{WW} (x_{10}^2, x_{21}^2, z's)$: this is why we performed this replacement in \eq{eq:UV_log_cancel_1}. This is the neighbor-dipole counterpart of the object defined in \eq{GWW_def0}, integrated over all impact parameters per \eq{eq:integrate_impact_parameter}. The lifetimes of the gluons that generate the subsequent evolution of $\Gamma^{WW} (x_{10}^2, x_{21}^2, z's)$ are cut off by $z' \, x_{21}^2$ from above, that is, they depend on the transverse size $x_{21}$ of another object.

For the transverse coordinate integral in the third line of \eq{eq:final_result_WW}, using the kernel function $\mathcal{K}_0^i(\un{x}_{20}, \un{x}_{21})$ one can see that there is no UV-divergent logarithmic contribution when $\un{x}_2\rightarrow \un{x}_1$. There is no IR-divergent logarithmic contribution either. However, there is a UV-divergent logarithmic contribution coming from the kinematic region where $\un{x}_2\rightarrow \un{x}_0$.  Using Eq.~\eqref{eq:new_structure_1} we see that in the $\un{x}_2\rightarrow \un{x}_0$ limit, the corresponding Wilson line structure reduces to 
\begin{align}\label{DE2}
&\frac{1}{2N_c} \llangle \mathrm{tr}\left[V^{j \,\textrm{G}[2]}_{\un{x}_1}V^{\dagger}_{\un{x}_1} V_{\un{x}_0}V_{\un{x}_2}^{\dagger}\right]\rrangle\, S_{02} (z's) + \mbox{c.c.} \Bigg|_{\un x_2 \to \un x_0}
= x_{02}^l \frac{1}{2N_c}\llangle \mathrm{tr}\left[V^{j \,\textrm{G}[2]}_{x_1}V^{\dagger}_{x_1} V_{x_0}\partial^lV_{x_0}^{\dagger}\right] + \mbox{c.c.}\rrangle + \mathcal{O}(x_{20}^2) .
\end{align}
We have expanded the eikonal Wilson line at $\un{x}_2$ around $\un{x}_0$ and kept the terms linear in $\un x_{02}$. Note that the eikonal dipole $S$-matrix $S_{02} (z's)$ reduces to $1$ when $x_{02}\rightarrow 0$. 

Analyzing the integral kernel in the third line of \eq{eq:final_result_WW} multiplied by $x_{02}^l$ from \eq{DE2} while concentrating on the leading divergence in the $x_{02}\rightarrow 0$ limit, we obtain, after averaging over the angles of $\un x_{20}$,
\begin{equation}\label{DE3}
\epsilon^{ji}\frac{x_{10}^2}{x_{20}^2x_{21}^2} \left(-\frac{x_{10}^i}{x_{10}^2} + \frac{x_{20}^i}{x_{20}^2}\right)x_{02}^l \approx - \frac{1}{x_{20}^2} \, \epsilon^{jl}.
\end{equation}
Combining equations \eqref{DE2} and \eqref{DE3}, utilizing the definition \eqref{eq:WW_helicity_operator}, and integrating over the impact parameter, yields the contribution of the diagrams $D$ and $E$ from \fig{fig:helicity_WW_diagrams_2} in the DLA,
\begin{equation}\label{eq:UV_log_cancel_2}
\frac{\alpha_s N_c}{2\pi } \int\limits^z_{1/sx_{10}^2} \frac{dz'}{z'} \int\limits^{x_{10}^2}_{1/z's} \frac{dx_{20}^2}{x_{20}^2} \, \Gamma_W(x_{10}^2, x_{20}^2, z's).
\end{equation}
In these real-emission diagrams $D$ and $E$ one should also use the neighbor correlator for the WW gluon helicity operator. It is interesting to note that in this kinematic region, the Wilson line correlator involving three different transverse coordinates reduces to the WW gluon helicity operator.

Now we analyze the terms in lines 4 and 6 of \eq{eq:final_result_WW}. 
By studying the kernel functions 
\begin{subequations}
\begin{align}
&\mathcal{K}_1^i(\un{x}_{20}, \un{x}_{21}) - \mathcal{K}_1^i(\un{x}_{21}, \un{x}_{21})=\frac{x_{20}^i}{x_{20}^2} + \frac{2(\un{x}_{20}\times \un{x}_{21}) \epsilon^{ji}x_{21}^j}{x_{20}^2x_{21}^2}-\frac{x_{21}^i}{x_{21}^2}, \label{K1-K1} \\
&\mathcal{K}_2^i(\un{x}_{20}, \un{x}_{21}) - \mathcal{K}_2^i(\un{x}_{21}, \un{x}_{21})= - \frac{\epsilon^{ji}x_{20}^j}{x_{20}^2} + \frac{2(\un{x}_{20}\cdot \un{x}_{21}) \epsilon^{ji}x_{21}^j}{x_{20}^2x_{21}^2} - \frac{\epsilon^{ji}x_{21}^j}{x_{21}^2},
\end{align}
\end{subequations}
one can see that the corresponding transverse coordinate integrals do not contain UV-divergent logarithmic contributions in the $\un{x}_2 \rightarrow \un{x}_0$ or $\un{x}_2\rightarrow \un{x}_1$ regions. There is no IR logarithmic divergence either. (One can show that the $- x_{10}^i/x_{21}^2$ term appearing in \eq{K1-K1} in the $x_{21} \gg x_{10}$ limit results in the logarithmic integral over $x_{21}$ acted upon by ${\un x}_{10} \cdot {\un \pd}_0$, with the latter operator removing one logarithm: hence, this term is not DLA.)
We conclude that lines 4 and 6 of \eq{eq:final_result_WW} do not contribute in the DLA.

Next we consider line 5 of \eq{eq:final_result_WW}. In that term, the kernel function vanishes when one sets $\un{x}_{20} = \un{x}_{21}$, so that the transverse coordinate integral does not have an IR-divergent logarithmic contribution in the kinematic region where $x_{20} \approx x_{21} \gg x_{10}$.  In addition, there is no UV logarithmic divergence when $\un{x}_2\rightarrow \un{x}_0$. However, there exists a UV logarithmic divergence when $\un{x}_2\rightarrow \un{x}_1$. In such limit, the kernel in line 5 of \eq{eq:final_result_WW} becomes
\begin{equation}\label{GH2}
\left[\mathcal{K}^{i,j}_2(\un{x}_{20}, \un{x}_{21}) - \mathcal{K}^{i,j}_2(\un{x}_{21}, \un{x}_{21})\right]\Big|_{\un{x}_2\rightarrow \un{x}_1} = - \epsilon^{ji} \frac{2}{x_{21}^2}.
\end{equation}
The corresponding Wilson line structure, simplified in \eq{eq:new_structure_3}, and multiplied by $\epsilon^{ji}$ from \eq{GH2} becomes 
\begin{align}\label{GH3}
&\epsilon^{ji} \left[ \frac{1}{2} \frac{1}{N_c}\llangle \mathrm{tr}\left[V_{\un{x}_0}\partial^iV_{\un{x}_0}^{\dagger}V_{\un{x}_2}^{j \, \textrm{G}[2]} V_{\un{x}_1}^{\dagger}\right]\rrangle \, S_{21}( zs)+ \frac{1}{2}\frac{1}{N_c}\left\langle\mathrm{tr}\left[V_{\un{x}_0}\partial^iV_{\un{x}_0}^{\dagger} V_{\un{x}_2}V_{\un{x}_1}^{\dagger}\right] \right\rangle \, G_{21}^j ( zs)+\mbox{c.c.}\right]_{\un{x}_2 \to \un{x}_1} \\
& \simeq 2 \, G^{WW}_{10} (z's) \to 2 \, \Gamma^{WW}_{10, 21} (z's) . \notag
\end{align}
Here we have, again, employed the definition \eqref{eq:WW_helicity_operator} and transitioned to the neighbor correlation function, as required by the lifetime ordering in the diagrams $G$ and $H$ of \fig{fig:helicity_WW_diagrams_2}.

Combining Eqs.~\eqref{GH2} and \eqref{GH3} and integrating over the impact parameters, we see that 
line 5 of \eq{eq:final_result_WW} (corresponding to the diagrams $G$ and $H$ of \fig{fig:helicity_WW_diagrams_2}), with the complex conjugate added, contributes
\begin{equation}\label{eq:UV_log_cancel_3}
\frac{\alpha_sN_c}{2\pi } \int\limits^z_{1/sx_{10}^2}\frac{dz'}{z'} \int\limits_{1/z's}^{x_{10}^2} \frac{ dx_{21}^2}{x_{21}^2} \, \Gamma_W(x_{10}^2, x_{21}^2, z's)
\end{equation}
in the DLA.

From Eq.~\eqref{eq:UV_log_cancel_1}, Eq.~\eqref{eq:UV_log_cancel_2} and Eq.~\eqref{eq:UV_log_cancel_3}, it is clear these contributions from UV-divergent logarithmic terms cancel. Diagrammatically-speaking, in the DLA, the diagrams $A, B, C, D, E, G, H$ in Fig.~\ref{fig:helicity_WW_diagrams_1} and Fig.~\ref{fig:helicity_WW_diagrams_2} cancel out: 
\begin{align}
    A+B+C+D+E+G+H \bigg|_{DLA} = 0. 
\end{align}
This conclusion is similar to those observed in \cite{Kovchegov:2018znm, Borden:2024bxa} for the similar (but not identical) situations.

We see that only the diagrams $K, L$ in Fig.~\ref{fig:helicity_WW_diagrams_3} may contribute to the evolution of $G^{WW}_{10}$ in the DLA. They contribute lines 7-9 of \eq{eq:final_result_WW}. We continue by analyzing line 7 of \eq{eq:final_result_WW}.  The kernel function is
\begin{equation}
\mathcal{J}_1(\un{x}_{20}, \un{x}_{21})
=-\frac{4(\un{x}_{20}\cdot\un{x}_{21})^2}{x_{20}^4x_{21}^2} + \frac{2}{x_{20}^2} .
\end{equation}
When integrating over the transverse coordinate $\un{x}_2$, this kernel function does not have UV-divergent logarithmic contributions from $\un{x}_{2}\rightarrow \un{x}_{0}$ or $\un{x}_{2}\rightarrow \un{x}_1$.  
There is, however, an IR-divergent logarithmic contribution when $x_{20}\approx x_{21} \gg x_{10}$,
\begin{equation}
\mathcal{J}_1(\un{x}_{20}, \un{x}_{21})\Big|_{x_{20} \approx  x_{21} \gg x_{10}} \approx  -\frac{2}{x_{20}^2}. 
\end{equation} 
Employing \eq{eq:KL_structure_1} and working in the linearized regime ($S \approx 1$), we see that line 7 of \eq{eq:final_result_WW} contributes, after the impact-parameter integration 
\begin{equation}\label{eq:IR_log_1}
\frac{\alpha_sN_c}{\pi} \int\limits_{\Lambda^2/s}^{z} \frac{dz'}{z'} \int\limits^{\mathrm{min}\left\{\frac{z}{z'}x_{10}^2, \frac{1}{\Lambda^2}\right\}}_{\mathrm{max}\left\{x_{10}^2,\frac{1}{z's} \right\}} \frac{d x_{20}^2}{x_{20}^2} \, G(x_{20}^2, z's).
\end{equation}

In line 8 of \eq{eq:final_result_WW} we anticipate the impact parameter integration which will bring in the decomposition \eqref{decomp21}, with only the $G_2$-term contributing \cite{Kovchegov:2018znm, Cougoulic:2022gbk}. Therefore, we multiply the explicit expression for the kernel function 
\begin{equation}
\mathcal{J}_2^i (\un{x}_{20}, \un{x}_{21})
=\frac{8(\un{x}_{20}\times \un{x}_{21})(\un{x}_{20}\cdot\un{x}_{21})}{x_{20}^4x_{21}^2} \left[-\frac{x_{21}^i}{x_{21}^2} + 2\frac{x_{20}^i}{x_{20}^2}\right] - \frac{4(\un{x}_{20}\cdot\un{x}_{21})}{x_{20}^4x_{21}^2} \epsilon^{ij} \left[ x_{20}^j  + x_{21}^j \right] + \frac{4(\un{x}_{20}\times \un{x}_{21})}{x_{20}^4 x_{21}^2}[x_{20}^i - x_{21}^i]
\end{equation}
by $\epsilon^{im} x_{20}^m$, getting
\begin{equation}\label{J2i_1}
\mathcal{J}^i_2(\un{x}_{20}, \un{x}_{21}) \, \epsilon^{im}\, x_{20}^m\\
= \frac{4(\un{x}_{20}\cdot\un{x}_{21})}{x_{20}^2x_{21}^2} -\frac{8(\un{x}_{20}\cdot\un{x}_{21})^3}{x_{20}^4x_{21}^4} - \frac{8(\un{x}_{20}\cdot\un{x}_{21})^2}{x_{20}^4x_{21}^2} + \frac{4}{x_{20}^2}.
\end{equation}
This object is needed to study the contribution of the first trace in the double angle brackets of line 8 of \eq{eq:final_result_WW}. There is no UV-divergent logarithmic contribution when $\un{x}_{2}\rightarrow \un{x}_1$ in the expression \eqref{J2i_1}. The possible UV logarithmic contributions when $\un{x}_{2}\rightarrow \un{x}_0$ cancel out. However, there exists an IR logarithmic divergence: 
\begin{equation}
\mathcal{J}_2^{i}(\un{x}_{20}, \un{x}_{21}) \, \epsilon^{im} x_{20}^m\Big|_{x_{20} \approx x_{21} \gg x_{10}} \approx - \frac{8}{x_{20}^2}. 
\end{equation}

Similarly, one obtains 
\begin{equation}\label{eq:K+L_kernel_contract_x20}
\mathcal{J}_2^i(\un{x}_{20}, \un{x}_{21}) \, \epsilon^{im} x_{21}^m
= \frac{12(\un{x}_{20}\cdot\un{x}_{21})}{x_{20}^4} -\frac{16(\un{x}_{20}\cdot\un{x}_{21})^3}{x_{20}^6x_{21}^2}  - \frac{8(\un{x}_{20}\cdot\un{x}_{21})^2}{x_{20}^4x_{21}^2} + \frac{4}{x_{20}^2}
\end{equation}
needed for the second trace in the double angle brackets of line 8 of \eq{eq:final_result_WW}.
There is no UV-divergent logarithmic contribution when $\un{x}_2\rightarrow \un{x}_1$ in \eq{eq:K+L_kernel_contract_x20}. There is no UV-divergent logarithmic contribution when $\un{x}_2\rightarrow \un{x}_0$ either. The analysis here is more involved. First of all, one notices that UV-divergent logarithmic $x_{20} \to 0$ contributions in the last two terms of Eq.~\eqref{eq:K+L_kernel_contract_x20} cancel each other after averaging over the angles of $\un x_{20}$. We are left to analyze the first two terms in Eq.~\eqref{eq:K+L_kernel_contract_x20}. Using $\un{x}_{21} = \un{x}_{20}-\un{x}_{10}$, we expand around $x_{20} =0$, obtaining
\begin{equation}\label{eq:expandx0-x1}
\frac{12(\un{x}_{20}\cdot\un{x}_{21})}{x_{20}^4} -\frac{16(\un{x}_{20}\cdot\un{x}_{21})^3}{x_{20}^6x_{21}^2}
\approx  -12 \frac{\un{x}_{20}\cdot\un{x}_{10}}{x_{20}^4}+ 16 \frac{(\un{x}_{20}\cdot\un{x}_{10})^3}{x_{20}^6 x_{10}^2} + \mathcal{O} \left( \frac{1}{x_{20}} \right) .
\end{equation}
The integrals over azimuthal angle of $\un x_{20}$ are carried out when appropriate. The order-$1/x_{20}^2$ terms in the kernel cancel after integrating over azimuthal angles of $\un x_{20}$. We are left with the terms that are of order $1/x_{20}^3$, which cancel after angular integration. However, when counting the powers of $x_{20}$ in the limit $\un{x}_2\rightarrow \un{x}_0$, we also need to consider the expansion of the polarized Wilson line correlator
\begin{equation}\label{expansion36}
G_2(x_{21}^2, zs) = G_2(x_{01}^2, zs) + x_{02}^i \partial^i G_2(x_{01}^2, zs) + \ldots .
\end{equation}
The term linear in $\un{x}_{20}$ can combine with the terms of order $1/x_{20}^3$ in \eq{eq:expandx0-x1} to generate potential UV-divergent logarithmic contribution. However, in the $\un{x}_{20} \rightarrow 0$ limit
\begin{equation}\label{anlge0}
\left[-12 \frac{\un{x}_{20}\cdot\un{x}_{10}}{x_{20}^4}+ 16 \frac{(\un{x}_{20}\cdot\un{x}_{10})^3}{x_{20}^6 x_{10}^2}\right] x_{02}^i \partial^i G_2(x_{10}^2, zs)\Longrightarrow  \left[\frac{6}{x_{20}^2} x_{10}^i - 16 \, \frac{3}{8x_{20}^2}x_{10}^i\right]\partial^i G_2(x_{10}^2, zs)=0,
\end{equation}
where the double arrow denotes the averaging over the angles of $\un x_{20}$. Let us also note that the derivative term in \eq{expansion36}, if it was to survive, would have come in with the ${\un x}_{10} \cdot \un \partial_0$ structure (alredy seen in \eq{anlge0}) which removes one logarithm, and is, therefore, would not be a DLA term. 

We conclude that there is no UV-divergent logarithmic contribution when $\un{x}_2\rightarrow \un{x}_0$ in \eq{eq:K+L_kernel_contract_x20}. The IR logarithmic contribution in \eq{eq:K+L_kernel_contract_x20} is 
\begin{equation}
\mathcal{J}_2^n(\un{x}_{20}, \un{x}_{21}) \, \epsilon^{nm} x_{21}^m\Big|_{x_{20} \approx x_{21} \gg x_{10}} \approx   -\frac{8}{x_{20}^2}.
\end{equation}
Using \eq{eq:KL_structure_2} in the linearized regime, we see that, under DLA, line 8 of \eq{eq:final_result_WW} leads to the following expression:
\begin{equation}\label{eq:IR_log_2}
\frac{\alpha_s N_c}{\pi} \int\limits^z_{\Lambda^2/s}\frac{dz'}{z'} \int\limits^{\mathrm{min}\left\{\frac{z}{z'}x_{10}^2, \frac{1}{\Lambda^2}\right\}}_{\mathrm{max}\left\{x_{10}^2,\frac{1}{z's} \right\}} \frac{ dx_{20}^2}{x_{20}^2} 2 \, G_2(x_{20}^2, z's).
\end{equation}

Finally, in line 9 of \eq{eq:final_result_WW}, the kernel function is 
\begin{equation}
\mathcal{J}_2(\un{x}_{20}, \un{x}_{21})  =  \frac{4(\un{x}_{20}\times \un{x}_{21})(\un{x}_{20}\cdot\un{x}_{21})}{x_{20}^4 x_{21}^2} .  
\end{equation}
There is no UV- or IR-divergent logarithmic contributions when one integrates over $\un{x}_2$. Therefore, line 9 of \eq{eq:final_result_WW} does not contribute in the DLA. Recalling a similar conclusion for line 6 of \eq{eq:final_result_WW}, we observe that the polarized Wilson line of type-3, $U^{\textrm{G}[3]}_{\un{x}_2}$, does not contribute to the evolution of the operator \eqref{eq:WW_helicity_operator} in the DLA. 

Combining Eqs.~\eqref{eq:IR_log_1} and \eqref{eq:IR_log_2}, one obtains the final expression for the large-$N_c$ small-$x$ evolution equation (in the DLA and in the linearized regime) for the object \eqref{eq:WW_helicity_operator} related to the WW gluon helicity TMD,  
\begin{equation}\label{eq:WW_evol_DLA}
G^{WW} (x^2_{10}, zs) = G^{WW \, (0)}(x_{10}^2, zs) +\frac{\alpha_s N_c}{\pi} \int\limits^z_{\Lambda^2/s}\frac{dz'}{z'} \int\limits^{\mathrm{min}\left\{\frac{z}{z'}x_{10}^2, \frac{1}{\Lambda^2}\right\}}_{\mathrm{max}\left\{x_{10}^2,\frac{1}{z's} \right\}} \frac{ dx_{20}^2}{x_{20}^2} \Big[G(x_{20}^2, z's) + 2 \, G_2(x_{20}^2, z's)\Big].
\end{equation}
It is exactly the same small-$x$ evolution equation as for $G_2(x_{10}^2, zs)$ \cite{Cougoulic:2022gbk}: the only possible difference is in the inhomogeneous terms/initial conditions. However, since at high enough energy the right-hand side of \eq{eq:WW_evol_DLA} is dominated by the second term, the potential difference in the initial conditions become irrelevant, and we can conclude that  
\begin{align}
    G^{WW} (x^2_{10}, s) \approx G_2 (x^2_{10}, s). 
\end{align}

Note that we can easily generalize \eq{eq:WW_evol_DLA} to the case of the large-$N_c \& N_f$ evolution: in the large-$N_c \& N_f$ limit, we only need to replace $G \to {\widetilde G}$ in \eq{eq:WW_evol_DLA} (see \cite{Cougoulic:2022gbk} for the definition of $\widetilde G$), such that the evolution equation becomes
\begin{align}\label{WWeq2}
G^{WW} (x_{10}^2, zs) = G^{WW \, (0)} (x_{10}^2, zs) + \frac{  \as \, N_c}{\pi} \, \int\limits_\frac{\Lambda^2}{s}^{z} \frac{d z'}{z'} \, \int\limits_{\max \left\{ x_{10}^2, \frac{1}{z' s} \right\}}^{\min \{\frac{z}{z'} x_{10}^2, 1/\Lambda^2 \}} \frac{d x^2_{20}}{x^2_{20}} \, \left[ {\widetilde G} (x^2_{20}, z's) + 2 \, G_2 (x^2_{20}, z's) \right]. 
\end{align}
Comparing the WW gluon helicity TMD from \eq{gww_GW} to the dipole gluojn helicity TMD  
\begin{align}\label{glue_hel_TMD57}
g_{1L}^{G \, dip} (x, k_T^2) = \frac{N_c}{\as 2 \pi^4} \, \int d^2 x_{10} \, e^{- i \un{k} \cdot \un{x}_{10} } \,  G_2 \left(  x_{10}^2,  zs = \frac{Q^2}{x} \right)
\end{align}
(see Eq.~(41) in \cite{Cougoulic:2022gbk}), and employing \eq{eq:integrate_impact_parameter}, 
we conclude that at sufficiently small $x$, at large $N_c$, and outside the saturation region, the dipole and WW gluon helicity TMDs are identical, 
\begin{align}\label{WW=dip}
g_{1L}^{G \, WW} (x, k_T^2)  \approx g_{1L}^{G \, dip} (x, k_T^2). 
\end{align}
Hence, the measurement of the double-spin asymmetry in inclusive dijet production we proposed above can help us find either of those TMDs. We also conclude that the small-$x$ asymptotics of the WW gluon helicity TMD is the same as that for the dipole gluon helicity TMD found in \cite{Borden:2023ugd, Cougoulic:2022gbk}.

The physical origin of \eq{WW=dip} can be easily understood by comparing the operator definitions of the two gluon helicity TMDs given in Eqs.~\eqref{eq:Gammaij} and \eqref{eq:Gamma_dip}. At the leading order in the gluon fields one can set the Wilson-line gauge links to unity, making the correlators in those two equations equal. This feature is in direct analogy to the unpolarized case \cite{Dominguez:2010xd, Dominguez:2011wm}. Let us stress though that this conclusion is for linear evolution only, without saturation effects. The unpolarized dipole and WW gluon TMDs evolve with the same BFKL evolution in the linear regime. The difference between them arises due to saturation effects. We expect the dipole and WW gluon helcity TMDs to also be different inside the saturation region.

\section{Conclusions and Outlook}
\label{sec:conclusions}

In this paper, we have studied the WW gluon helicity distribution at small $x$. We first derived the operator expression for the small-$x$ limit of WW gluon helicity distribution in terms of polarized Wilson line correlators in \eq{eq:WW_G_final}. It is to be compared with the operator expression for the dipole gluon helicity distribution in \eq{eq:dipole_G_final}. We further derived the cross section for double-spin asymmetry of inclusive quark-antiquark dijet production in the longitudinally polarized electron-proton collisions at high energies (averaged over the angles of the electron's transverse momentum). We showed that in the back-to-back (correlation) limit, this cross section uniquely probes the WW gluon helicity distribution, as exhibited in \eq{DSA106}. Using the LCOT method, we also constructed the small-$x$ helicity evolution equation \eqref{eq:final_result_WW} for the correlator $G^{WW}$ related to the WW gluon helicity distribution. The evolution equation was simplified to \eq{eq:WW_evol_DLA} in the linearized regime, at large $N_c$, and under the DLA: \eq{eq:WW_evol_DLA} was found to be the same as the evolution equation for the dipole amplitude $G_2$ related to the dipole gluon helicity distribution. Since the relation between the WW gluon helicity TMD and $G^{WW}$, \eq{gww_GW}, is equivalent to the relation between the gluon dipole helicity TMD and $G_2$, \eq{glue_hel_TMD57}, we concluded that the dipole and WW gluon helicity TMDs are identical outside the saturation region and at DLA. We conclude that the double-spin asymmetry for inclusive dijet production, particularly in the back-to-back limit, directly measures the WW gluon TMD, which is related to the gluon spin content of proton at small $x$. Practically speaking, it can help constrain the initial conditions of the small-$x$ helicity evolution equations in the phenomenological applications like those conducted in \cite{Adamiak:2021ppq,  Adamiak:2023yhz, Adamiak:2025dpw}.

In the nonlinear (dense) regime, when gluon saturation effects are important, the small-$x$ evolution equations for the WW gluon helicity distribution and the dipole gluon helicity distribution are very different. However, strictly speaking, saturation effects enter evolution equations at the single-logarithmic level. Identification of all single logarithmic contributions is still an active area of study \cite{Kovchegov:2021lvz}, even for the dipole gluon helicity distribution outside the saturation region. The single logarithmic contributions are even less explored in the WW gluon helicity TMD. Therefore, it appears difficult at this point to discuss the differences between the dipole and WW gluon helicity TMDs in the saturation region. 

In the future, one would like to construct the complete helicity-dependent JIMWLK evolution equation (see \cite{Cougoulic:2019aja} for the first steps in this direction). The evolution equation for the WW gluon helicity distribution derived in this paper, \eq{eq:WW_evol_DLA}, is expected to then be obtained as a limit of the helicity JIMWLK evolution applied to the color-quadrupoles containing one polarized Wilson line. 

In this work we have only isolated the leading-order expression in the back-to-back limit for the double-spin asymmetry of inclusive dijet production cross section. For phenomenological studies, it will also be useful to obtain the next-to-leading order (linear in $\Delta_\perp$) terms in the back-to-back limit. In addition to the WW gluon helicity distribution, these terms are expected to probe various other quark and gluon helicity distributions, like the twist-3 gluon helicity-flip TMD  $h^{\perp\, WW}_{3L}(x, \Delta^2_{\perp})$ discussed in Appendix~\ref{sec:twist-3TMD} 
(see the recent similar study \cite{Altinoluk:2024zom} for unpolarized collisions). Note that, as we mentioned above, the elastic dijet production in polarized $e+p$ collisions was recently shown to probe the orbital angular momentum distributions in \cite{Hatta:2016aoc,Bhattacharya:2022vvo, Bhattacharya:2023hbq, Bhattacharya:2024sck, Kovchegov:2024wjs}: the relevant terms in the dijet production cross section are linear in $\Delta_\perp$.  For inclusive dijet production in the back-to-back limit, large double logarithms of type $\alpha_s \ln^2(p_T/\Delta_{\perp})$ need to be resummed \cite{Mueller:2012uf, Mueller:2013wwa, Taels:2022tza, Caucal:2022ulg}. These Sudakov double logarithms in longitudinal double-spin asymmetry for dijet production will be left for future investigations. 

Longitudinal double-spin asymmetry for inclusive dijet production in polarized proton-proton collisions was measured at RHIC \cite{STAR:2016kpm, STAR:2019yqm, STAR:2021mfd, STAR:2021mqa, RHICSPIN:2023zxx}. The calculational techniques developed in the present paper for DIS along with those developed by the authors in \cite{Kovchegov:2024aus} can be applied to study the double-spin asymmetry of inclusive dijet production in polarized proton-proton collisions where data are available. The data from RHIC for the dijet double-spin asymmetry can help put more constraints on the global analysis of proton's spin structure at small $x$ initiated in \cite{Adamiak:2021ppq,  Adamiak:2023yhz, Adamiak:2025dpw}, ultimately resulting in more precise theoretical predictions for the EIC.

While we cannot comment on the feasibility of measuring $g_{1L}^{G \, WW} (x, k_T^2)$ at the EIC using the process we have considered here, inclusive dijet production in the longitudinally polarized $e+p$ collisions, without performing a dedicated study, let us note again that such a study for the same inclusive dijet production process, albeit in the collinear factorization framework with the goal of extracting the gluon helicity PDF $\Delta G (x, Q^2)$, was carried out in \cite{Page:2019gbf} (see also Chapter~7 in \cite{AbdulKhalek:2021gbh}).  The study considers $x$ values down to $5 \times 10^{-3}$ for $10\, \mathrm{GeV}^2 < Q^2 < 100\, \mathrm{GeV}^2$, thus including a part of the small-$x$ region. The study notes that a challenge for the measurement is in correctly differentiating the quark-antiquark dijet, which originates from the gluon-initiated process, from the background quark-gluon dijet, which originates from the quark-initiated process (see Fig.~1 in \cite{Page:2019gbf}). The measurement also will require a substantial integrated luminosity. The overall conclusion in \cite{Page:2019gbf} is that $\Delta G (x, Q^2)$ extraction from the inclusive dijet production at EIC will be possible, and will be complementary to the $\Delta G (x, Q^2)$ extraction from the proton $g_1$ structure function measurements. We anticipate that $g_{1L}^{G \, WW} (x, k_T^2)$ extraction for the inclusive dijet production discussed in this work will face similar challenges: the quark-gluon jet background and the need to achieve high integrated luminosity. It remains to be seen whether the measurement will be possible at the EIC.


\section*{Acknowledgments}

The authors are grateful to Elke Aschenauer, Alexei Prokudin, Vladimir Skokov, Anselm Vossen, and Feng Yuan for informative discussions. 

This material is based upon work supported by
the U.S. Department of Energy, Office of Science, Office of Nuclear
Physics under Award Number DE-SC0004286 and within the framework of the Saturated Glue (SURGE) Topical Theory Collaboration. \\

\appendix

\section{Small-$x$ limit of twist-3 WW gluon helicity-flip TMD}
 \label{sec:twist-3TMD}

In this Appendix, we derive the small $x$ limit of the twist-3 WW gluon helicity-flip TMD. This analysis is parallel to that for the twist-3 \textit{dipole} gluon helicity-flip TMD performed in \cite{Kovchegov:2024aus}. We begin by considering the following gluon-field correlator
\begin{equation}\label{Gamma_L_T3}
\Gamma_{WW}^{ij; l+} (k; P, S_L) = \int \frac{d^4\xi}{(2\pi)^4} \, e^{ik\cdot\xi} \, \langle P, S_L| \tr \left[F^{ij}(0) \, \mathcal{U}^{[+]} [0, \xi] \, F^{l+}(\xi) \, \mathcal{U}^{[+]} [\xi, 0] \right]  |P, S_L\rangle 
\end{equation}
with both gauge links being future-pointing staples. For the longitudinally polarized proton state with helicity $S_L$ at hand, according to \cite{Mulders:2000sh}, one has the decomposition
\begin{equation}\label{MGamma_ijl}
    M_p \, \Gamma_{WW}^{ij, l}(x, \un{k}) \equiv \int dk^- \, \Gamma_{WW}^{ij; l+}(k; P, S_L) \equiv - \frac{i}{4} S_L \, \epsilon^{ij} \, k^l \, h_{3L}^{\perp\, WW}(x, k_T^2)\end{equation}
with $k^+ = x P^+$ and  $M_p$ being the mass of the proton, as in the main text. Equation \eqref{MGamma_ijl} defines $h_{3L}^{\perp\, WW}(x, k_T^2)$, the twist-3 WW gluon helicity-flip TMD. 

Let us denote
\begin{equation}
\widetilde{\Gamma}_{WW}^{l}(x, \un{k}) \equiv 2i\epsilon^{ij} \sum_{S_L} \frac{1}{2} \, S_L \, M_p \, \Gamma^{ij, l}_{WW}(x, \un{k}) = k^l \,  h_{3 L}^{\perp\, WW}(x, k_T^2).
\end{equation}
This new object can be simplified at small $x$ as follows: 
\begin{equation}\label{Gamma_tilde_WW}
\begin{split}
\widetilde{\Gamma}_{WW}^{l}(x, \un{k}) = &\frac{2i}{V^-(2\pi)^3}\epsilon^{ij} \sum_{S_L} \frac{1}{2} S_L \int d\xi^- d^2 \xi  \, d \zeta^- d^2 \zeta \,  e^{ixP^+(\xi^--\zeta^-)}e^{ -i \un{k} \cdot( \un{\xi}- \un{\zeta})}\\
&\qquad \times \frac{1}{2} \Bigg[ \left\langle P, S_L\left| \mathrm{tr}\left[F^{ij}(\zeta) \mathcal{U}^{[+]} [\zeta, \xi] F^{l+}(\xi) \mathcal{U}^{[+]} [\xi, \zeta] \right]\right |P, S_L\right\rangle_{\xi^+=\zeta^+=0}\\
&\qquad \qquad - \left\langle P, S_L\left| \mathrm{tr}\left[F^{l+}(\zeta) \mathcal{U}^{[+]} [\zeta,\xi] F^{ij}(\xi) \mathcal{U}^{[+]} [ \xi, \zeta] \right]\right |P, S_L\right\rangle_{\xi^+=\zeta^+=0} \Bigg] \\
=&\frac{2}{V^-(2\pi)^3} \sum_{S_L} \frac{1}{2} S_L \int d^2 \xi  d^2 \zeta e^{ -i \un{k} \cdot( \un{\xi}- \un{\zeta})}\,  \langle P, S_L| \mathrm{tr}\left[B_z(x, \un{\zeta}) E^l(x, \un{\xi}) + \left[E^l(x,\un{\zeta})\right]^{\dagger}\left[B_z(x, \un{\xi})\right]^{\dagger}\right]|P, S_L\rangle\\
=&\frac{4}{(2\pi)^3}  \frac{1}{ig^2} \int d^2 \xi\,  d^2 \zeta\, e^{ -i \un{k} \cdot( \un{\xi}- \un{\zeta})} \, \llangle \mathrm{tr}\left[ V_{\un{\zeta}}^{\textrm{G}[1]}V_{\un{\zeta}}^{\dagger} V_{\un{\xi}}\partial^l V^{\dagger}_{\un{\xi}}\right] - \mathrm{tr}\left[V_{\un{\zeta}}\partial^lV^{\dagger}_{\un{\zeta}} V_{\un{\xi}}^{\textrm{G}[1]}V_{\un{\xi}}^{\dagger}\right]\rrangle\\
=&\frac{4i}{(2\pi)^3}  \frac{1}{g^2} \int d^2 \xi  \, d^2 \zeta\, e^{ -i \un{k} \cdot( \un{\xi}- \un{\zeta})} \, \llangle \mathrm{tr}\left[ V_{\un{\xi}}^{\textrm{G}[1]}V_{\un{\xi}}^{\dagger} V_{\un{\zeta}}\partial^l V^{\dagger}_{\un{\zeta}}\right] \rrangle + \mbox{c.c.} . \\
\end{split}
\end{equation}
Here $E^l(x, \un{\xi})$ is defined in \eq{eq:Ej_full}: we have employed its expansion in the powers of $x$ given in \eq{Exp} above. We have also defined the longitudinal chromo-magnetic field, with its expansion in $x$,
\begin{equation}
B_z(x, \un{\zeta}) \equiv \frac{i}{2} \, \epsilon^{ij} \int\limits_{-\infty}^\infty d\zeta^- e^{-ixP^+ \zeta^-} V_{\un{\zeta}}[\infty, \zeta^-] F^{ij}(\zeta^-, \un{\zeta}) V_{\un{\zeta}}[\zeta^-, \infty] = \frac{s}{g P^+} V_{\un{\zeta}}^{\textrm{G}[1]} V_{\un{\zeta}}^{\dagger} + \mathcal{O}(x^1)
\end{equation}
and with $V_{\un{\zeta}}^{\textrm{G}[1]}$  defined in \eq{VG1} above. It is also easy to verify that 
\begin{equation}
\left(V_{\un{\zeta}}^{\textrm{G}[1]}V_{\un{\zeta}}^{\dagger}\right)^{\dagger} = V_{\un{\zeta}}V_{\un{\zeta}}^{\textrm{G}[1]\dagger} =  - V_{\un{\zeta}}^{\textrm{G}[1]}V_{\un{\zeta}}^{\dagger},
\end{equation}
which clarifies the last step in \eq{Gamma_tilde_WW}.
 To conclude, the small-$x$ limit of twist-3 WW gluon helicity-flip TMD is 
\begin{equation}
h^{\perp\, WW}_{3L}(x, k_T^2) = \frac{i}{\as \, 8 \pi^4} \, \frac{k^l}{k_T^2} \int d^2\xi  d^2\zeta  e^{-i\un{k}\cdot(\un{\xi}-\un{\zeta})} \llangle \mathrm{tr}\left[ V_{\un{\xi}}^{\textrm{G}[1]}V_{\un{\xi}}^{\dagger} V_{\un{\zeta}}\partial^l V^{\dagger}_{\un{\zeta}}\right] \rrangle + \cc , \end{equation}
with the right-hand side proportional to \eq{eq:relate_twist_3} in the main text, as desired.

\section{An operator derivation of the small-$x$ evolution equation for the unpolarized WW gluon distribution}
\label{appendix:reproduce_unpolWW_evol}

In this Appendix, we reproduce the small-$x$ evolution equation for the WW gluon distribution that was derived in  \cite{Dominguez:2011gc} using the LCOT method. This exercise serves as baseline and a cross check of the operator method we use.  In \cite{Dominguez:2011gc}, the small-$x$ evolution equation for WW gluon distribution is obtained from the corresponding evolution equation for color quadrupole by further taking transverse derivatives of the quadrupole equation and then collapsing the four transverse coordinates to two transverse coordinates. Our operator method is implemented by directly calculating the relevant Feynman diagrams contributing to one step of small-$x$ evolution. 

\begin{figure}[h]
    \centering
    \includegraphics[width=0.35\textwidth]{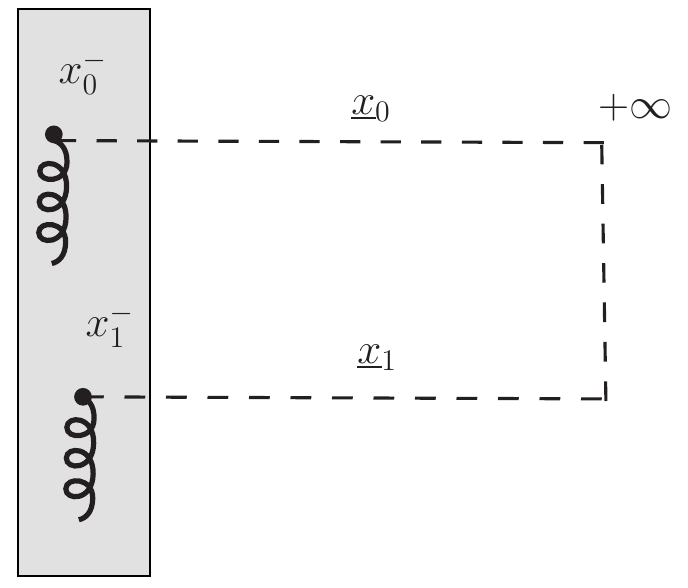}
    \caption{Diagrammatic representation of the operator $\mathcal{O}^{ij}_{01}$ related to the WW gluon distribution using semi-infinite Wilson lines in the adjoint representation, denoted by the dashed lines. The thick gluon lines inside the shockwave represent the gluon field $F^{i+} = \partial^i A^+$ at the eikonal order. }
	\label{fig:WW_Operator}
\end{figure}

The (unpolarized) WW gluon TMD can be extracted from the following Wilson line structure 
\begin{equation}\label{eq:Oij_fundamental}
\mathcal{O}^{ij}_{01} = \frac{1}{N_c} \Big\langle \mathrm{tr}\left[V_{\un{x}_0}\partial^i V_{\un{x}_0}^{\dagger} V_{\un{x}_1} \partial^j V_{\un{x}_1}^{\dagger} \right]\Big\rangle .
\end{equation}
Note that we work in the light-cone gauge of the projectile, $A^-=0$. Equation~\eqref{eq:Oij_fundamental} can be rewritten in terms of semi-infinite Wilson lines in the adjoint representation as
\begin{equation}\label{eq:Oij_adjoint}
\mathcal{O}^{ij}_{01} = - \frac{g^2}{2N_c}  \int\limits_{-\infty}^{\infty} dx_0^- \, dx_1^- \, \left\langle U^{da}_{\un{x}_0}[\infty, x_0^-]U^{dc}_{\un{x}_1}[\infty, x_1^-]  \partial^i A^{a +} (x_0^-, \un{x}_0) \partial^j A^{c +} (x_1^-, \un{x}_1) \right\rangle .
\end{equation}
This object is diagrammatically represented in Fig.~\ref{fig:WW_Operator}. Eq.~\eqref{eq:Oij_adjoint} and Fig.~\ref{fig:WW_Operator} are the starting point for deriving the small-$x$ evolution equation for $\mathcal{O}^{ij}_{01}$. 

The diagrams contributing to the small-$x$ evolution of $\mathcal{O}^{ij}_{01}$ are shown in Figs.~\ref{fig:WW_diagrams_1}, ~\ref{fig:WW_diagrams_2}, and \ref{fig:WW_diagrams_3}. They are organized according to the number of how many thick gluon lines (insertions of the eikonal background gluon field $F^{i+} = \partial^i A^+$) are inside the shockwave. For example, in Fig.~\ref{fig:WW_diagrams_1} both background $F^{i+}$ fields are inside the shockwave. These three diagrams are the virtual diagrams in which gluon radiation, propagation, and absorption happen outside of the shockwave. In Fig.~\ref{fig:WW_diagrams_2}, one of the two background fields ($F^{j+}$) is inside the shockwave, while the other ($F^{i+}$) field becomes a quantum fluctuating field. In Fig.~\ref{fig:WW_diagrams_3}, both $F^{i+}$ and $F^{j+}$ fields become quantum fluctuating fields, either before or after the shockwave.


\subsection{Diagrams $A', B', C'$}

\begin{figure}[h]
    \centering
    \includegraphics[width=0.8\textwidth]{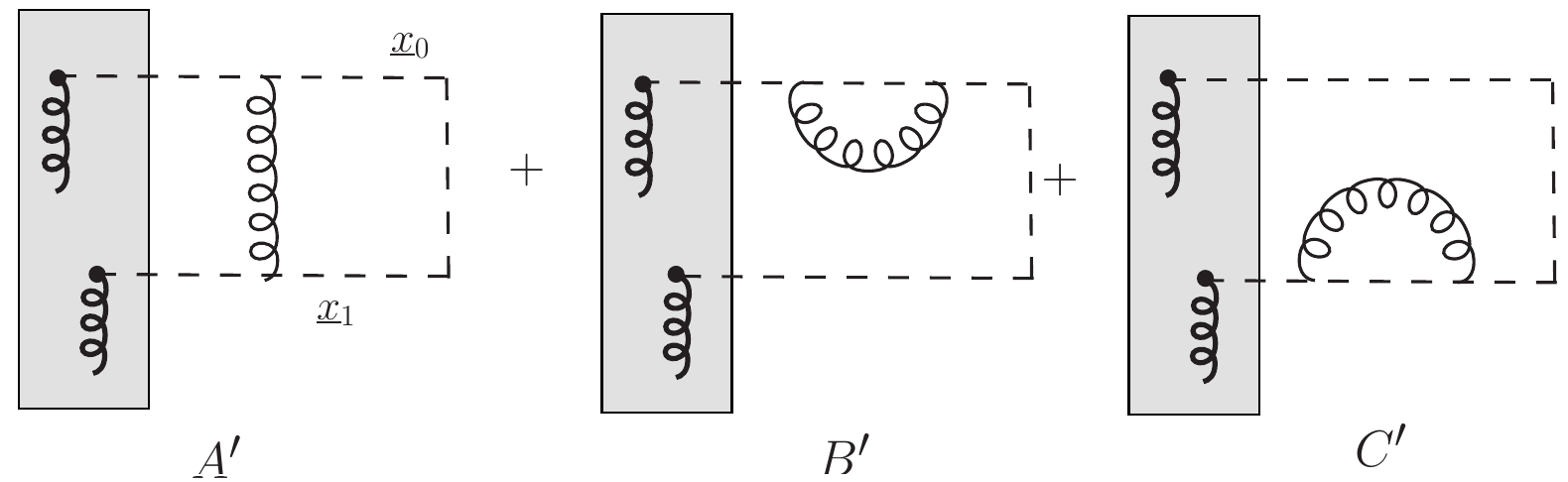}
    \caption{Diagrams for one step of evolution for $\mathcal{O}^{ij}_{01}$ resulting from virtual gluon radiation. In this subset of diagrams, the two background gluon fields stay inside the shockwave.}
	\label{fig:WW_diagrams_1}
\end{figure}

The three diagrams in Fig.~\ref{fig:WW_diagrams_1} are similar to the virtual diagrams in the derivation of the BK/JIMWLK equation.  We just quote the result here:
\begin{equation}\label{eq:final_A+B+C}
A'+ B'+ C' = - \frac{\alpha_sN_c}{2\pi^2} \int\limits^z_{\Lambda^2/s} \frac{dz'}{z'} \int d^2x_2 \, \frac{x_{10}^2}{x_{20}^2x_{21}^2} \, O^{ij}_{01} (z') .
\end{equation}


\subsection{Diagrams $D', E', F', G', H', I'$}

\begin{figure}[h]
    \centering
    \includegraphics[width=0.9\textwidth]{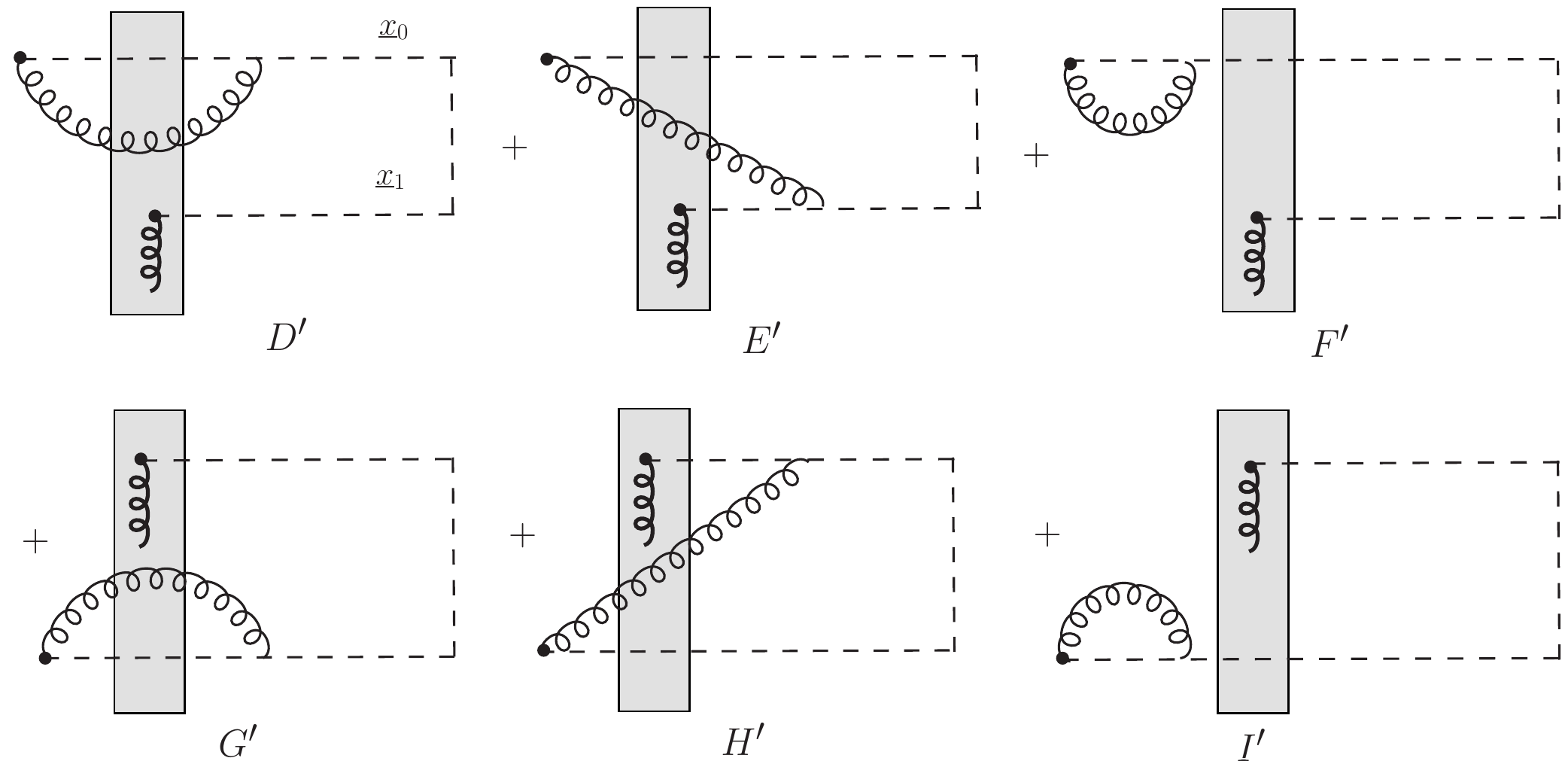}
    \caption{Diagrams for one step evolution of $\mathcal{O}^{ij}_{01}$. In this subset of diagrams, one of the two background gluon field strength is located inside the shockwave, while the other one is located outside the shockwave.}
	\label{fig:WW_diagrams_2}
\end{figure}

For diagram $E'$, the background field at $\un{x}_0$ becomes a quantum field and it is located to the left (before) the shockwave. The other quantum field comes from the eikonal Wilson line at transverse position $\un{x}_1$ and is located to the right (after) the shockwave (the shockwave is understood to be at $x^- =0$). The contribution of the diagram $E'$ is calculated using the LCOT method, employing the background-field gluon propagator in \eq{eik}. This yields   
\begin{equation}
E'= \frac{-g^2}{2N_c} \, \frac{-i g}{4\pi^3} \int\limits^z_{\Lambda^2/s} \frac{dz'}{z'} \int d^2x_2 \, \frac{2x_{20}^ix_{20}^l-\delta^{il}x_{20}^2}{x_{20}^4} \, \frac{-x_{21}^l}{x_{21}^2} \, \int_{-\infty}^{\infty} dx_1^- \left\langle U^{hc}_{\un{x}_1}[\infty, x_1^-]\partial^j A^{c +} (x_1^-,\un{x}_1)  \, \mathrm{Tr}\left[U_{\un{x}_2}^{\dagger} T^a U_{\un{x}_0}\right] \right\rangle (z') .
\end{equation}
Diagram $D'$ can be obtained from the diagram $E'$ by replacing $\un x_1 \to \un x_0$ in the kernel and changing the overall sign.  
Summing up the diagrams $D'$ and $E'$, one obtains
\begin{equation}\label{eq:D+E_exp}
\begin{split}
D'+E'= &\frac{-g^2}{2N_c}\, \frac{-i g}{4\pi^3} \int\limits^z_{\Lambda^2/s} \frac{dz'}{z'} \int d^2x_2 \, \frac{2x_{20}^ix_{20}^l-\delta^{il}x_{20}^2}{x_{20}^4} \left( -\frac{x_{21}^l}{x_{21}^2} +\frac{x_{20}^l}{x_{20}^2}\right) \,  \int_{-\infty}^{\infty} dx_1^- \, \left\langle U^{hc}_{\un{x}_1}[\infty, x_1^-]\partial^j A^{c +} (x_1^-,\un{x}_1) \right. \\
&\qquad \times \left. \mathrm{Tr}\left[U_{\un{x}_2}^{\dagger} T^a U_{\un{x}_0}\right] \right\rangle (z') . 
\end{split}
\end{equation}

One can further simplify the Wilson line structure in \eq{eq:D+E_exp}. First note that
\begin{equation}
2 \, \mathrm{tr}\left[V_{\un{x}_1}\partial^j V^{\dagger}_{\un{x}_1} t^d\right] = -ig\int_{-\infty}^{\infty} dx_1^- U_{\un{x}_1}^{\dagger cd}[\infty, x_1^-] \, \partial^j A^{c +} (x_1^-, \un{x}_1) .
\end{equation}
Therefore, the Wilson-line structure in Eq.~\eqref{eq:D+E_exp} is proportional to
\begin{equation}
\mathrm{tr}\left[V_{\un{x}_1}\partial^j V^{\dagger}_{\un{x}_1} t^a\right]\mathrm{Tr}\left[U_{\un{x}_2}^{\dagger} T^a U_{\un{x}_0}\right]
=\frac{1}{2}\mathrm{tr}\left[V_{\un{x}_1}\partial^j V^{\dagger}_{\un{x}_1}V_{\un{x}_0}V_{\un{x}_2}^{\dagger}\right]\mathrm{tr}\left[V_{\un{x}_0}^{\dagger}V_{\un{x}_2}\right] - \frac{1}{2}\mathrm{tr}\left[V_{\un{x}_1}\partial^j V^{\dagger}_{\un{x}_1} V_{\un{x}_2}V^{\dagger}_{\un{x}_0}\right] \mathrm{tr}\left[V_{\un{x}_2}^{\dagger}V_{\un{x}_0}\right].
\end{equation}
The kernel function in Eq.~\eqref{eq:D+E_exp} can also be rearranged as 
\begin{equation}
 \frac{2x_{20}^ix_{20}^l-\delta^{il}x_{20}^2}{x_{20}^4}\left( -\frac{x_{21}^l}{x_{21}^2} +\frac{x_{20}^l}{x_{20}^2}\right)=\frac{x_{10}^2}{x_{20}^2x_{21}^2} \left(-\frac{x_{10}^i}{x_{10}^2} + \frac{x_{20}^i}{x_{20}^2}\right).
\end{equation}
Using these expressions, we see that Eq.~\eqref{eq:D+E_exp} becomes
\begin{equation}\label{eq:final_D+E}
\begin{split}
D'+E'= &\frac{-\alpha_s}{2\pi^2N_c} \, \int\limits^z_{\Lambda^2/s} \frac{dz'}{z'} \int d^2x_2 \, \frac{x_{10}^2}{x_{20}^2\, x_{21}^2} \left(-\frac{x_{10}^i}{x_{10}^2} + \frac{x_{20}^i}{x_{20}^2}\right)\\
&\times \left\langle \mathrm{tr}\left[V_{\un{x}_1}\partial^j V^{\dagger}_{\un{x}_1}V_{\un{x}_0}V_{\un{x}_2}^{\dagger}\right]\mathrm{tr}\left[V_{\un{x}_0}^{\dagger}V_{\un{x}_2}\right] -\mathrm{tr}\left[V_{\un{x}_1}\partial^j V^{\dagger}_{\un{x}_1} V_{\un{x}_2}V^{\dagger}_{\un{x}_0}\right] \mathrm{tr}\left[V_{\un{x}_2}^{\dagger}V_{\un{x}_0}\right] \right\rangle (z') .
\end{split}
\end{equation}
Diagrams $G'$ and $H'$ can be obtained from the expression for diagrams $D'+E'$ by interchanging the transverse positions $\un{x}_0\leftrightarrow \un{x}_1$ and spatial indices $i\leftrightarrow j$,  
\begin{equation}\label{eq:final_G+H}
\begin{split}
G'+H'= &\frac{-\alpha_s}{2\pi^2N_c} \int\limits^z_{\Lambda^2/s} \frac{dz'}{z'} \int d^2x_2 \, \frac{x_{10}^2}{x_{20}^2\, x_{21}^2} \left(\frac{x_{10}^j}{x_{10}^2} + \frac{x_{21}^j}{x_{21}^2}\right)\\
&\qquad \times \left\langle \mathrm{tr}\left[V_{\un{x}_0}\partial^i V^{\dagger}_{\un{x}_0}V_{\un{x}_1}V_{\un{x}_2}^{\dagger}\right]\mathrm{tr}\left[V_{\un{x}_1}^{\dagger}V_{\un{x}_2}\right] - \mathrm{tr}\left[V_{\un{x}_0}\partial^i V^{\dagger}_{\un{x}_0} V_{\un{x}_2}V^{\dagger}_{\un{x}_1}\right] \mathrm{tr}\left[V_{\un{x}_2}^{\dagger}V_{\un{x}_1}\right]\right\rangle (z').
\end{split}
\end{equation}
Diagram $F'$ and diagram $I'$, along with the other diagrams where one field strength tensor ($F^{i+}$) is located to the right of the shock wave while another one is in the shock wave, vanish because of the vanishing color factors.


\subsection{Diagrams $J', K', L', M'$}

\begin{figure}[!t]
    \centering
    \includegraphics[width=0.9\textwidth]{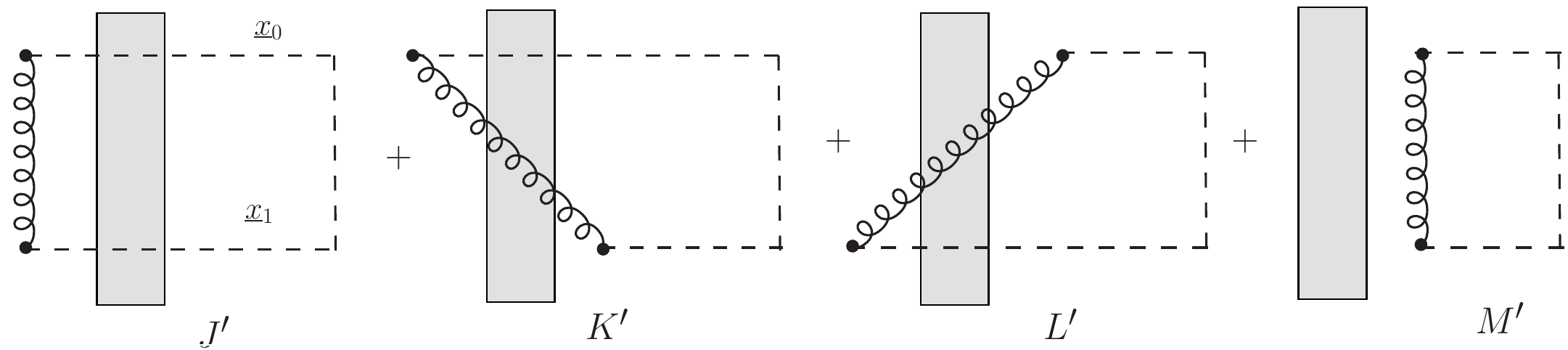}
    \caption{Diagrams for one step evolution of $\mathcal{O}^{ij}_{01}$. In this subset of diagrams, both background gluon field strengths are located outside the shockwave.}
	\label{fig:WW_diagrams_3}
\end{figure}

For the diagram $K'$, both field strengths become quantum fields. The one with the transverse position $\un{x}_0$ is located before the shockwave, while the one with the transverse position $\un{x}_1$ is located after the shockwave. Once again, employing the propagator \eqref{eik}, we obtain
\begin{equation}\label{eq:diagramK_exp}
\begin{split}
K' =&\frac{\alpha_s}{2\pi^2 N_c} \int\limits^z_{\Lambda^2/s} \frac{dz'}{z'} \int d^2x_2\, \left( \frac{\delta^{il}}{x_{20}^2} - \frac{2x_{20}^i x_{20}^l}{x_{20}^4}\right) \left( \frac{\delta^{jl}}{x_{21}^2} - \frac{2x_{21}^jx_{21}^l}{x_{21}^4}\right) \left\langle \mathrm{Tr}\left[U^{\dagger}_{\un{x}_2}U_{\un{x}_0}\right] \right\rangle (z') \\
=&\frac{\alpha_s}{2\pi^2 N_c} \int\limits^z_{\Lambda^2/s} \frac{dz'}{z'} \int d^2x_2 \, \left( \frac{\delta^{il}}{x_{20}^2} - \frac{2x_{20}^i x_{20}^l}{x_{20}^4}\right) \left( \frac{\delta^{jl}}{x_{21}^2} - \frac{2x_{21}^jx_{21}^l}{x_{21}^4}\right) \left\langle \mathrm{tr}\left[V_{\un{x}_2}V^{\dagger}_{\un{x}_0}\right]\mathrm{tr}\left[V_{\un{x}_0} V^{\dagger}_{\un{x}_2}\right] - 1 \right\rangle (z') .
\end{split}
\end{equation}
Here we have expressed the trace of two Wilson lines in adjoint representation in terms of the Wilson lines in fundamental representation using
\begin{equation}
\mathrm{Tr}\left[U^{\dagger}_{\un{x}_2}U_{\un{x}_0}\right]=\mathrm{tr}\left[V_{\un{x}_2}V^{\dagger}_{\un{x}_0}\right]\mathrm{tr}\left[V_{\un{x}_0} V^{\dagger}_{\un{x}_2}\right] - 1.
\end{equation}
The expression for the diagram $L'$ can be obtained by interchanging transverse positions $\un{x}_0\leftrightarrow \un{x}_1$ and spatial indices $i\leftrightarrow j$ in the diagram $K'$ from Eq.~\eqref{eq:diagramK_exp}:
\begin{equation}
L' = \frac{\alpha_s}{2\pi^2 N_c} \int\limits^z_{\Lambda^2/s} \frac{dz'}{z'} \int d^2x_2\, \left( \frac{\delta^{il}}{x_{20}^2} - \frac{2x_{20}^i x_{20}^l}{x_{20}^4}\right) \left( \frac{\delta^{jl}}{x_{21}^2} - \frac{2x_{21}^jx_{21}^l}{x_{21}^4}\right) \left\langle \mathrm{tr}\left[V_{\un{x}_2}V^{\dagger}_{\un{x}_1}\right]\mathrm{tr}\left[V_{\un{x}_1} V^{\dagger}_{\un{x}_2}\right] - 1\right\rangle (z') .
\end{equation}

We now calculate diagrams $J'$ and $M'$. Since the gluon in those diagrams does not interact with the shock wave, its propagator can be obtained from that in \eq{eik} by replacing $U^{ba}_{{\un x}_2} \to \delta^{ba}$ in it and by modifying the $x^-$-integration limits, which results in the overall sign change. Specifically, the gluon propagator in the diagrams $J'$ is
\begin{align}\label{eik2}
\int\limits_{-\infty}^0 dx_{2'}^- \,  
\int\limits^0_{-\infty} dx_2^- \, 
\contraction[2ex]
{}
{a}{^{+ \, a}
(x_{2'}^- , \ul{x}_1) \Big] \:}
{a}
\: 
a^{+ \, a} (x_{2'}^- , \ul{x}_1) \:
a^{\, + \, b} (x_2^- , \ul{x}_0)   = \delta^{ba} \, \frac{1}{4 \pi^3} \, \int\limits^z_{\Lambda^2/s} \frac{d z'}{z'} \, \int d^2 x_2 \, \frac{{\un x}_{21} \cdot {\un x}_{20}}{x_{21}^2 \, x_{20}^2} .
\end{align}
Contribution of the diagram $J'$ is (cf. \eq{eq:diagramK_exp})
\begin{equation}\label{J'}
\begin{split}
J' = -\frac{\alpha_s }{2\pi^2N_c} \int\limits^z_{\Lambda^2/s} \frac{dz'}{z'} \int d^2x_2 \left( \frac{\delta^{il}}{x_{20}^2} - \frac{2x_{20}^i x_{20}^l}{x_{20}^4}\right) \left( \frac{\delta^{jl}}{x_{21}^2} - \frac{2x_{21}^jx_{21}^l}{x_{21}^4}\right) \left\langle \mathrm{tr}\left[V_{\un{x}_0}V^{\dagger}_{\un{x}_1}\right]\mathrm{tr}\left[V_{\un{x}_1} V^{\dagger}_{\un{x}_0}\right] - 1\right\rangle . 
\end{split}
\end{equation}
Diagram $M'$ can be similarly computed (or it can be obtained from \eq{J'} by replacing all Wilson lines in it by unity operators): this yields
\begin{equation}
\begin{split}
M' =
-\frac{\alpha_s }{2\pi^2N_c} \int\limits^z_{\Lambda^2/s} \frac{dz'}{z'} \int d^2x_2 \left( \frac{\delta^{il}}{x_{20}^2} - \frac{2x_{20}^i x_{20}^l}{x_{20}^4}\right) \left( \frac{\delta^{jl}}{x_{21}^2} - \frac{2x_{21}^jx_{21}^l}{x_{21}^4}\right) (N_c^2-1).
\end{split}
\end{equation}

Adding the four diagrams together, one gets
\begin{equation}\label{eq:final_J+K+L+M}
\begin{split}
&J'+K'+L'+M' = \frac{\alpha_s}{2\pi^2 N_c} \int\limits^z_{\Lambda^2/s} \frac{dz'}{z'} \, \int d^2x_2 \left( \frac{\delta^{il}}{x_{20}^2} - \frac{2x_{20}^i x_{20}^l}{x_{20}^4}\right) \left( \frac{\delta^{jl}}{x_{21}^2} - \frac{2x_{21}^jx_{21}^l}{x_{21}^4}\right)\\
&\qquad \times \left\langle \mathrm{tr}\left[V_{\un{x}_2}V^{\dagger}_{\un{x}_0}\right]\mathrm{tr}\left[V_{\un{x}_0} V^{\dagger}_{\un{x}_2}\right] + \mathrm{tr}\left[V_{\un{x}_2}V^{\dagger}_{\un{x}_1}\right]\mathrm{tr}\left[V_{\un{x}_1} V^{\dagger}_{\un{x}_2}\right] - \mathrm{tr}\left[V_{\un{x}_0}V^{\dagger}_{\un{x}_1}\right]\mathrm{tr}\left[V_{\un{x}_1} V^{\dagger}_{\un{x}_0}\right]  - N_c^2 \right\rangle (z') .
\end{split}
\end{equation}


\subsection{Evolution equation}

Collecting Eqs.~\eqref{eq:final_A+B+C},~\eqref{eq:final_D+E},~\eqref{eq:final_G+H},~\eqref{eq:final_J+K+L+M}, and differentiating both sides of the resulting equation with respect to rapidity $Y \sim \ln (z s)$, one obtains the final result for the small-$x$ evolution of $\mathcal{O}^{ij}_{01}$, the correlator related to the unpolarized WW gluon distribution:  
\begin{align}\label{eq:final_WW_eovlution}
\frac{d}{dY}O^{ij}_{01} (Y) = &- \frac{\alpha_sN_c}{2\pi^2}  \int d^2x_2 \, \frac{x_{10}^2}{x_{20}^2\,x_{21}^2} \, O^{ij}_{01} (Y) \\ 
&-\frac{\alpha_s}{2\pi^2N_c}\int d^2x_2 \, \frac{x_{10}^2}{x_{20}^2\, x_{21}^2} \left(-\frac{x_{10}^i}{x_{10}^2} + \frac{x_{20}^i}{x_{20}^2}\right)\notag \\
&\qquad \times \left\langle \mathrm{tr}\left[V_{\un{x}_1}\partial^j V^{\dagger}_{\un{x}_1}V_{\un{x}_0}V_{\un{x}_2}^{\dagger}\right]\mathrm{tr}\left[V_{\un{x}_0}^{\dagger}V_{\un{x}_2}\right] -\mathrm{tr}\left[V_{\un{x}_1}\partial^j V^{\dagger}_{\un{x}_1} V_{\un{x}_2}V^{\dagger}_{\un{x}_0}\right] \mathrm{tr}\left[V_{\un{x}_2}^{\dagger}V_{\un{x}_0}\right]\right\rangle (Y) \notag \\ 
&-\frac{\alpha_s}{2\pi^2N_c} \int d^2x_2 \, \frac{x_{10}^2}{x_{20}^2\, x_{21}^2} \left(\frac{x_{10}^j}{x_{10}^2} + \frac{x_{21}^j}{x_{21}^2}\right)\notag\\
&\qquad \times \left\langle \mathrm{tr}\left[V_{\un{x}_0}\partial^j V^{\dagger}_{\un{x}_0}V_{\un{x}_1}V_{\un{x}_2}^{\dagger}\right]\mathrm{tr}\left[V_{\un{x}_1}^{\dagger}V_{\un{x}_2}\right] - \mathrm{tr}\left[V_{\un{x}_0}\partial^j V^{\dagger}_{\un{x}_0} V_{\un{x}_2}V^{\dagger}_{\un{x}_1}\right] \mathrm{tr}\left[V_{\un{x}_2}^{\dagger}V_{\un{x}_1}\right]\right\rangle (Y) \notag \\ 
&+\frac{\alpha_s}{2\pi^2 N_c} \int d^2x_2 \left( \frac{\delta^{il}}{x_{20}^2} - \frac{2x_{20}^i x_{20}^l}{x_{20}^4}\right) \left( \frac{\delta^{jl}}{x_{21}^2} - \frac{2x_{21}^jx_{21}^l}{x_{21}^4}\right) \notag \\
&\qquad \times \left\langle \mathrm{tr}\left[V_{\un{x}_2}V^{\dagger}_{\un{x}_0}\right]\mathrm{tr}\left[V_{\un{x}_0} V^{\dagger}_{\un{x}_2}\right] + \mathrm{tr}\left[V_{\un{x}_2}V^{\dagger}_{\un{x}_1}\right]\mathrm{tr}\left[V_{\un{x}_1} V^{\dagger}_{\un{x}_2}\right] - \mathrm{tr}\left[V_{\un{x}_0}V^{\dagger}_{\un{x}_1}\right]\mathrm{tr}\left[V_{\un{x}_1} V^{\dagger}_{\un{x}_0}\right]  - N_c^2 \right\rangle (Y) . \notag
\end{align}
This equation is exactly the same as the evolution equation obtained in \cite{Dominguez:2011br, Dominguez:2011gc}. The advantage of our direct approach here is that now each term is clearly associated with the corresponding Feynman diagram.

Equation \eqref{eq:final_WW_eovlution} is not closed. While on the left-hand side it contains the operator $O^{ij}_{01}$, on the right-hand side of the evolution equation,  
there appear new objects like
\begin{equation}\label{obj}
\mathrm{tr}\left[V_{\un{x}_1}\partial^j V^{\dagger}_{\un{x}_1}V_{\un{x}_0}V_{\un{x}_2}^{\dagger}\right]
\end{equation}
with three different transverse coordinates. They are neither dipoles nor quadrupoles. In order to attempt to obtain a closed set of evolution equations, one should also derive a small-$x$ evolution equation for the new object in \eq{obj}: it is likely that such equation, combined with \eq{eq:final_WW_eovlution} and the (BK) evolution for the unpolarized dipole amplitude will form a closed set of equation at large $N_c$. For instance, in \cite{Jalilian-Marian:2004vhw} an equation for a (different) 3-point function made out of light-cone Wilson lines was constructed and, when combined with the dipole and quadrupole evolution, yielded a closed system of equations (at large $N_c$).

One can readily verify that \eq{eq:final_WW_eovlution} is UV-finite and, therefore, contains no double-logarithmic contributions, unlike the helicity evolution considered in the main text. We note that the integral over $\un{x}_2$ in the second term on the right of \eq{eq:final_WW_eovlution}
seems to be dominated by the UV-divergent region $\un{x}_{2}\rightarrow \un{x}_0$.  Using the Taylor expansions for the corresponding Wilson line structure we write
\begin{align}
& \left\langle \mathrm{tr}\left[V_{\un{x}_1}\partial^j V^{\dagger}_{\un{x}_1}V_{\un{x}_0}V_{\un{x}_2}^{\dagger}\right]\mathrm{tr}\left[V_{\un{x}_0}^{\dagger}V_{\un{x}_2}\right] - \mathrm{tr}\left[V_{\un{x}_1}\partial^j V^{\dagger}_{\un{x}_1} V_{\un{x}_2}V^{\dagger}_{\un{x}_0}\right] \mathrm{tr}\left[V_{\un{x}_2}^{\dagger}V_{\un{x}_0}\right] \right\rangle \notag\\
= &N_c \left\langle - \mathrm{tr}\left[V_{\un{x}_1}\partial^j V^{\dagger}_{\un{x}_1}V_{\un{x}_0}\partial^l V_{\un{x}_0}^{\dagger}\right] \, {x}_{20}^l + \mathrm{tr}\left[V_{\un{x}_1}\partial^j V^{\dagger}_{\un{x}_1} \partial^lV_{\un{x}_0}V^{\dagger}_{\un{x}_0}\right] \, {x}_{20}^l \right\rangle  + \mathcal{O}(x_{20}^2) \notag\\
=&- 2N_c \, x_{20}^l\, \left\langle  \mathrm{tr}\left[V_{\un{x}_1}\partial^j V^{\dagger}_{\un{x}_1}V_{\un{x}_0}\partial^l V_{\un{x}_0}^{\dagger}\right] \right\rangle + \mathcal{O}(x_{20}^2) \notag\\
=&- 2N_c^2 \, x_{20}^l \, O^{lj}_{01}+ \mathcal{O}(x_{20}^2).
\end{align}
This gives, for the leading $\un{x}_{2}\rightarrow \un{x}_0$ divergence in the second term on the right of \eq{eq:final_WW_eovlution},
\begin{equation}
-\frac{\alpha_s}{2\pi^2N_c}\int d^2x_2\, \frac{x_{10}^2}{x_{20}^2 x_{21}^2} \left(-\frac{x_{10}^i}{x_{10}^2} + \frac{x_{20}^i}{x_{20}^2}\right) \, (-2 \, N_c^2 ) \, x_{20}^l \, O^{lj}_{01} (Y) \Bigg|_{\un{x}_{2}\to \un{x}_0}
=\frac{\alpha_sN_c}{2\pi^2} \int d^2x_2 \, \frac{1}{x_{20}^2}\,  O^{ij}_{01} (Y) ,
\end{equation}
where we have averaged over the angles of $\un x_{20}$. This expression cancels the potentially UV-divergent (when $\un{x}_{2}\rightarrow \un{x}_0$) double-logarithmic contribution from the first term on the right of \eq{eq:final_WW_eovlution}. The same analysis and conclusion apply to the UV region $\un{x}_2\rightarrow \un{x}_1$, where the divergent terms in the first and third terms on the right of \eq{eq:final_WW_eovlution} cancel each other. We conclude that the evolution equation \eqref{eq:final_WW_eovlution} has no UV divergences in its kernel: therefore, it is entirely within the single-logarithmic approximation.

\bibliography{softgluon,references}
\end{document}